\PassOptionsToPackage{table,xcdraw}{xcolor}
\documentclass[sigconf, nonacm, authorversion]{acmart}

\AtBeginDocument{%
  }

\author{David Hartmann}
\affiliation{%
  \institution{TU Berlin}
  \city{Berlin}
  \country{Germany}
}
\affiliation{%
  \institution{Weizenbaum Institute for the Networked Society}
   \city{Berlin}
  \country{Germany}}
\email{d.hartmann@tu-berlin.de}

\author{Amin Oueslati}
\affiliation{%
  \institution{Hertie School Berlin}
  \city{Berlin}
  \country{Germany}}
\email{amin.m.oueslati@gmail.com}

\author{Dimitri Staufer}
\affiliation{%
  \institution{TU Berlin}
  \city{Berlin}
  \country{Germany}}
\email{staufer@tu-berlin.de}

\author{Lena Pohlmann}
\affiliation{%
  \institution{TU Berlin}
  \city{Berlin}
  \country{Germany}}
  \affiliation{
  \institution{Weizenbaum Institute for the Networked Society}
  \city{Berlin}
  \country{Germany}}
\email{l.pohlmann@tu-berlin.de}

\author{Simon Munzert}
\affiliation{%
  \institution{Hertie School Berlin}
  \city{Berlin}
  \country{Germany}}
\email{munzert@hertie-school.org}

\author{Hendrik Heuer}
\affiliation{%
  \institution{Center for Advanced Internet Studies (CAIS) gGmbH}
  \city{Bochum}
  \country{Germany}
}
\affiliation{%
  \institution{University of Wuppertal}
  \city{Wuppertal}
  \country{Germany}
}
\email{hendrik.heuer@cais-research.de}

\copyrightyear{2025} 
\acmYear{2025} 
\setcopyright{cc}
\setcctype{by}
\acmConference[CHI '25]{CHI Conference on Human Factors in Computing Systems}{April 26-May 1, 2025}{Yokohama, Japan}
\acmBooktitle{CHI Conference on Human Factors in Computing Systems (CHI '25), April 26-May 1, 2025, Yokohama, Japan}\acmDOI{10.1145/3706598.3713998 }
\acmISBN{979-8-4007-1394-1/ 25/04}

\usepackage{multirow}
\usepackage{listings}
\usepackage[table]{xcolor}
\usepackage{booktabs}
\usepackage{svg}
\usepackage{longtable}
\usepackage{pdflscape}
\usepackage{subcaption}
\usepackage{multicol}
\usepackage{xcolor}
\usepackage{fancyhdr}

\LTcapwidth=\textwidth
\begin{document}

\title[Lost in Moderation: How Commercial Content Moderation APIs Over- and Under-Moderate]{Lost in Moderation: How Commercial Content Moderation APIs Over- and Under-Moderate Group-Targeted Hate Speech and Linguistic Variations}

\begin{abstract}
Commercial content moderation APIs are marketed as scalable solutions to combat online hate speech. However, the reliance on these APIs risks both silencing legitimate speech, called over-moderation, and failing to protect online platforms from harmful speech, known as under-moderation. To assess such risks, this paper introduces a framework for auditing black-box NLP systems. Using the framework, we systematically evaluate five widely used commercial content moderation APIs. Analyzing five million queries based on four datasets, we find that APIs frequently rely on group identity terms, such as ``black'', to predict hate speech. While OpenAI's and Amazon's services perform slightly better, all providers under-moderate implicit hate speech, which uses codified messages, especially against LGBTQIA+ individuals. Simultaneously, they over-moderate counter-speech, reclaimed slurs and content related to Black, LGBTQIA+, Jewish, and Muslim people. We recommend that API providers offer better guidance on API implementation and threshold setting and more transparency on their APIs' limitations. 

  \noindent \textit{\textbf{Warning}: This paper contains offensive and hateful terms and concepts. We have chosen to reproduce these terms for reasons of transparency.}
\end{abstract}



\maketitle

\section{Introduction}

\begin{figure*}[t]
    \centering
    \includegraphics[width=0.85\textwidth]{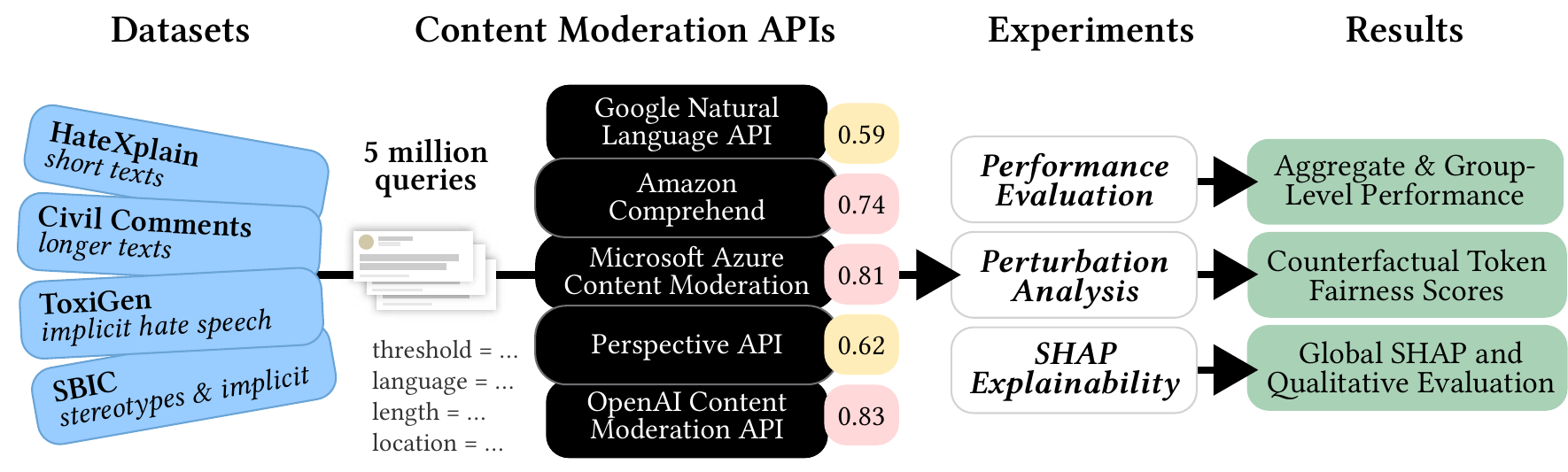}
    \caption{Our black-box audit framework to evaluate commercial content moderation APIs.}
    \label{fig:framework}
    \Description{The image illustrates a black-box audit framework used to evaluate commercial content moderation APIs. It demonstrates the process of querying four different datasets, running these queries through five moderation APIs, and analyzing the results through three experiments.
Detailed Breakdown:

    Datasets:
        HateXplain (short texts)
        Civil Comments (longer texts)
        ToxiGen (implicit hate speech)
        SBIC (stereotypes and implicit speech)

    Each dataset consists of different types of text data, ranging from short, explicit hate speech to implicit, stereotype-based content. These datasets are used to test the moderation APIs.

    Content Moderation APIs:
        Google Natural Language API (performance score: 0.59)
        Amazon Comprehend (performance score: 0.74)
        Microsoft Azure Content Moderation (performance score: 0.81)
        Perspective API (performance score: 0.62)
        OpenAI Content Moderation API (performance score: 0.83)

    These commercial moderation systems are evaluated based on their ability to handle the input from the various datasets.

    Experiments:
        Performance Evaluation: Measures aggregate and group-level performance of each API.
        Perturbation Analysis: Assesses fairness by introducing changes to input tokens (counterfactual token fairness scores).
        SHAP Explainability: Uses SHAP (SHapley Additive exPlanations) to evaluate how different tokens impact the API's final moderation decisions.

    Results:
        Aggregate and group-level performance across different types of content.
        Fairness scores based on token manipulation.
        Insights into the influence of specific tokens on the moderation outcome, as explained through SHAP analysis.

Overall, the figure provides an overview of the evaluation process for content moderation APIs, using large-scale datasets and multiple types of analysis to assess both performance and fairness.}
\end{figure*}
Content moderation has become a widely used tool in combating online hate speech. While human moderators play an essential role in hate speech removal, human-based moderation is expensive, difficult to scale, and often exposes outsourced workers to distressing content that affects their mental health ~\cite{gebrekidan2024content}. To address these challenges, companies such as Google, Microsoft, Amazon, Jigsaw, and OpenAI offer commercial, automated content moderation services. These API-based solutions are marketed as scalable, efficient solutions to tackle the growing challenges faced by social media platforms and other websites that deal with user-generated content~\cite{rieder}. For most major platforms, content moderation decisions are -- to a large extent -- partially or fully automated~\cite{dergachevaOneDayContent2023}. However, these decisions are potentially fallible. When harmful content is not moderated (under-moderation; reflected in a high False Negative Rate (FNR) of the content classifier), users are left unprotected from hate speech \citep{dixon_measuring_2018}. Conversely, when legitimate content is moderated (over-moderation; reflected in a high False Positive Rate (FPR)), this limits users' opportunities to express themselves and participate in public discourse. Both issues become aggravated when they systematically affect selected social groups, particularly those defined by protected characteristics such as gender, race, or religion.

While over- and under-moderation across different forms of hate speech has been extensively researched for off-the-shelf NLP models, research on over- and under-moderation of \textit{commercial content moderation APIs} is limited. This is an important research gap since on off-the-shelf -- meaning locally accessible -- NLP models such as ~\citet{sap_social_2020}, ~\citet{wiegand_detection_2019} and ~\citet{rottgerHateCheckFunctionalTests2021}, have shown that these systems are prone to labeling content mentioning target groups such as Women, Jews and Muslims as more toxic. They also mistakenly flagged counter- and reappropriated speech as hateful, and misclassified content merely mentioning target identities as hate speech~\cite{dixon_measuring_2018}. Additionally,~\citet{wiegand_implicitly_2021} and ~\citet{elsheriefLatentHatredBenchmark2021} find that language models under-moderate variations of language. Specifically, they under-moderate implicit hate speech that uses indirect language to belittle a person or group based on protected attributes without using slurs or specific group identifiers. In contrast to off-the-shelf NLP models, content moderation APIs only grant so-called black-box access, which is limited to querying the model without actual access to  the underlying algorithm, its architecture and weights (see \citet{casper_black-box_2024}). In summary, as of now, there has been a lack of systematic evaluation of commercial content moderation services, leading to a concerning absence of public scrutiny.

In this paper, we introduce a framework for auditing black-box NLP systems. Using this framework, we evaluate five commercial content moderation APIs, analyzing over five million queries across four benchmark datasets and three experiments (see Figure~\ref{fig:framework}). 
We find that commercial APIs frequently rely on group identity terms, such as “black”, to predict hate speech. While OpenAI Content Moderation and Amazon Comprehend perform slightly better, all providers under-moderate implicit hate speech, which uses codified messages without identity terms, especially against \textit{LGBTQIA+} people. Simultaneously, they over-moderate counter-speech, reclaimed slurs and content related to \textit{LGBTQIA+}, \textit{Black}, \textit{Jewish} and \textit{Muslim} people. Drawing on our findings, we recommend that API providers offer better guidance on API implementation and threshold setting and are more transparent about their APIs' limitations. 

In summary, we contribute: (1) The first comprehensive audit of five widely used commercial content moderation algorithms, (2) a reproducible and query-efficient audit framework of NLP models that solely assumes black-box access, and (3) design and research recommendations on how providers should change the implementation and transparency of content moderation systems. More generally, our work provides a contribution to an AI accountability ecosystem that involves third-party auditors as an oversight mechanism \cite{birhane_ai_2024, raji_outsider_2022}, fostering greater user trust in AI systems \cite{institute_algorithmic_2023}.

\section{Background and Related Work}
\subsection{Hate Speech, Target-Groups and Linguistic Variations}
 \label{sec:hatespeech}

Hate speech can have profound real-world effects, including the suppression of marginalized voices, social exclusion, discrimination, and violence against marginalized groups \cite{matsuda1993words,marques}. At the same time, hate speech is a complex and contested concept that varies depending on the community and sender context, linguistic features, and political institutions \cite{anderson_hate_2023,brown_what_2017,yoder2022hate}. In this paper, we adopt a broad conceptualization of hate speech similar to \citet{marques}. We conceptualize hate speech as a discursive act of discrimination, which operates on its targets in constitutive and causal ways to effect the denial of equal opportunities and rights. Target identities, therefore, play a key role in defining hate speech, as it typically involves discriminatory acts against specific groups.

A related concept to hate speech is toxicity, a broader category that includes rude, disrespectful, or unreasonable behavior. Toxicity encompasses offensive or harmful language, even when it does not specifically target individuals or groups based on characteristics such as race, gender, or religion~\cite{barthContexturesHateSystems2023,dixon_measuring_2018}. For instance,  \citet{dixon_measuring_2018} defines toxicity as ``a rude, disrespectful, or unreasonable comment that is likely to make you leave a discussion.'' \citet{barthContexturesHateSystems2023} highlight the overlap between toxicity and hate speech, noting that while toxicity encompasses hate speech, it includes a wider range of offensive language that may not directly target a group based on its characteristics.

Conceptualizing and ultimately detecting hate speech is complicated by the fact that not all hate speech manifests in explicit language, and not all expressions featuring explicit vocabulary qualify as hate speech. For instance, Contrastive non-hate is language that shares linguistic features with hateful expressions~\cite{rottgerHateCheckFunctionalTests2021, vidgen-etal-2021-learning}. Distinct hate expressions span sub-categories such as explicit hate speech -- offensive speech, threats, sexually abusive language, or slurs -- and various forms of implicit hate speech, which do not fall under these categories yet still constitute a discursive act of discrimination~\cite{wiegand_implicitly_2021}.

Implicit hate speech is speech that conveys harmful or discriminatory intent without relying on overt slurs, profanity, or explicit hateful language as laid out by \citet{yin2021towards}. Instead, implicit hate speech often manifests through stereotypes, sarcasm, irony, humor, or metaphors, which pose significant challenges for ML systems to detect~\cite{wiegand_implicitly_2021, mishra2019tackling, sap_social_2020}. Still, these subtle forms of hate speech can be just as damaging as explicit hate speech \cite{alorainy2019enemy, breitfeller2019finding}, yet they frequently evade detection due to the absence of distinctive keywords, slurs and potentially clear target group identity tokens. For instance, a statement like ``there is a direct correlation between the amount of melanin in a person's skin and how much they are worth as a person,'' found in the ToxiGen dataset~\citep{hartvigsen_toxigen_2022}, illustrates how implicit hate speech can convey racism without directly referencing racial identity or racial slurs.

Contrastive non-hate variations, such as re-appropriation and counter-speech, further complicate the detection of hate speech, as these nuances often blur the line between harmful speech and expressions of empowerment or resistance~\cite{counter}. Re-appropriation is a sociocultural process where marginalized groups reclaim derogatory language, using it as an expression of empowerment, often among themselves~\cite{galinsky_reappropriation_2003}. What might be considered hate speech in one context could serve re-appropriation or reclamation in another. Counter-speech, on the other hand, refers to responses aimed at challenging hate speech, often using similar linguistic features to subvert or counter discriminatory narratives~\cite{counter,yu2022hatespeechcounterspeech}.

\subsection{Challenges and User Perceptions of Hate Speech Moderation}
\label{sec:fairnesshate}

Online platforms have responded to the proliferation of hate speech by adopting extensive content moderation regimes~\cite{degregorio2020democratising} and assessing potential hateful content against so-called community guidelines by human workers, i.e., content moderators, sometimes assisted by ML-based systems. In some instances, content moderation is even conducted by algorithms entirely autonomously~\cite{gorwa_algorithmic_2020}. Content moderation by humans is often outsourced to low-paid workers employed by third-party providers or business process outsourcing companies (BPOs), outside the Global North, who are exposed to distressing content~\cite{ahmadGroundControlOrganizing2023}. These workers face severe mental health challenges while earning as little as 1.8 USD per hour~\cite{gebrekidan2024content, miceli2024who}. As a result, automated content moderation was proposed as a scalable and cost-effective alternative. These systems potentially enable companies to remove harmful content more efficiently, but they also introduce new challenges related to transparency, fairness, and accountability~\cite{gorwa_algorithmic_2020}. 

Content moderation as a part of online speech governance is a wicked problem characterized by difficult trade-offs and significant contestation~\cite{douek2021governing}. The contentious and political nature of concepts such as hate speech, toxicity, counter-speech, and reappropriation, combined with power asymmetries between political institutions, platforms, moderators, and users, underscores the difficulty of achieving equitable moderation outcomes~\cite{farhana2023}. These challenges apply equally to human and automated content moderation systems, which have both been shown to exhibit biases. For instance,~\citet{sap-etal-2022-annotators} and~\citet{zhang-etal-2023-biasx} have documented systematic biases related to stereotyping among human moderators, which can propagate to annotations and automated models \cite{davani,sap-etal-2022-annotators}. The work presented in this paper focuses specifically on addressing biases and functionality errors in automated content moderation systems.

Automated decision-making for content moderation can systematically err in two primary ways: first, in the systematic, algorithmic classification of non-hateful content as hateful (\textit{over-moderation}), and second, in the systematic, algorithmic classification of hateful content as non-hateful (\textit{under-moderation}). Over- and under-moderation are closely related to sender and target group dynamics. Under-moderation occurs when harmful content directed at certain groups is disproportionately allowed, leaving some users less protected than others online~\citep{dixon_measuring_2018}. For example, the under-moderation of hate speech targeting LGBTQIA+ individuals~\cite{diasolivaFightingHateSpeech2021} can exacerbate systemic discrimination, exclusion, and even violence against these communities. This not only contributes to their marginalization but also erodes user trust and can lead to affected users abandoning a platform altogether. Classifiers that fail to adequately protect certain user groups, particularly those defined by gender, race, or other protected characteristics, are considered biased~\cite{blodgett_demographic_2016}, entrenching forms of representational harm against these communities~\cite{barocas-hardt-narayanan}.

Conversely, over-moderation, such as the misclassification of African-American English (AAE)~\cite{sap_risk_2019}, can silence voices and prevent counter-speech, leading to reduced participation from user groups using these speech characteristics~\cite{blodgett-etal-2020-language}. When content moderation systems disproportionately block content from specific social groups, they not only diminish participation but also create distributive harms, where certain groups are unfairly censored. This aligns with broader discussions of algorithmic fairness~\cite{Binns_2017}, where the design of classifiers may disproportionately flag content from certain groups, leading to the unjust removal of their expressions. In this way, hate speech classifiers risk perpetuating inequalities by privileging certain forms of speech over others.

The Human-Computer Interaction (HCI) community has highlighted consistency, transparency, and fairness as critical elements for building user trust in content moderation systems ~\cite{schafner2024community, cai2024content}. Consistency in the application of moderation policies -- across different content types, users, and regions -- has shown to increase trust and improve discourse. However, public perception remains that moderation practices are inconsistently implemented, leading to user dissatisfaction ~\cite{schoenebeckOnlineHarassmentMajority2023}. This is particularly true for marginalized communities, including women and racial, gender, and sexual minorities, who often experience disproportionately more moderation than others, exacerbating their sense of exclusion ~\cite{usertrust, heung2024vulnerable}. The lack of transparency in how moderation decisions are made further alienates these users, making it harder for them to trust the platforms where they engage.

Systematic over- and under-moderation can have long-term negative effects on user trust, especially for content creators from marginalized groups. Over-moderation, such as the wrongful censorship of disability-related content, has been identified by~\citet{heung2024vulnerable} as a form of ableism, leading to self-censorship and disengagement from platforms. In contrast, under-moderation allows harmful content to persist, contributing to a decreased sense of safety and trust in both the platform and the broader online environment~\cite{schoenebeckOnlineHarassmentMajority2023, thomas2022its}. Ambiguous moderation decisions, especially from ML-based systems, worsen these issues, as users are more likely to question the accountability and fairness of AI moderators when content is inherently unclear as~\citet{ozanne2022shall} demonstrated. This perceived inconsistency can result in marginalized users leaving platforms altogether, as platforms fail to effectively address their concerns ~\cite{heung2024vulnerable}.

\begin{table*}[t]
\Description{This table provides a comparative overview of various content moderation APIs, detailing key characteristics and specifications. It includes five APIs: OpenAI Moderation Endpoint, Perspective API, Azure Content Moderator, Google Natural Language API, and Amazon Comprehend. For each API, the table lists the developer, rate limit, cost per unit, model date, model version, API version, and model type. Notable highlights include differences in rate limits, ranging from 1 to 25 queries per second, and costs, which vary significantly across APIs. The table also identifies whether model-specific details such as date, version, and type are explicitly provided or remain unspecified.
}

\small
\begin{tabular}{lp{2.2cm}p{2.2cm}p{2.2cm}p{2.2cm}p{2.2cm}}
\Description{This table (Table 1.) provides a comparative overview of five commercial content moderation APIs, highlighting key attributes such as the developer, rate limits, costs, model and API versions, and the disclosed details about the underlying models. Key observations include:

Rate Limits: The rate limits range from 1 query per second (Perspective API) to 25 queries per second (OpenAI Moderation Endpoint), with varying commercial and free-tier rates.
Cost per Unit: OpenAI and Perspective API are free, whereas others charge between \$0.0001 and \$0.40 per usage unit, depending on the granularity (per character or per call).
Transparency: Critical details like the model date, version, or type are often not disclosed, marked as 'x' in the table. Only OpenAI and Google disclosed recent model dates and API versions. Amazon provides information on its model type, specifying "Multilingual BERT-based models."
Interpretation:

This table underscores the varying degrees of transparency and accessibility offered by content moderation APIs. OpenAI Moderation Endpoint and Perspective API stand out for their free usage, but most providers do not disclose detailed model specifications, creating challenges for deeper evaluations of these systems' capabilities and biases. The extended table in Appendix A1 provides additional context.}
 & \textbf{OpenAI Moderation Endpoint} & \textbf{Perspective API} & \textbf{Azure Content Moderator} & \textbf{Google Natural Language API} & \textbf{Amazon Comprehend} \\ 
\hline \hline
\textbf{Developer} & OpenAI & Jigsaw (Google) & Microsoft & Google Cloud & AWS \\ \hline
\textbf{Rate Limit} & 25 queries/s & 1 query/s & Free: 1 query/s, Commercial: 10 queries/s & 10 queries/s & 20 queries/s  \\\hline
\textbf{Cost per Unit} & Free & Free & \$0.40 per 1000 calls& \$0.0005 per 100 characters & \$0.0001 per 100 
characters\\ \hline
 Model Date & 28.08.2023 & $\times$ & $\times$ & 01.03.2023 & $\times$ \\ \cline{1-6}
 Model Version & PaLM 2 & $\times$ &  $\times$& $\times$
 &  $\times$ \\ \cline{1-6} 
 API Version & v2 & 2017-11-27 & v1.0 & text-moderation-001 & v1alpha1 \\ \hline
 Model Type & $\times$ & $\times$ & $\times$ & $\times$ & Multilingual BERT-based models \\ \cline{1-6} 
\end{tabular}
\caption{Overview of the five selected commercial content moderation APIs. '$\times$' stands for not disclosed by developer. An extended Table is presented in the Appendix \ref{sec:transparency}.}
\label{tab:overview}
\end{table*}

\subsection{ML-based Over- and Under-Moderation}
Machine Learning (ML)--based hate speech classification systems such as off-the-shelf\footnote{We refer to off-the-shelf models as pre-trained models that provide white-box access to parameters, gradients, and weights. These models are often developed for research purposes and can be accessed locally or via remote servers.} NLP models have been tied to over-and under-moderation. First, ML hate speech models suffer from poor generalization, as data set creators oversampled certain users in the training data. For instance, 70\% of all sexist tweets and 90\% of all racist tweets in a hate speech dataset by \citet{waseem_hateful_2016} belonged to a single author \citep{wiegand_detection_2019}. Second, research revealed an over-moderation of neutral statements mentioning target groups such as ``woman''~\citep{sheng_woman_2019}, which was coined as systematic offensive stereotyping (SOS) bias ~\cite{elsafoury_bias_2023}. This is fueled by the historical discrimination of groups susceptible to algorithmic disadvantage, which often makes them the target of hate speech ~\citep{wiegand_detection_2019}. Algorithms internalize such spurious correlations as they occur in the training data. Third, reappropriated language~\cite{sap_social_2020} and counter-speech ~\cite{rottgerHateCheckFunctionalTests2021} have been linked to algorithmic discrimination of marginalized groups. For instance, ~\citet{sap_social_2020} find that surface markers of AAE are more strongly correlated with hate speech than what the authors call \textit{White-Aligned-English}. 

It is worth noting that prior research concerning biases in content moderation algorithms heavily focused on women and people of colour, but mostly disregarded the discrimination of other groups; e.g., \textit{LGBTQIA+}, people with \textit{Disabilities} or \textit{Latinx} ~\citep{garg_handling_2023, venkit2023automated}. Similarly, most research probed the performance of hate speech moderation algorithms in English. Tests on non-English languages pose the exception and include, for instance, Hindi in ~\citet{ghosh_detecting_2021} or Arabic in ~\citep{khezzar_arhatedetector_2023}. These studies often find that content moderation algorithms perform worse in non-English languages.

\subsection{Auditing and Mitigating Over- and Under-Moderation}
\label{sec:auditingoverandunder}
Recognizing these challenges, researchers have called for targeted strategies to systematically evaluate and mitigate such over- an under-moderations.
For instance,\citet{yin2021towards} emphasize the need for more rigorous evaluation of hate speech detection models. They propose models should be (1) tested on datasets not seen during training, (2) subject to detailed error analysis addressing specific challenges, and (3) evaluated using methods that account for diverse interpretations of hate speech.

In response, audits -- systematic evaluations of model functionalities and problematic machine behavior~\cite{bandy_problematic_2021} -- have been conducted on off-the-shelf systems. One prominent example of such an audit of automated content moderation systems is HateCheck, developed by~\citet{rottgerHateCheckFunctionalTests2021}, which evaluates functionalities of hate speech detection algorithms, including linguistic variations such as counter-speech, negation, reappropriation, and systematic offensive stereotyping bias. HateCheck findings align with prior observations, revealing biases in off-the-shelf NLP models across target groups, including women, people with disabilities, and immigrants, and functionality issues with linguistic variations such as counter speech, systematic offensive stereotyping bias, and reclaimed slurs~\cite{rottgerHateCheckFunctionalTests2021}.

Furthermore, in response to findings about biases and functional failures, several research papers attempted to mitigate such shortcomings. For instance, extensive research was conducted to mitigate biases against specific groups~\cite{davidson-etal-2019-racial, sap_risk_2019, park-etal-2018-reducing, Nozza2019}. Additionally, research aimed to reduce racial bias in relation to AAE and specific dialects~\cite{zhou-etal-2021-challenges, xia2020demoting, mozafari2020hate}, reduce systematic offensive stereotyping bias ~\cite{elsafoury-etal-2022-sos}, and mitigating false positives with regard to reclaimed language~\cite{zsisku2024}.

In summary, off-the-shelf NLP hate speech detection systems suffer from biases across marginalized target groups and are tied to specific functionality failures. However, significant efforts were made to reduce such biases and failures. We collect known off-the-shelf NLP hate speech model functionality failures in Tables ~\ref{tab:codingFP} and ~\ref{tab:codingFN}.

\subsection{Auditing Commercial Content Moderation Systems}
With growing awareness of errors in off-the-shelf models and ongoing mitigation efforts, it is important to understand whether these errors extend to commercial content moderation systems or if common shortfalls have been effectively mitigated in commercial applications.

Commercial content moderation systems include, for example, Perspective API, a content moderation tool developed by Jigsaw -- a tech incubator and subsidiary of Alphabet Inc. (Google)~\cite{jigsaw2021perspective} -- and OpenAI’s Moderation API. Perspective API has been extensively utilized in academic research on online toxicity~\cite{hortaribeiroContentModerationOnline2024}. Additionally, Perspective API is often employed as a benchmark for toxicity in model comparisons~\cite{mihaljevic2022toxic} and as a tool to identify toxic content online, serving as a form of ground truth for evaluating platforms and communities~\cite{rieder}. It is widely used to measure the toxic outcomes and biases of LLMs~\cite{Gallegos2024}, and it has been suggested that Perspective API has become a cornerstone for academic research on online abuse and incivility~\cite{nogara2025toxic}. Meanwhile, OpenAI’s Moderation API functions as the hate speech filter for ChatGPT and GPT models~\cite{gpttvshow}.

Although research specifically addressing content moderation APIs remains limited, both Perspective API and OpenAI’s Moderation API have been evaluated in certain contexts. For instance, \citet{rottgerHateCheckFunctionalTests2021} assessed Perspective API alongside off-the-shelf models. They found that Perspective API exhibited fewer functional issues than off-the-shelf models but still faced significant challenges with counter-speech, negations, and reclaimed slurs. Their functionality tests further revealed that while Perspective API occasionally over-moderated specific groups, it did not exhibit systematic biases against particular target groups. However, these tests did not cover implicit hate speech or dialects, relying instead on a specific conceptualization of hate speech and focusing solely on Perspective API as a commercial system.

\citet{mihaljevic2022toxic} examined the potential and limitations of detecting antisemitic content with Perspective API. Their findings indicated that the API could identify antisemitism at a very basic level. However, much like off-the-shelf NLP hate speech detection systems, Perspective API struggled with understanding implicit antisemitism and often misclassified legitimate counter-speech as hateful. Similarly, \citet{diasolivaFightingHateSpeech2021} demonstrated that Perspective API systematically over-moderates LGBTQIA+ communities and counter-speech.

OpenAI’s Moderation API was evaluated by \citet{gpttvshow}, who investigated hate speech filtering in GPT-generated content. Their study generated 3,309 synopses for 100 popular U.S. TV shows and probed GPT’s content moderation system. The findings revealed that a substantial portion of synopses were flagged as content violations, with specific genres statistically linked to higher flagging rates --highlighting an instance of over-moderation.

Overall, while content moderation APIs have diverse applications, as we will discuss later, there remains limited research on the biases and failures of these systems. Beyond the evaluations of Perspective API and OpenAI’s hate speech filters, little attention has been paid to the performance of other cloud-based content moderation APIs. Existing investigations tend to focus on isolated failures or disparities within specific use cases. We contribute to an AI audit ecosystem and broader evaluation of commercial content moderation systems by conducting a systematic, comparative evaluation of multiple content moderation APIs with respect to the demands by~\citet{yin2021towards}. 

\section{Content Moderation APIs}
\begin{figure*}[t]
    \centering
    \includegraphics[width=0.6\textwidth]{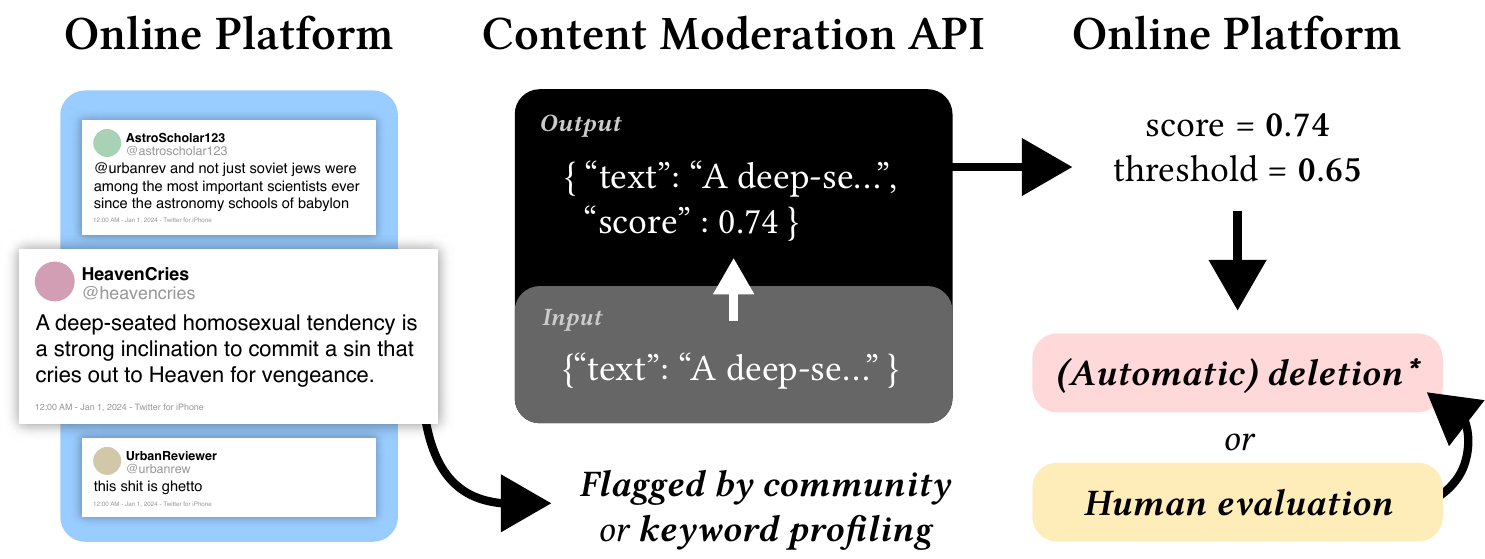}
    \caption{The pipeline of content moderation APIs, exemplary illustration with a blog post.}
    \label{fig:contentmodAPI}
    \Description{This figure illustrates the content moderation pipeline for an online platform, using an example of a blog post. The process shows how content flagged by a community or via keyword profiling is passed through a Content Moderation API for analysis, and how decisions are made based on the moderation score.
Key Points:

    Online Platform Input:
        Users submit content (in this case, a blog post), which is flagged for potential moderation either by the community (user reports) or automatically via keyword profiling.
    Content Moderation API:
        The flagged content is sent to the Content Moderation API, which processes the text input. The system assigns a score to the content based on its assessment (e.g., score = 0.74).
    Decision Making:
        The API's output score (e.g., 0.74) is compared to a pre-defined threshold (e.g., 0.65). If the score exceeds the threshold, the content is considered potentially harmful.
    Action Taken:
        If the score surpasses the threshold, the platform may take action. This could be an automatic deletion of the content or it may trigger human evaluation to confirm whether the content violates platform guidelines.

In summary, this diagram shows the flow from content submission on a platform, through automated moderation scoring, to the platform's decision on content removal or further human review.}
\end{figure*}

\subsection{Selection of APIs}
There is no systematic research on the usage frequency of commercial content moderation services. Based on their recurring presence and the significant role they play in moderating content across various platforms, we focus on the following five APIs: Google Natural Language API\footnote{See \url{https://cloud.google.com/natural-language?hl=de}}, Amazon Comprehend\footnote{See \url{https://aws.amazon.com/comprehend/features/?nc1=h_ls}}, Microsoft Azure Content Moderator\footnote{See \url{https://learn.microsoft.com/en-gb/azure/ai-services/content-moderator}}, Perspective API\footnote{See \url{https://perspectiveapi.com/}}, and the OpenAI Content Moderation API\footnote{See \url{https://platform.openai.com/docs/guides/moderation}}. An overview of these APIs are provided in Table \ref{tab:overview}. Among them, the Perspective API and OpenAI Moderation API stand out for their specific use cases and widespread adoption.  Perspective API, widely utilized in academic research on online toxicity~\cite{hortaribeiroContentModerationOnline2024}, is also employed by prominent organizations such as the New York Times, Vox Media, and OpenWeb. In 2021, it processed 500 million requests daily~\cite{jigsaw2021perspective}. Similarly, OpenAI's Moderation API, which acts as the hate speech filter for ChatGPT and GPT models~\cite{gpttvshow}, supports more than 200 million weekly users~\cite{roth_chatgpt_2024}.

\subsection{Integration and Application}
Figure~\ref{fig:contentmodAPI} shows the basic usage of these APIs for moderation of text input. A comment, post, prompt, or any other text is sent to the API. In addition, users provide information on language, sentence length, and location data, and specify a decision threshold. The length of this text can vary. However, most APIs have a maximal text length, and OpenAI specifies that for maximal performance, a text chunk's optimal length is 2,000 characters. Most of these services have prices per 1,000 tokens; as of today, only Perspective API and OpenAI Moderation are free to use. Free tiers of other APIs come with severe rate limits. 

After providing the text input, the API returns an array with various confidence scores depending on the specific service. Where most earlier published classifiers and datasets have cast the detection problem of hate speech to be a binary classification problem~\cite{sap_social_2020}, current work and all of the audited content moderation APIs classify hate speech along distinct sub-categories. An overview of the characteristics and transparency of APIs is provided in Appendix Tables~\ref{tab:api_comparison} and~\ref{tab:configurations}.


The designers and community organizers of end-user applications are responsible for determining how to handle the output, and there are various ways to manage the output of a content moderation API based on the application and integration of the API~\cite{rieder}. Score thresholds that define the occurrence of hate speech or one of its sub-categories can vary, just as the action taken in response. Possible sanctions do not only include deletion; instead, platforms rely on a range of interventions such as implementing age barriers, geo-blocking, or temporary holds, appending fact-checking labels and trigger warnings, and not recommending the content to anyone~\cite{gillespie}. 

To illustrate, \citet{rieder} describes for Perspective API how potential moderation decisions can be processed from the API's output. First, the output can be sent to a content moderator in the case of online moderation or a \textit{human in the loop} (HINL) that is guided in their decision by the output score. Second, a hybrid or cascade solution is one where only specific content that is close to the threshold is given to a human. Other algorithms provide guidance on which content needs human review, similar to the active learning strategies in ML ~\cite{lykouris2024learningdefercontentmoderation}. Third, blogs or deployers with limited resources could decide to just define a threshold and moderate content with rules for sub-categories, which would be equivalent to automatic deletion of the displayed comment in Figure ~\ref{fig:contentmodAPI}. Jigsaw, the company behind Perspective API, discourages full automation, recognizing that ML models are prone to errors. They further recommend human oversight to mitigate potential mistakes and ensure that moderation actions remain contextually appropriate~\cite{rieder}.

\section{Audit Framework}
\label{sec:blackbox}

Third-party audits are critical for ensuring accountability and transparency in ML systems~\citep{raji_actionable_2019, hartmann2024addressing}. Unlike internal or contracted audits, independent audits offer greater scrutiny as they operate without company interference~\citep{birhane_ai_2024,raji_outsider_2022}. However, third-party auditors commonly have limited access to the systems they evaluate ~\cite{hartmann2024addressing}. This is also true for the commercial content moderation systems under study, where access to internal components like weights, gradients, and training data are restricted. This limitation makes the auditing process more challenging, as the analysis must rely on input-output behavior without insight into the model’s inner workings \cite{casper_black-box_2024,jarke20245}. In face of these challenges, third-party audits are crucial for an AI accountability ecosystem.

We propose a black-box audit for content moderation APIs\footnote{Code: \url{https://anonymous.4open.science/r/Content-Moderation-API-Pipeline-874B/}}, as shown in Figure~\ref{fig:framework}. The audit comprises a set of experiments with whom we seek to answer three research questions:
\begin{enumerate}
    \item \textbf{Comparative Performance:} How do the selected APIs perform in terms of hate speech detection, particularly across different target groups and linguistic variations?
    \item \textbf{Over- and Under-Moderation:} Where do these APIs tend to over-moderate or under-moderate specific target groups and linguistic variations?
    \item \textbf{Failures:} Why do these APIs over- and under-moderate in certain contexts, and what do these misclassifications reveal about their general functionality and operational design?
\end{enumerate}

We apply this framework to evaluate the five mentioned commercial content moderation APIs. Overall, we collect and analyze over five million queries using four benchmark datasets.

We design three experiments addressing the research questions: (1) comparative aggregate and target-group performance, (2) perturbation sensitivity analysis (PSA) with counterfactual fairness scores, and (3) Shapley Additive exPlanations (SHAP) explainability. In total, we explore over- and under-moderation of eight marginalized groups, including (\textit{Asian}, \textit{Black}, \textit{Disability}, \textit{Female}, \textit{Jewish}, \textit{Latinx}, \textit{LGBTQIA+}, \textit{Muslim}), which can be mapped across all datasets with limited assumptions. We only use these eight to create maximum comparability.

In Experiment~\ref{sec:experiment1}, we assess each API's performance across four benchmark datasets in aggregate and across the eight target groups. By comparing performance on four datasets (ToxiGen, Civil Comments, HateXplain, and SBIC; introduced below), we aim to identify areas where APIs over- or under-moderate specific groups and linguistic variations. This directly addresses \textbf{RQ1} and \textbf{RQ2} by exploring how these moderation systems function across datasets and target-groups. 

In Experiment~\ref{sec:experiment2}, we test the robustness of each API's classification decisions with Perturbation Sensitivity Analysis. By introducing minor changes to input text that should not affect classification, we evaluate the over-and under-moderation of the target group by the API in relation to the dominant groups. Biases in the output confidence scores would not only indicate over- or under-moderation but also provide insights into the shortcomings of these models. Therefore, this experiment addresses \textbf{RQ2} and \textbf{RQ3} by examining the failures in API functionality.

SHAP in Experiment~\ref{sec:experiment3} provides an explanation about the contribution of tokens to the APIs confidence scores. We assess these contributions together with a qualitative evaluation of model failures, which helps us understand the specific causes of over- and under-moderation. By providing an interpretable visualisation of how the models weigh different features, this experiment addresses \textbf{RQ3} by offering a deeper understanding of the APIs’ internal workings. 

We use four benchmark datasets, each offering specific linguistic variations of hate speech:

\begin{itemize}
    
    \item \textbf{Civil Comments} — Comprised of longer, human-written examples of hate speech, this dataset averages 48.3 words per sequence and contains annotations from multiple human annotators~\cite{jigsaw_jigsaw_2019}.
    \item \textbf{HateXplain} — This dataset includes human-written hate speech examples specifically labeled for target groups by at last three annotators and accompanying hate speech rationales~\cite{mathew_hatexplain_2021}.
    \item \textbf{Social Bias Inference Corpus (SBIC)} — Focusing on implicit hate speech and stereotypes, SBIC offers another lens for examining hate speech detection~\cite{sap_social_2020}.
    \item \textbf{ToxiGen} — This dataset contains implicit and adversarial hate speech, using synthetic data (genAI) to diversify hate speech corpora~\cite{hartvigsen_toxigen_2022}.
\end{itemize}

The diversity in these datasets ensures that we capture both explicit and implicit forms of hate speech, and the variety of target groups and text types allows us to assess the potential limitations of the content moderation APIs. Table~\ref{tab:datasets_descriptive_statistics} in the Appendix offers a descriptive overview of the four datasets. All datasets are balanced on hate and non-hate speech, both at the aggregate and the group-level, to avoid distortion of performance metrics.

Conducting these experiments was resource-intensive due to the high number of API calls, some limited to one query per second. This amounted to 5 million queries, which had to be parallelized for some APIs. For instance, SHAP analyses of the Perspective API and Microsoft Content Moderator required two weeks of continuous requests. Experiments 1 and 2 were conducted between March 3, 2024, and April 15, 2024, and later repeated from June 15, 2024, to July 11, 2024, yielding consistent results. Experiment 3 was performed from July 12, 2024, to August 18, 2024.
We acknowledge that these content moderation systems are dynamic, and their performance may vary over time. However, our framework provides a reusable methodology for re-assessing these APIs as they evolve.

\section{Experiments}


\subsection{Experiment 1: Comparative Aggregate and Target-Group Performance Evaluations on Four Data Sets}
\label{sec:experiment1}
\subsubsection{Method}
We evaluate all cloud-based content moderation algorithms across multiple datasets using both threshold-variant and threshold-invariant performance metrics established in the ML field \cite{borkan, muller2016introduction}. A scale-variant metric changes depending on the specific decision threshold set in a classification model, affecting metrics like Precision, Recall, and F1 score. In contrast, a threshold-agnostic metric evaluates model performance across all possible thresholds, providing a more comprehensive measure, such as the ROC AUC \cite{elsafoury_bias_2023}. These metrics are assessed at the aggregate level and for specific target groups. Specifically, we compute metrics such as the F1 score, True Positive Rate (TPR), False Positive Rate (FPR), and the threshold-invariant ROC AUC Score. A key part of this evaluation is measuring these metrics not only in aggregate but also at the group level, with the aim of identifying any performance disparities that might affect certain groups. This analysis is grounded in the principle of Equality of Odds, as theorized by \citet{hardt_equality_2016} and implemented by \citet{dixon_measuring_2018}, which seeks to ensure that models perform consistently across all groups.

At the group level, we use a pinned ROC AUC, a metric introduced by \citet{dixon_measuring_2018}, to allow for more robust, scale-invariant comparisons across sub-groups. This metric works by "pinning" the data for each sub-group to the same baseline distribution, creating a dataset with 50\% of the data belonging to the sub-group and 50\% randomly drawn from the overall dataset. The authors acknowledge certain limitations with this approach in later research. However, it remains the most reliable scale-invariant metric for addressing group-level performance variation \cite{borkan}. 

As noted by \citet{fortuna_toxic_2020}, the definitions and sub-categories of hate speech vary widely across datasets and services. These definitions do not always align with the sub-category titles, and some services fail to provide clear definitions for their categories (see Appendix Section \ref{tab:configurations}). To ensure fair and consistent comparisons, we follow a three-step process: (1) gather all available category values for each API, (2) determine which values best align with each dataset’s outcomes, and (3) select the maximum value for each API to optimize comparison accuracy.

\subsubsection{Results}
\paragraph{Aggregate-Level Performance}

\begin{table*}[t]

    \centering
    \small
    
    \begin{tabular}{l|l|cccccccc}
        \toprule
        Dataset & Characteristics& Moderation Service & ACC & AUC & F1 & FPR & FNR \\
        \midrule
        \multirow{5}{*}{ToxiGen \cite{hartvigsen_toxigen_2022}} & \multirow{2}{*}{Implicit / Synthetic: }
        & Amazon & \colorbox[RGB]{182.12686567164178, 197.91044776119404, 255}{\makebox[6mm][c]{70.4\%}} & 79.2\% &\colorbox[RGB]{182.12686567164178, 197.91044776119404, 255}{\makebox[6mm][c]{68.9\%}} & 7.2\% & 51.9\% \\
        && Google & 62.7\% & \colorbox[RGB]{255, 172.0285714285714, 197.4}{\makebox[6mm][c]{65.0\%}} & 62.7\% & \colorbox[RGB]{255, 172.0285714285714, 197.4}{\makebox[6mm][c]{39.1\%}} & 35.5\% \\
        &\multirow{2}{*}{7,800 samples}& OpenAI & 70.3\% & \colorbox[RGB]{182.12686567164178, 197.91044776119404, 255}{\makebox[6mm][c]{87.2\%}} & 68.1\% & 5.6\% & \colorbox[RGB]{182.12686567164178, 197.91044776119404, 255}{\makebox[6mm][c]{33.2\%}} \\
        && Microsoft & \colorbox[RGB]{255, 172.0285714285714, 197.4}{\makebox[6mm][c]{59.8\%}} & 67.0\% & 57.4\% & 16.4\% & 63.9\% \\
        && Perspective & 61.6\% & 82.9\% & \colorbox[RGB]{255, 172.0285714285714, 197.4}{\makebox[6mm][c]{55.5\%}} & \colorbox[RGB]{182.12686567164178, 197.91044776119404, 255}{\makebox[6mm][c]{1.2\%}} & \colorbox[RGB]{255, 172.0285714285714, 197.4}{\makebox[6mm][c]{75.4\%}} \\
        \midrule
        \multirow{5}{*}{Civil Comments \cite{borkan}} & \multirow{2}{*}{Explicit / Long:}
        & Amazon & \colorbox[RGB]{182.12686567164178, 197.91044776119404, 255}{\makebox[6mm][c]{92.2\%}} & \colorbox[RGB]{182.12686567164178, 197.91044776119404, 255}{\makebox[6mm][c]{97.4\%}} & \colorbox[RGB]{182.12686567164178, 197.91044776119404, 255}{\makebox[6mm][c]{92.2\%}} & 7.5\% & 8.1\% \\
        && Google & \colorbox[RGB]{255, 172.0285714285714, 197.4}{\makebox[6mm][c]{69.9\%}} & \colorbox[RGB]{255, 172.0285714285714, 197.4}{\makebox[6mm][c]{67.4\%}} & \colorbox[RGB]{255, 172.0285714285714, 197.4}{\makebox[6mm][c]{67.2\%}} & \colorbox[RGB]{255, 172.0285714285714, 197.4}{\makebox[6mm][c]{58.4\%}} & \colorbox[RGB]{182.12686567164178, 197.91044776119404, 255}{\makebox[6mm][c]{1.8\%}} \\
        &\multirow{2}{*}{50,000 samples}& OpenAI & 78.6\% & 86.9\% & 78.6\% & 17.1\% & 25.6\% \\
        && Microsoft & 75.8\% & 81.7\% & 75.7\% & 20.4\% & \colorbox[RGB]{255, 172.0285714285714, 197.4}{\makebox[6mm][c]{28.1\%}} \\
        && Perspective & 87.8\% & 97.2\% & 87.7\% & \colorbox[RGB]{182.12686567164178, 197.91044776119404, 255}{\makebox[6mm][c]{3.3\%}} & 20.9\% \\
        \midrule
        \multirow{4}{*}{HateXplain \cite{mathew_hatexplain_2021}} & \multirow{2}{*}{Explicit / Short}
        & Amazon &70.2\%& 75.1\% & 69.9\% & 44.2\% &20\% \\
        && Google &\colorbox[RGB]{255, 172.0285714285714, 197.4}{\makebox[6mm][c]{66.4\%}}&\colorbox[RGB]{255, 172.0285714285714, 197.4}{\makebox[6mm][c]{52.2\%}} & \colorbox[RGB]{255, 172.0285714285714, 197.4}{\makebox[6mm][c]{58.9\%}} & \colorbox[RGB]{255, 172.0285714285714, 197.4}{\makebox[6mm][c]{76.8\%}} & \colorbox[RGB]{182.12686567164178, 197.91044776119404, 255}{\makebox[6mm][c]{4\%}} \\
        &\multirow{2}{*}{14,000 samples}& OpenAI & \colorbox[RGB]{182.12686567164178, 197.91044776119404, 255}{\makebox[6mm][c]{78\%}} &   \colorbox[RGB]{182.12686567164178, 197.91044776119404, 255}{\makebox[6mm][c]{87.3\%}}&\colorbox[RGB]{182.12686567164178, 197.91044776119404, 255}{\makebox[6mm][c]{77.2\%}} &\colorbox[RGB]{182.12686567164178, 197.91044776119404, 255}{\makebox[6mm][c]{41.1\%}} & 8.86\% \\
        && Microsoft &69\%& 66.5\% & 66.6\% & 61.2\% & 10.3\% \\
        && Perspective & 70.6\% &70.2\%&75.1\%&42.3 \%& \colorbox[RGB]{255, 172.0285714285714, 197.4}{\makebox[6mm][c]{20.6\%}}\\
        \midrule
        \multirow{5}{*}{SBIC \cite{sap_social_2020}} &\multirow{3}{*}{Implicit / Real:}  & Amazon & \colorbox[RGB]{182.12686567164178, 197.91044776119404, 255}{\makebox[6mm][c]{80.1\%}} &72.7\% & \colorbox[RGB]{182.12686567164178, 197.91044776119404, 255}{\makebox[6mm][c]{ 80.6\%}} &12.9\%& 25.1 \%\\
        && Google & 64.4\% & \colorbox[RGB]{255, 172.0285714285714, 197.4}{\makebox[6mm][c]{66.2\%}} & 63.4\% & \colorbox[RGB]{255, 172.0285714285714, 197.4}{\makebox[6mm][c]{50.34\%}} &\colorbox[RGB]{182.12686567164178, 197.91044776119404, 255}{\makebox[6mm][c]{22.39\%}} \\
        &\multirow{2}{*}{33,000 samples}& OpenAI & 67.4\% & \colorbox[RGB]{182.12686567164178, 197.91044776119404, 255}{\makebox[6mm][c]{81.1\%}} & 65.9\% & \colorbox[RGB]{182.12686567164178, 197.91044776119404, 255}{\makebox[6mm][c]{9\%}} & 53.4\% \\
       & & Microsoft & 62.7\% &68.1\% & 62.7\% & 32\% & 42\% \\
       & & Perspective & \colorbox[RGB]{255, 172.0285714285714, 197.4}{\makebox[6mm][c]{60.1\%}} & 73.9\% & \colorbox[RGB]{255, 172.0285714285714, 197.4}{\makebox[6mm][c]{60\%}} & 21.8\% & \colorbox[RGB]{255, 172.0285714285714, 197.4}{\makebox[6mm][c]{55\%}} 
                \Description{This table (Table 2.) provides a comprehensive comparison of performance metrics for five commercial content moderation services—Amazon, Google, OpenAI, Microsoft, and Perspective—evaluated on four datasets (ToxiGen, Civil Comments, HateXplain, and SBIC). Metrics include Accuracy (ACC), Area Under the Curve (AUC), F1-score (F1), False Positive Rate (FPR), and False Negative Rate (FNR). The datasets vary in characteristics such as implicit/explicit bias, synthetic/real data, and length.
    
        Key Observations:
        
        Performance Highlights:
        
        Best Performers: Metrics highlighted in blue represent the best performance across moderation services for a given dataset and metric.
        Worst Performers: Metrics highlighted in red indicate the worst performance.
        Dataset-Specific Results:
        
        ToxiGen Dataset: OpenAI achieved the highest AUC (87.2\%) and lowest FNR (33.2\%), while Microsoft had the lowest accuracy (59.8\%) and Perspective had the highest FPR (39.1\%).
        Civil Comments Dataset: Amazon excelled across most metrics (e.g., 92.2\% ACC and F1), while Google performed worst with 67.4\% AUC and the highest FPR (58.4\%).
        HateXplain Dataset: OpenAI performed best for most metrics, including AUC (87.3\%) and F1 (77.2\%), while Google had the lowest F1 (58.9\%) and highest FPR (76.8\%).
        SBIC Dataset: Amazon demonstrated superior performance with the highest accuracy (80.1\%) and F1 (80.6\%), whereas Perspective struggled with the lowest accuracy (60.1\%) and F1 (60\%).
        Service Insights:
        
        Amazon generally outperformed competitors in detecting explicit and implicit bias in both synthetic and real datasets.
        Google and Perspective exhibited inconsistent results, with notable struggles on implicit datasets like SBIC and ToxiGen.
        Interpretation:
        
        This table highlights the variability in performance across moderation services, revealing strengths and weaknesses in detecting toxic content depending on dataset characteristics. While some services like Amazon and OpenAI excel across multiple datasets, others like Google and Perspective display limitations, particularly in addressing implicit or nuanced bias. The blue and red color coding further emphasizes the comparative strengths and shortcomings for each metric and service.}
    \end{tabular}
    \caption{Performance metrics by moderation service and dataset. Blue shading signals the best performance, while red shading indicates the worst performance. All datasets are balanced on toxic and non-toxic phrases.}
    \label{tab:peformance-metrics-aggreagte-outcomes}
\end{table*}

Table \ref{tab:peformance-metrics-aggreagte-outcomes}
presents the aggregated performance results for selected benchmark datasets. The term ``aggregated'' refers to a representative sample of the entire dataset, rather than focusing on specific hate-targeted groups. Our findings reveal significant performance variations across moderation APIs and datasets. Overall, Amazon Comprehend and OpenAI Moderation demonstrate the most consistent performance across the datasets. OpenAI’s content moderation algorithm performs best on ToxiGen and HateXplain, generalizing well across different datasets. For Civil Comments and SBIC, Amazon Comprehend yields the most accurate results, although Perspective API also performs comparatively well on Civil Comments. In contrast, Google Natural Language API consistently exhibits the lowest performance, primarily due to a relatively high FPR, suggesting an over-moderation tendency.

We also observe notable cross-dataset performance variations. Civil Comments and HateXplain exhibit similar performance ranges, while ToxiGen and SBIC show somewhat lower results. In general, FPR is elevated across all explicit hate speech datasets, whereas the FNR is particularly high for implicit hate speech datasets. This discrepancy is especially evident with the higher FNR observed across moderation services on ToxiGen and SBIC. Interestingly, the longer text sequences in Civil Comments appear to enhance the performance of all APIs, whereas performance declines on HateXplain. Additionally, AUC scores tend to be higher than accuracy (ACC) scores, indicating that threshold settings may be contributing to misclassification, an important finding that we will discuss later.
\paragraph{Group-Level Performance}

Table~\ref{tab:performance_metrics_group_outcomes} performance metrics at the group level, highlighting noticeable disparities across moderation services and marginalized groups. Group \textit{Female} consistently received the most accurate moderation across datasets, evidenced by high Pinned ROC AUC values, followed by group \textit{Black}. In contrast, groups \textit{LGBTQIA+}, \textit{Disability}, and \textit{Jewish} experience significantly less accurate moderation, with particularly poor performance observed in specific datasets. 

At the service level, Amazon Comprehend exhibits strong, consistent performance across most marginalized groups, often achieving the highest accuracy, followed by OpenAI Content Moderation. Perspective API also ranks among the top services, particularly on the \textit{Jewish} and \textit{LGBTQIA+} groups. Microsoft Azure Content Moderation, however, continues to underperform, aligning with its results in table \ref{tab:peformance-metrics-aggreagte-outcomes}, and shows the weakest performance for several groups, notably \textit{Black} and \textit{Muslim}. Meanwhile, Google Text Moderation struggles considerably, with stark underperformance for groups \textit{Disability} and \textit{Jewish}, especially in datasets like CivilComments and ToxiGen.

In terms of FPR, the analysis reveals that content moderation tends to over-moderate speech related to the \textit{Black} and \textit{LGBTQIA+} communities. For example, on ToxiGen, Google Text Moderation severely over-moderates content targeting the \textit{Jewish} and \textit{Disability} groups, leading to a FPR as high as 99\%. Put more tangibly, while the data contains 200 toxic and 200 non-toxic examples for group \textit{Jewish}, Google Text Moderation predicts toxic speech in 385 instances.

On the flip side, FNRs underscore the challenge of detecting implicit hate speech, particularly on ToxiGen and SBIC. However, caution is warranted when interpreting results from ToxiGen, as it is partially synthetically created, as discussed in Section \ref{sec:limitations}. Nevertheless, this finding is consistent with SBIC and across most services, the \textit{Disability} group experiences significant under-moderation, with hate speech frequently going undetected. Additionally, groups like \textit{Asian} and \textit{Latinx} appear to be more prone to under-moderation, especially when using Microsoft Azure Content Moderation.

In summary, Amazon Comprehend and Perspective API show the most balanced performance across marginalized groups, while Google Text Moderation and Microsoft Azure exhibit significant inconsistencies, particularly for underrepresented and vulnerable groups like \textit{Disability} and \textit{Jewish}.
\begin{table*}[t]
\centering
\Description{This table presents the Pinned ROC AUC performance of different content moderation services (Amazon, Google, Microsoft, OpenAI, and Perspective API) across three datasets—Civil Comments, HateXplain, and ToxiGen—and various marginalized groups (e.g., Asian, Black, Disabled, Female, Jewish, Latinx, LGBTQIA+, Muslim).
Key Points:

    Civil Comments Dataset:
        Amazon performs the best across all groups, achieving the highest ROC AUC scores (ranging from 85\% to 96\%).
        Google and Microsoft show lower scores, with Google especially underperforming for certain groups (e.g., only 61\% for Black users).
        OpenAI and Perspective API also perform well, with Perspective API reaching the second-best scores in most categories, especially for Disabled (95\%) and LGBTQIA+ (86\%).

    HateXplain Dataset:
        OpenAI demonstrates the strongest performance in most categories (84\%-93\%), while Amazon and Perspective also do well, though Amazon's results are more varied.
        Google and Microsoft perform worse, with scores ranging from 57\%-83\%.

    ToxiGen Dataset:
        Amazon and OpenAI show strong performance, consistently achieving higher scores for groups like Female (93\% and 82\%, respectively) and Disabled (80\% and 87\%).
        Microsoft has the lowest performance, particularly for groups like Latinx and LGBTQIA+ (49\% and 47\%, respectively).
        Perspective API maintains strong performance, with scores reaching 85\%-87\% for various groups.

Overall, Amazon tends to perform well across most datasets and marginalized groups, while Google and Microsoft often lag behind. Perspective API shows solid performance across all datasets, especially for the Civil Comments dataset.}
\begin{tabular}{l|rrrrrrrr}
\toprule
\textbf{}              & \textit{Asian} & \textit{Black} & \textit{Disabled} & \textit{Female} & \textit{Jewish} & \textit{Latinx} & \textit{LGBTQIA+} &\textit{Muslim} \\ \midrule
\multicolumn{9}{|c|}{\textbf{CivilComments}}                                                                       \\
\midrule
Amazon        & \textbf{92\%}  & \textbf{86\%}  & \textbf{96\%}  & \textbf{95\%}  & \textbf{91\%}  & \textbf{95\%}  & \textbf{85\%} & \textbf{89\%}  \\ 
Google        & 72\%  & 66\%  & 45\%  & 61\%  & 58\%  & 73\%  &60\%& 57\%  \\ 
Microsoft     & 70\%  & 58\%  & 67\%  & 73\%  & 67\%  & 64\%  &60\%& 63\%  \\ 
OpenAI        & 80\%  & 73\%  & \textbf{83\%}  & 79\%  & 77\% & \textbf{87\%}  &72 \%& 74\%  \\ 
Perspective API & \textbf{91\%}  & \textbf{87\%}  & \textbf{95\%}  & \textbf{94\%}  & \textbf{93\%}  &98 \%& \textbf{86\%}  & \textbf{89\%}  \\ \midrule
\multicolumn{9}{|c|}{\textbf{HateXplain}}                                                                         \\ \midrule
Amazon        & 70\%  & \textbf{80\%}  & --- & \textbf{86\%}  & \textbf{83\%}  & ---  & 90\%& \textbf{88\%}  \\ 
Google        & \textbf{82\%}  & 68\%  & ---  & 68\%  & 31\%  & --- &\textbf{83} \% & 42\%  \\ 
Microsoft     & \textbf{81\%}  & 72\%  & ---  & \textbf{76\%}  & 78\%  & --- &75 \% & 78\%  \\ 
OpenAI        & \textbf{84\%}  & \textbf{88\%}  & ---  & \textbf{86\%}  &\textbf{93\%}    & ---  & \textbf{95\%}&\textbf{93\%}  \\ 
Perspective API & 76\%  & \textbf{77\%}  & ---  & \textbf{89\%}  & \textbf{90\%}  &  ---& \textbf{89\%}&\textbf{90\%}  \\ \midrule
\multicolumn{9}{|c|}{\textbf{ToxiGen}}                                                                             \\ \midrule
Amazon        & \textbf{80\%}  & 68\%  & \textbf{80\%}  & \textbf{86\%}  & 77\%  & 71\%  &69 \%& \textbf{75\%}  \\ 
Google        & 75\%  & 66\%  & 41\%  & 76\%  & 49\%  & 73\%&57\%  & 50\%  \\ 
Microsoft     & 65\%  & 66\%  & 68\%  & 71\%  & 64\%  & 57\%  &49\%& 61\%  \\ 
OpenAI        & \textbf{86\%}  & \textbf{87\%}  & \textbf{93\%}  & \textbf{82\%}  & \textbf{86\%}  & \textbf{86\%}  &\textbf{86\%}& \textbf{85\%}  \\ 
Perspective API & \textbf{85\%}  & 75\%  & \textbf{82\%}  & \textbf{90\%}  & \textbf{81\%}  & 72\% &73\% & \textbf{86\%}  

\Description{This table (Table 3.) presents the pinned ROC AUC scores for five commercial content moderation services—Amazon, Google, Microsoft, OpenAI, and Perspective API—evaluated across three datasets (CivilComments, HateXplain, and ToxiGen) and eight marginalized groups (Asian, Black, Disabled, Female, Jewish, Latinx, LGBTIQ+, and Muslim). The datasets differ in size and characteristics, and performance is measured as the ability of models to detect hate or bias associated with each group.

Key Observations:

CivilComments Dataset:

Perspective API and Amazon demonstrate strong performance, with AUC scores above 90\% for most groups.
Google and Microsoft generally underperform, with scores ranging from 31\% (Google for Jewish) to 73\% (Google for Latinx).
OpenAI achieves balanced performance, with scores ranging from 73\% (Black) to 87\% (Latinx).
HateXplain Dataset:

OpenAI leads for most groups, achieving the highest scores, including 93\% for Jewish and 95\% for LGBTIQ+.
Amazon and Perspective API perform well, with scores above 80\% for many groups.
Google struggles for some groups, with a particularly low score of 31\% for Jewish.
ToxiGen Dataset:

OpenAI exhibits consistent strong performance across all groups, with scores above 85\%.
Perspective API performs well, with scores above 80\% for most groups.
Microsoft and Google show weaker performance, with some scores below 50\% (e.g., 41\% for Disabled by Google).
Interpretation:

The table highlights disparities in the performance of moderation services for different marginalized groups and datasets. While OpenAI and Perspective API tend to excel across datasets, Google and Microsoft frequently struggle, particularly with certain groups like Disabled and Jewish. The results emphasize the variability in moderation services' ability to handle diverse types of hate speech and bias, revealing potential gaps in fairness and inclusivity.}
\end{tabular}
\caption{Pinned ROC AUC is presented per moderation service, dataset and marginalized group. ToxiGen includes 4,268 observations, HateXplain includes 1,748, Civil Comments consists of 19,228 observations, and SBIC is comprised of 5,806. For Civil Comments, target groups and hate labels were coded by human annotators. For ToxiGen we constrain our analysis to a subset of 10,000 observations, which were annotated by human reviewers. For both Civil Comments and ToxiGen, when running aggregate-level tests, we only include phrases for which all annotators are assigned the same hate label. 
}
\label{tab:performance_metrics_group_outcomes}
\end{table*}

\subsection{Experiment 2: Perturbation Analysis with Counterfactual Fairness}
\label{sec:experiment2}

\subsubsection{Method}
Perturbation Sensitivity Analysis (PSA) offers an additional, arguably more robust evaluation of group-level biases by using counterfactual fairness evaluation \cite{prabhakaran-etal-2019-perturbation}. With the target-group performance evaluation, we derive conclusions about group-level biases from variations in False Negative and False Positive Rates. This grounds on two assumptions. First, the validity of the ground truth in the data. A moderation service would appear to over-moderate or under-moderate if this over- or under-moderations is actually encapsulated in our benchmark datasets that we use for evaluation. Second, we assume that data on all groups comes from the same underlying distributions. If one group were to contain many more edge cases than other groups, this would aggravate False Positive and FNRs, compromising our ability to infer biases.    

PSA avoids such pitfalls by solely exchanging group-specific identity tokens while holding the remainder of the sentence constant \citep{prabhakaran_perturbation_2019}. Table \ref{tab:psa_methodology_illustration} displays the fundamental logic. We follow prior research in defining an anchor group against which other groups are compared \citep{prabhakaran_perturbation_2019}. Using the dominant group as baseline, Counterfactual Token Fairness scores are computed as the difference in hate speech between the baseline and the corresponding marginalized group.

PSA makes two assumptions: (1) counterfactual pairs should convey the same or neutral meaning, avoiding any implicit biases or derogatory connotations. While constructing toxic counterfactuals is theoretically possible, it is methodologically demanding and exceeds the scope of this project. Instead, we construct 34 \textit{neutral} counterfactual pairs. Importantly, each marginalized group is represented by multiple tokens, reflecting its different semantic representations. For instance, the marginalized group \textit{female} also manifests as \textit{woman} and \textit{women}. Furthermore, (2) there should be no unique interactions between a particular marginalized token and the context of the sentence that would skew the analysis. This is challenging in real-world applications, as certain combinations might evoke stereotypes or specific cultural connotations. Thus, the project uses data consisting largely of short and explicit statements. Furthermore, CTF scores are calculated separately for toxic and non-toxic statements, with the latter generally supporting the assumption of counterfactual symmetry more consistently.

PSA experiments are conducted using two distinct datasets. First, the synthetic \textit{Identity Phrase Templates} from \citet{dixon_measuring_2018} are used. The set contains 77,000 synthetic examples of which 50\% are toxic. These avoid stereotypes and complex sentence structures by design, which ensures that the symmetric counterfactual assumption is met. Mapping the dataset, which contains a broader set of identities, to the 34 tokens for marginalized identities which fit to the eight relevant marginalized groups that we use in study results in 25,738 sentence pairs (see Table \ref{tab:minority_majority_token}). Second, by applying the same logic, 9,190 sentence pairs are derived from the HateXplain dataset. Specifically, we identified  occurrences of these 34 tokens for marginalized identities in HateXplain sentences and replaced them with counterfactual dominant tokens. Statistics about these sentences -- synthetic and non-synthetic -- are presented in Table \ref{tab:psa_datasets_descriptive_statistics}. 
\begin{figure*}[t]
\Description{This figure displays a Perturbation Sensitivity Analysis comparing synthetic and real-world data from two sources: Identity Phrase Templates (synthetic) and the HateXplain dataset (real). The analysis focuses on Counterfactual Token Fairness (CTF) scores, which measure the difference in toxicity between a phrase containing a dominant baseline token (representing a dominant group) and its marginalized perturbation (representing a marginalized group).
Key Elements:

    Synthetic Data (Left Plot):
        This plot shows CTF scores for synthetic data derived from Identity Phrase Templates (based on Dixon et al., 2019).
        Groups analyzed include Muslim, LGBTQ+, Latinx, Jewish, Female, Black, and Asian.
        Positive CTF scores indicate a higher toxicity score for the marginalized group (i.e., bias), while negative scores suggest lower toxicity.
        Amazon and Perspective show the highest variability, particularly for groups like Muslim and LGBTQ+, while other services like Microsoft tend to cluster closer to zero, indicating more fairness.

    Real Data (Right Plot):
        This plot shows CTF scores based on real data from HateXplain.
        The analysis focuses on similar marginalized groups as in the synthetic data, but the scores generally trend closer to zero, particularly for groups like Disability and Female.
        However, significant biases appear for Latinx and Muslim groups in certain services, with Perspective showing the highest mean bias for these groups.

    Service Comparison:
        The figure compares five services (Amazon, Google, OpenAI, Microsoft, and Perspective), with their CTF scores represented by different colored points and error bars.
        Each point represents the average score for a specific group, with error bars indicating a 95
        Overall, Perspective and Amazon exhibit higher variability and more positive CTF scores, while Google and Microsoft show more stable, generally neutral scores.

Interpretation:

This analysis illustrates the extent of unintended bias in toxicity classification, where certain marginalized groups receive disproportionately higher toxicity scores (positive CTF) when their identity terms are substituted for dominant group terms. The real-world data shows somewhat lower bias overall compared to the synthetic data, but certain services still exhibit significant biases depending on the group and the content moderation model.}
   \includegraphics[width=\linewidth]{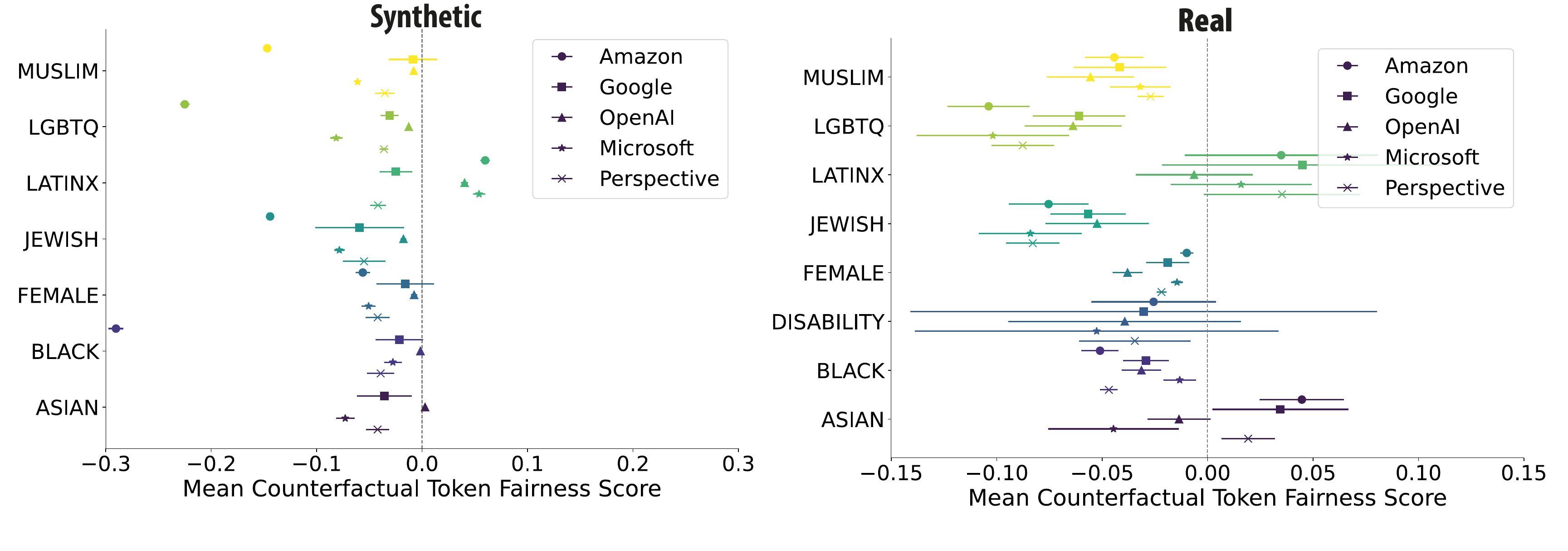}
\caption{Perturbation Sensitivity Analysis on synthetic data from the Identity Phrase Templates in \citet{dixon_measuring_2018} and non-synthetic data from HateXplain. Counterfactual Token Fairness (CTF) scores are computed as the difference in toxicity between the phrase containing the baseline dominant token and its marginalized perturbation. Counterfactual Token Fairness scores per marginalized group and service are averaged and reported for non-toxic . Besides a point estimate, the figure also includes a 95\% confidence interval assuming a student-t distribution.}
\label{fig:psa}
\end{figure*}
\subsubsection{Results}
Figure \ref{fig:psa} displays PSA results on the synthetic and non-synthetic datasets, created from the \citet{dixon_measuring_2018} Sentence Templates and HateXplain respectively. A negative CTF score indicates that, on average, tokens for marginalized identities are associated with higher hate speech scores than their counterfactual dominant tokens. Upon general inspection, two observations arise. First, differences in confidence scores by and large are more pronounced on non-hate speech than hate speech data. Intuitively this makes sense, as scores are generated non-linearly with a definite upper bound. Thus, when other elements in a sentence induce high confidence score, the marginal effect from identity tokens is comparably lower. Second, we notice greater variation in the mean Counterfactual Token Fairness scores in non-synthetic than in synthetic data. This was to be expected, as the sentences from HateXplain contain more contextual information which interacts with the tokens. Depending on the context, the difference in hate speech scores between majority and marginalized token thus varies to a greater extent. By contrast, within non-synthetic examples, the greater variation of CTF mean scores associated with non-toxic data is likely an artifact of the former's smaller sample size. This observation in itself is revealing: When analyzing HateXplain, we found that 85\% of the sentences containing neutral tokens for marginalized identities from our PSA were labeled as hate speech.

Overall, the results suggest that most minorities get associated with higher levels of hate speech scores than dominant majorities, although these effects appear relatively small for some groups, and vary in general across groups and services. Group \textit{LGBTQIA+} is associated with the strongest negative bias, occurring for all samples and services. By contrast, we observe limited negative bias against groups \textit{Latinx} and \textit{Asian}. In fact, in most samples and services, tokens for marginalized identities related to group \textit{Latinx} incur, on average, lower toxicity scores than their \textit{White} counterfactual. Similarly, groups \textit{Female} and \textit{Black} are linked to comparably small negative biases. Unfortunately, our results allow for no robust inference for group \textit{Disability}. While group \textit{Disability} is entirely absent from the synthetic data, as no examples are included in the \citet{dixon_measuring_2018} Sentence Templates, the non-synthetic data only contains very few instances, resulting in large variation for mean CTF scores.

Other patterns emerge that are more confined to particular services. For instance, Amazon Comprehend is associated with the overall strongest negative bias, linked to groups \textit{Muslim} and \textit{Jewish}. Notably, when applied to hate speech data, this effect vanishes. Interestingly, for Google we are presented with findings that are somewhat difficult to reconcile with the previous section's result. More specifically, we would have expected a large negative bias placed by Google Text Moderation onto group \textit{Muslim}, which does not seem to be the case. Potentially, the extreme FPRs were not driven by \textit{Muslim} identity terms themselves, but rather by other words co-occurring such as \textit{Islam} or paraphrases of identity-related terms. However, we find that Google Text Moderation as higher variance also for the synthetic data set. Further research is required to explain the disparity observed. Lastly, it is worth noting that OpenAI Content Moderation seems the least biased, across groups and datasets.

\subsection{Experiment 3: SHAP Explainability and Qualitative Evaluation}
\label{sec:experiment3}
\begin{table*}[t]
    \centering
    \small
    \begin{tabular}{p{1.2cm}p{0.8cm}|p{0.6cm}p{5cm}|p{0.6cm}p{4.7cm}}
        \textbf{API} & \textbf{FN/FP} & \textbf{Rate} & \textbf{ToxiGen Tokens} & \textbf{Rate} & \textbf{HateXplain Tokens} \\ \midrule
        \multirow{2}{*}{\textbf{Amazon}} & FN & 51.9\% & \textit{steel}, \textit{waiting}, \textit{brave}, \textit{play}, \textit{psychology} & 20\% & \textit{source}, \textit{history}, \textit{corner}, \textit{welcome}, \textit{think} \\
               & FP & 7.2\% & \textit{dumb}, \textit{asshole}, \textit{homosexuality}, \textit{caucasian}, \textit{rites} & 44.2\% & \textit{coward}, \textit{uts}, \textit{porn}, \textit{penis}, \textit{rats} \\ \midrule
        \multirow{2}{*}{\textbf{OpenAI}}& FN & 33.2\% & \textit{lucky}, \textit{between}, \textit{send}, \textit{health}, \textit{natural} & 8.9\% & \textit{words}, \textit{penguin}, \textit{phobia}, \textit{text}, \textit{international} \\
               & FP & 5.6\% & \textit{gay}, \textit{kill}, \textit{chinese}, \textit{muslim}, \textit{christians} & 41.1\% & \textit{ded}, \textit{traitor}, \textit{n*gger}, \textit{n*gga}, \textit{jews} \\ \midrule
        \multirow{2}{*}{\textbf{Google}} & FN & 35.5\% & \textit{lucky}, \textit{chapter}, \textit{playing}, \textit{character}, \textit{openend} & 4\% & \textit{cat}, \textit{feminine}, \textit{asian}, \textit{girls}, \textit{illo} \\
               & FP & 39.1\% & \textit{weapon}, \textit{mormon}, \textit{drunk}, \textit{hindu}, \textit{bullying} & 76.8\% & \textit{tax}, \textit{christians}, \textit{gay}, \textit{slut}, \textit{raped} \\ \midrule
        \multirow{2}{*}{\textbf{Microsoft}}& FN & 63.9\% & \textit{ist}, \textit{ple}, \textit{unce}, \textit{health}, \textit{episode} & 10.3\% & \textit{ro}, \textit{st}, \textit{lings}, \textit{bathroom}, \textit{ists} \\
                  & FP & 16.4\% & \textit{dumb}, \textit{crap}, \textit{child}, \textit{breast}, \textit{holy} & 61.2\% & \textit{bitch}, \textit{gay}, \textit{loser}, \textit{violence}, \textit{arabic} \\ \midrule
        \multirow{2}{*}{\textbf{Perspective}} & FN & 75.4\% & \textit{sau}, \textit{lucky}, \textit{define}, \textit{protect}, \textit{greatest} & 20.6\% & \textit{salute}, \textit{rights}, \textit{statistics}, \textit{door}, \textit{season} \\
                    & FP & 1.2\% & \textit{terrorist}, \textit{n*gger}, \textit{queer}, \textit{black}, \textit{are} & 42.3\% & \textit{gay}, \textit{fucked}, \textit{loser}, \textit{coward}, \textit{balls} \\
    \Description{This table (Table4) presents the top five most contributing global SHAP values for False Positives (FP) and False Negatives (FN) for five commercial content moderation APIs (Amazon, OpenAI, Google, Microsoft, and Perspective) across two datasets: Toxigen (left) and HateXplain (right).

Key Observations:

False Negatives (FN):

Positive or neutral tokens such as steel, natural, and lucky contribute to high FN rates, leading to misclassification of hate speech as non-hate speech.
FN rates range from 33.2\% (OpenAI for Toxigen) to 75.4\% (Perspective for Toxigen) across datasets.
Microsoft has the highest FN rate for Toxigen (63.9\%) and exhibits difficulty with tokens like ist, ple, and episode.
False Positives (FP):

Identity tokens like gay, jews, and christians frequently contribute to FPs, where non-hate speech is misclassified as hate speech.
FP rates vary widely, from as low as 1.2\% (Perspective for Toxigen) to as high as 76.8\% (Google for HateXplain).
Perspective demonstrates a low FP rate for Toxigen (1.2\%) but struggles with HateXplain, where gay, loser, and balls are significant contributors.
Dataset-Specific Insights:

Toxigen: Amazon exhibits relatively low FP rates (7.2\%) but struggles with FN rates (51.9\%). Perspective has the lowest FP rate (1.2\%) but the highest FN rate (75.4\%).
HateXplain: OpenAI achieves the lowest FN rate (8.9\%) but still suffers from high FP rates (41.1\%) with tokens like ded and traitor contributing significantly.
Interpretation:

The table highlights the challenges faced by different APIs in moderating hate speech, particularly in distinguishing between neutral, identity-based, and harmful tokens. Tokens like gay and jews are frequent contributors to FPs, suggesting a bias in how APIs interpret identity-related terms. Conversely, neutral or contextually positive terms like lucky and health contribute to FNs, indicating difficulty in capturing nuanced hate speech.}
    \end{tabular}
    \caption{ToxiGen (left) and HateXplain (right) five most contributing Global SHAP Values for False Positives (FP) and False Negatives (FN) per API. For example, identity tokens like \textit{gay} and \textit{jews} contribute to OpenAI misclassifying non-hate speech as hate speech (FP), while positive words like \textit{lucky} and \textit{health} contribute to misclassifying hate speech as non-hate speech (FN).}
    \label{tab:combinedSHAP}
\end{table*}

\subsubsection{Method}
To understand where APIs fail to classify hate speech and to get insights into the causes of these failures, we extend our analysis with SHAP \cite{lundberg2017unified}. The method is based on so-called Shapley values, which are derived from cooperative game theory and give insight into the feature-contribution to the output \cite{rozemberczki_2022} and make the inner mechanisms of the model understandable to humans \cite{murdoch}. We selected SHAP since it has the advantage of having an off-the-shelf available NLP text evaluation \cite{mosca-etal-2022-shap}, being one of the black-box access methods that can be used without access to weights or gradients \cite{Lyu}, is theoretically grounded and, in contrary to LIME, is not susceptible to instability due to the non-random selection of perturbations \cite{Lyu,mosca-etal-2022-shap}. Furthermore, SHAP can be used for local and global explanations and is also model-agnostic \cite{mosca-etal-2022-shap}. However, the disadvantage of SHAP -- especially in the NLP case -- is that the calculation is NP-hard \cite{IntroductionExplainableAI} and, thus, computationally expensive. 

We, therefore, had to use two datasets, namely, HateXplain and ToxiGen, to understand where the APIs over- and under-moderate and the reason for not analyzing the whole data set but the false positives and negatives. This alone led to more than one million queries for all APIs and, thus, a lot of computational resources. We, however, tried to minimize these resources by parallelizing queries of queries dependent on rate limits per minute of each API. We represent features $x$ as word tokens \( x = \{w_1, w_2, \ldots, w_n\} \) to ensure interpretability. SHAP computes the marginal contribution of each token \( w_i \) within \( x \) to the model output \( f(x) \), where \( f: \mathcal{X} \to [0,1] \) and whereby assesses each token's impact on the decision \( \mathbb{P}(y = 1 \mid x) > 0.5 \), indicating whether the probability of the positive class exceeds 0.5 based on \( f(x) \) \cite{mosca-etal-2022-shap}. A BERT-based tokenizer was used to retrieve the tokens for each misclassified sentence. Shapley values were calculated through model-agnostic SHAP by using a ``[MASK]'' token as a perturbation. 

\begin{table*}[t]
\begin{minipage}[c]{0.49\linewidth}
\centering
\vspace{3.88em}
\begin{tabular}{p{2.55cm}p{0.55cm}p{0.55cm}p{0.55cm}p{0.55cm}p{0.6cm}}

\multicolumn{5}{c}{\textbf{Over-Moderation (False Positives)}} \\
\textbf{Category}     & \textbf{AMZ} & \textbf{GPT} & \textbf{GOO} & \textbf{MIC} & \textbf{PER} \\ 
\hline
    Counter-speech & \colorbox[RGB]{255, 221.51941747572815, 231.75728155339806}{9\% \hspace{0.8mm}} & \colorbox[RGB]{255, 230.66480446927375, 238.10614525139664}{6\% \hspace{0.8mm}} & \colorbox[RGB]{255, 218.89005235602093, 229.93193717277487}{9\% \hspace{0.8mm}} & \colorbox[RGB]{255, 228.11111111111111, 236.33333333333334}{7\% \hspace{0.8mm}} &   \colorbox[RGB]{255, 219.46153846153845, 230.32867132867133}{9\% \hspace{0.8mm}}   \\ 
    Dialects & \colorbox[RGB]{255, 228.56796116504853, 236.6504854368932}{7\% \hspace{0.8mm}} & \colorbox[RGB]{255, 226.60893854748605, 235.2905027932961}{7\% \hspace{0.8mm}} & \colorbox[RGB]{255, 251.19895287958116, 252.36125654450262}{1\% \hspace{0.8mm}} & \colorbox[RGB]{255, 232.5925925925926, 239.44444444444446}{6\% \hspace{0.8mm}} & \colorbox[RGB]{255, 234.69230769230768, 240.9020979020979}{5\% \hspace{0.8mm}}    \\ 
    Descriptive-FP & \colorbox[RGB]{255, 193.3252427184466, 212.18446601941747}{16\%} & \colorbox[RGB]{255, 208.35754189944134, 222.62011173184356}{12\%} & \colorbox[RGB]{255, 125.76439790575915, 165.28272251308903}{35\%} & \colorbox[RGB]{255, 219.14814814814815, 230.11111111111111}{9\% \hspace{0.8mm}} &\colorbox[RGB]{255, 196.6153846153846, 214.46853146853147}{16\%}    \\ 
    OM profanity & \colorbox[RGB]{255, 246.18932038834953, 248.88349514563106}{2\% \hspace{0.8mm}} & \colorbox[RGB]{255, 241.81843575418995, 245.84916201117318}{3\% \hspace{0.8mm}} & \colorbox[RGB]{255, 237.89528795811518, 243.12565445026178}{4\% \hspace{0.8mm}} & \colorbox[RGB]{255, 241.55555555555554, 245.66666666666666}{3\% \hspace{0.8mm}} & \colorbox[RGB]{255, 232.15384615384616, 239.13986013986013}{6\% \hspace{0.8mm}}    \\ 
    Negation-FP & \colorbox[RGB]{255, 251.4757281553398, 252.55339805825244}{0\% \hspace{0.8mm}} & \colorbox[RGB]{255, 253.98603351955308, 254.29608938547486}{0\% \hspace{0.8mm}} & \colorbox[RGB]{255, 255.0, 255.0}{0\% \hspace{0.8mm}} & \colorbox[RGB]{255, 246.03703703703704, 248.77777777777777}{2\% \hspace{0.8mm}} & \colorbox[RGB]{255, 234.69230769230768, 240.9020979020979}{5\% \hspace{0.8mm}}   \\ 
    Non-protected & \colorbox[RGB]{255, 237.37864077669903, 242.76699029126215}{4\% \hspace{0.8mm}} & \colorbox[RGB]{255, 233.70670391061452, 240.21787709497207}{5\% \hspace{0.8mm}} & \colorbox[RGB]{255, 245.4973821989529, 248.40314136125653}{2\% \hspace{0.8mm}} & \colorbox[RGB]{255, 219.14814814814815, 230.11111111111111}{9\% \hspace{0.8mm}} &\colorbox[RGB]{255, 247.3846153846154, 249.71328671328672}{2\% \hspace{0.8mm}}    \\
    OM slurs & \colorbox[RGB]{255, 193.3252427184466, 212.18446601941747}{16\%} & \colorbox[RGB]{255, 204.30167597765364, 219.80446927374302}{13\%} & \colorbox[RGB]{255, 234.09424083769633, 240.4869109947644}{5\% \hspace{0.8mm}} & \colorbox[RGB]{255, 196.74074074074073, 214.55555555555554}{16\%} & \colorbox[RGB]{255, 171.23076923076923, 196.84615384615384}{23\%}    \\
    Re-appropriation & \colorbox[RGB]{255, 230.33009708737865, 237.873786407767}{6\% \hspace{0.8mm}} & \colorbox[RGB]{255, 217.4832402234637, 228.95530726256982}{10\%} & \colorbox[RGB]{255, 253.09947643979058, 253.6806282722513}{0\% \hspace{0.8mm}} & \colorbox[RGB]{255, 237.07407407407408, 242.55555555555554}{4\% \hspace{0.8mm}} & \colorbox[RGB]{255, 255.0, 255.0}{0\% \hspace{0.8mm}}     \\ 
    SOS & \colorbox[RGB]{255, 179.22815533980582, 202.39805825242718}{20\%} & \colorbox[RGB]{255, 200.24581005586592, 216.98882681564245}{15\%} & \colorbox[RGB]{255, 186.58115183246073, 207.5026178010471}{18\%} & \colorbox[RGB]{255, 156.40740740740742, 186.55555555555554}{27\%} &\colorbox[RGB]{255, 229.6153846153846, 237.37762237762237}{6\% \hspace{0.8mm}}    \\ 
    Unsure & \colorbox[RGB]{255, 217.99514563106797, 229.31067961165047}{10\%} & \colorbox[RGB]{255, 206.3296089385475, 221.21229050279328}{13\%} & \colorbox[RGB]{255, 192.28272251308903, 211.4607329842932}{17\%} & \colorbox[RGB]{255, 219.14814814814815, 230.11111111111111}{9\% \hspace{0.8mm}} & \colorbox[RGB]{255, 214.3846153846154, 226.8041958041958}{11\%}  \\
    Hate & \colorbox[RGB]{255, 242.66504854368932, 246.4368932038835}{3\% \hspace{0.8mm}} & \colorbox[RGB]{255, 218.49720670391062, 229.65921787709496}{10\%} & \colorbox[RGB]{255, 241.69633507853402, 245.76439790575915}{3\% \hspace{0.8mm}} & \colorbox[RGB]{255, 246.03703703703704, 248.77777777777777}{2\% \hspace{0.8mm}} & \colorbox[RGB]{255, 206.76923076923077, 221.5174825174825}{13\%} \\ \hline \hline
\textbf{$\sum$ Codes}   & 206 
               & 358 
               & 191 
               & 81 & 92 \\  
\Description{This table (Table 5) summarizes the results of the qualitative evaluation of SHAP explanations from Experiment 3. It categorizes False Positives (FP) and False Negatives (FN) into different operationalized codes for over-moderation and under-moderation. The operationalizations of the codes are provided in Tables 14 and15.

Key Observations:

Over-Moderation (FP):

Categories such as counter-speech, descriptive FP, and SOS have the highest rates of over-moderation across models.
Amazon and Perspective API exhibit high FP rates for descriptive FP (16\%) and slurs (16\% for Amazon and 23\% for Perspective).
Dialects, re-appropriation, and non-protected group tokens show lower FP rates, with values around 5\% to 7\% across all APIs.
Negation-based errors are rare, with FP rates of 0\% for OpenAI, Google, and Perspective, and 2\% for Microsoft.
Under-Moderation (FN):

Categories such as implicit hate and descriptive FN have the highest FN rates, with implicit hate reaching up to 28\% for Microsoft and 24\% for Google.
Negation FN rates are consistently low, ranging from 2\% to 4\% across all APIs.
Positive terms, paraphrased targets, and spelling variations contribute moderately to FN rates, with values between 8\% and 15\% for most APIs.
The category “unsure” indicates coder uncertainty, accounting for 12\% to 18\% of FN cases depending on the API.
Model-Specific Insights:

Perspective API and Amazon tend to over-moderate categories such as slurs and SOS more aggressively than other APIs.
Microsoft and OpenAI display relatively balanced behavior across categories but struggle with implicit hate and descriptive FN.
Google shows the highest rates of under-moderation for implicit hate (24\%) and descriptive FN (19\%).
Interpretation:

The table highlights the varying strengths and weaknesses of the APIs in addressing nuanced cases of hate speech and moderation. Over-moderation errors (FP) are commonly associated with slurs, counter-speech, and descriptive text, while under-moderation errors (FN) arise predominantly in implicit hate and descriptive contexts. The “unsure” category suggests that some errors may require additional context or interpretation beyond the model’s capabilities.}
\end{tabular}

\end{minipage}
\hfill
\begin{minipage}[t]{0.49\linewidth}
\centering
\begin{tabular}{p{2.55cm}p{0.55cm}p{0.55cm}p{0.55cm}p{0.55cm}p{0.6cm}}

\multicolumn{5}{c}{\textbf{Under-Moderation (False Negatives)}} \\   \textbf{Category} & \textbf{AMZ} & \textbf{GPT} & \textbf{GOO} & \textbf{MIC} & \textbf{PER} \\ 
\hline
    Counter-speech-FN & \colorbox[RGB]{252.11946902654867, 252.7433628318584, 255}{0\% \hspace{0.8mm}} & \colorbox[RGB]{251.87019230769232, 252.54807692307693, 255}{0\% \hspace{0.8mm}} & \colorbox[RGB]{241.6840909090909, 244.5681818181818, 255}{4\% \hspace{0.8mm}} & \colorbox[RGB]{253.20165745856355, 253.59116022099448, 255}{0\% \hspace{0.8mm}} & \colorbox[RGB]{255.0, 255.0, 255}{0\% \hspace{0.8mm}}   \\ 
    Descriptive-FN & \colorbox[RGB]{185.86725663716814, 200.84070796460176, 255}{21\%} & \colorbox[RGB]{205.96634615384616, 216.58653846153845, 255}{15\%} & \colorbox[RGB]{191.37954545454545, 205.1590909090909, 255}{19\%} & \colorbox[RGB]{181.26795580110496, 197.23756906077347, 255}{22\%} & \colorbox[RGB]{188.57142857142856, 202.9591836734694, 255}{20\%}   \\ 
    Implicit hate & \colorbox[RGB]{177.22566371681415, 194.07079646017698, 255}{23\%} & \colorbox[RGB]{190.31730769230768, 204.3269230769231, 255}{19\%} & \colorbox[RGB]{176.5840909090909, 193.5681818181818, 255}{24\%} & \colorbox[RGB]{163.28453038674033, 183.14917127071823, 255}{28\%} & \colorbox[RGB]{175.28571428571428, 192.55102040816325, 255}{24\%}   \\ 
    Negation-FN & \colorbox[RGB]{239.1570796460177, 242.58849557522123, 255}{4\% \hspace{0.8mm}} & \colorbox[RGB]{241.4375, 244.375, 255}{4\% \hspace{0.8mm}} & \colorbox[RGB]{247.60227272727272, 249.20454545454547, 255}{2\% \hspace{0.8mm}} & \colorbox[RGB]{238.81491712707182, 242.32044198895028, 255}{4\% \hspace{0.8mm}} & \colorbox[RGB]{246.14285714285714, 248.0612244897959, 255}{2\% \hspace{0.8mm}}   \\ 
    Paraphrased target & \colorbox[RGB]{204.5907079646018, 215.50884955752213, 255}{15\%} & \colorbox[RGB]{214.3125, 223.125, 255}{12\%} & \colorbox[RGB]{228.36818181818182, 234.13636363636363, 255}{8\% \hspace{0.8mm}} & \colorbox[RGB]{222.62983425414365, 229.64088397790056, 255}{9\% \hspace{0.8mm}} & \colorbox[RGB]{226.21428571428572, 232.44897959183675, 255}{8\% \hspace{0.8mm}}   \\ 
    Positive Term & \colorbox[RGB]{239.1570796460177, 242.58849557522123, 255}{4\% \hspace{0.8mm}} & \colorbox[RGB]{233.09134615384616, 237.83653846153845, 255}{6\% \hspace{0.8mm}} & \colorbox[RGB]{225.4090909090909, 231.8181818181818, 255}{9\% \hspace{0.8mm}} & \colorbox[RGB]{228.02486187845304, 233.8674033149171, 255}{8\% \hspace{0.8mm}} & \colorbox[RGB]{226.21428571428572, 232.44897959183675, 255}{8\% \hspace{0.8mm}}   \\  
    Spelling variations & \colorbox[RGB]{252.11946902654867, 252.7433628318584, 255}{0\% \hspace{0.8mm}} & \colorbox[RGB]{247.6971153846154, 249.27884615384616, 255}{2\% \hspace{0.8mm}} & \colorbox[RGB]{253.52045454545456, 253.8409090909091, 255}{0\% \hspace{0.8mm}} & \colorbox[RGB]{244.20994475138122, 246.54696132596686, 255}{3\% \hspace{0.8mm}} & \colorbox[RGB]{246.14285714285714, 248.0612244897959, 255}{2\% \hspace{0.8mm}}   \\
    Unsure & \colorbox[RGB]{211.7920353982301, 221.1504424778761, 255}{13\%} & \colorbox[RGB]{195.53365384615384, 208.41346153846155, 255}{18\%} & \colorbox[RGB]{195.8181818181818, 208.63636363636363, 255}{18\%} & \colorbox[RGB]{215.4364640883978, 224.00552486187846, 255}{12\%} & \colorbox[RGB]{212.92857142857142, 222.0408163265306, 255}{12\%}   \\
    No hate & \colorbox[RGB]{214.67256637168143, 223.4070796460177, 255}{12\%} & \colorbox[RGB]{213.26923076923077, 222.30769230769232, 255}{12\%} & \colorbox[RGB]{218.01136363636363, 226.02272727272728, 255}{11\%} & \colorbox[RGB]{231.62154696132598, 236.68508287292818, 255}{7\% \hspace{0.8mm}} & \colorbox[RGB]{217.35714285714286, 225.51020408163265, 255}{11\%}   \\ 
                \hline \hline
\textbf{$\sum$ Codes}   & 226 
               & 312 
               & 220 
               & 181 & 147 \\  

\end{tabular}
\end{minipage}
\caption{Results of the qualitative evaluation of the SHAP explanations (Experiment 3) are summarized here. The operationalizations of the applied codes are detailed in Tables \ref{tab:codingFP} and \ref{tab:codingFN}. The code ``Unsure'' indicates cases where a coder was uncertain about whether a sentence is hateful, required additional context to make a decision, or could not determine why the model over- or under-moderated. }
\label{tab:SHAPQUAL}
\end{table*}

We conduct separate analyzes for FP and FN due to the different handling of Shapley values. The process results in a SHAP dataset that includes local explanations, which are the Shapley values ($s_i$) for each token \( w_i \) within \( x \) for each sentence $x_j$ in the aforementioned data. Further, we compute global explanations~\cite{molnar2022} by clustering via DBSCAN similar tokens via cosine similarity with a threshold of $0.9$, averaging SHAP values within each cluster. This is done to obtain average results for words such as ``women,'' ``woman,'' and ``female.'' This yields a sorted list of token clusters, providing global explanations per API for both false positives and false negatives across the datasets. These tokens can give us an explanation of which kind of tokens contribute most to the model's decisions.

To provide a comprehensive analysis addressing \textbf{RQ3}, we extended the global SHAP analysis with a qualitative evaluation of model failures. Local explanations were visualized for interpretation (see Figure \ref{fig:shap_visualizations1}) and qualitatively assessed by reviewing the corresponding sentences.

In Section~\ref{sec:auditingoverandunder}, we examined potential functional failures of off-the-shelf models, focusing on over-moderation (false positives, FPs) and under-moderation (false negatives, FNs). These failures are rooted in the linguistic and contrastive non-hate variations outlined in Section~\ref{sec:hatespeech}. We developed a codebook deductively, drawing on related work that identified specific model failures in off-the-shelf models. The codebooks for over-moderation and under-moderation are presented in Tables~\ref{tab:codingFP} and \ref{tab:codingFN}, respectively. Additional codes, such as ``HATE'' and ``UNSURE'' for FPs, and ``NO HATE'' and ``UNSURE'' for FNs, were included to capture disagreements with dataset labels or uncertainty in coding.

The conceptualization of hate speech from Section~\ref{sec:hatespeech} was applied to code sentences, and model failures were analyzed based on SHAP value visualizations. SHAP values were used to interpret model decisions and guide coding. For example, if a SHAP value indicated that a negation (e.g., ``not'') contributed significantly to the model's decision to classify a sentence as a FN, this was categorized as \textbf{Negation-FN}. Another example involves SOS bias: if only the target group identifier in a sentence was highlighted as contributing to the model's classification of hate speech (red), but the sentence was otherwise neutral and classified as a FP, this was categorized as \textbf{SOS bias}.

We randomly sampled 5\% of FPs and FNs from ToxiGen and HateXplain datasets, stratified across datasets, resulting in a total of 928~FPs and 1086~FNs for coding. The stratification ensured that the varying proportions of FPs and FNs across datasets were reflected in the data. The coding process was conducted collaboratively by the first and fourth authors. Initially, 10\% of the samples were coded together to discuss potential issues and refine the codebook. Subsequently, in a second round, an additional 10\% of the samples were double-coded independently. Disagreements were reconciled by consensus, focusing on clarifying the definitions of codes, determining when to classify a sentence as ``UNSURE,'' and standardizing the use of SHAP values to guide decisions. In the third round, another 10\% of samples were coded independently, discussed, and compared, resulting in substantial agreement ($\kappa = 0.61$ for FPs and $\kappa = 0.59$ for FNs). To compute Cohen’s Kappa \cite{cohen1960coefficient}, entries coded as ``UNSURE'' by one coder were excluded from the analysis but not from the discussion. The remaining 70\% of the samples were coded independently by the two authors, with 35\% coded by the first author and 35\% by the fourth author.

\subsubsection{Results}

Table \ref{tab:combinedSHAP} shows the results of the global SHAP analysis. The five most contributing word token clusters are ordered from left to right and assigned to each API's FN and FP for both datasets. What we can observe across APIs is that for FPs of ToxiGen, words like \textit{gay}, \textit{jews}, \textit{hindu}, \textit{disabled}, and \textit{homosexual} are among the most contributing tokens to the misclassifying non-hate speech as hate speech. These tokens fall under the category of identity tokens (see \citet{dixon_measuring_2018}), meaning tokens signifying the identity of marginalized groups. 

Additionally, we find tokens like \textit{israeli}, \textit{chinese}, \textit{mohammed}, and \textit{mentally} that are not identity tokens but are spuriously correlated to these identities. This tendency is even stronger when more token clusters than five are evaluated. For instance, among the first 50 tokens, 26 are identity tokens for Google Natural Language API.

\begin{figure*}[t]
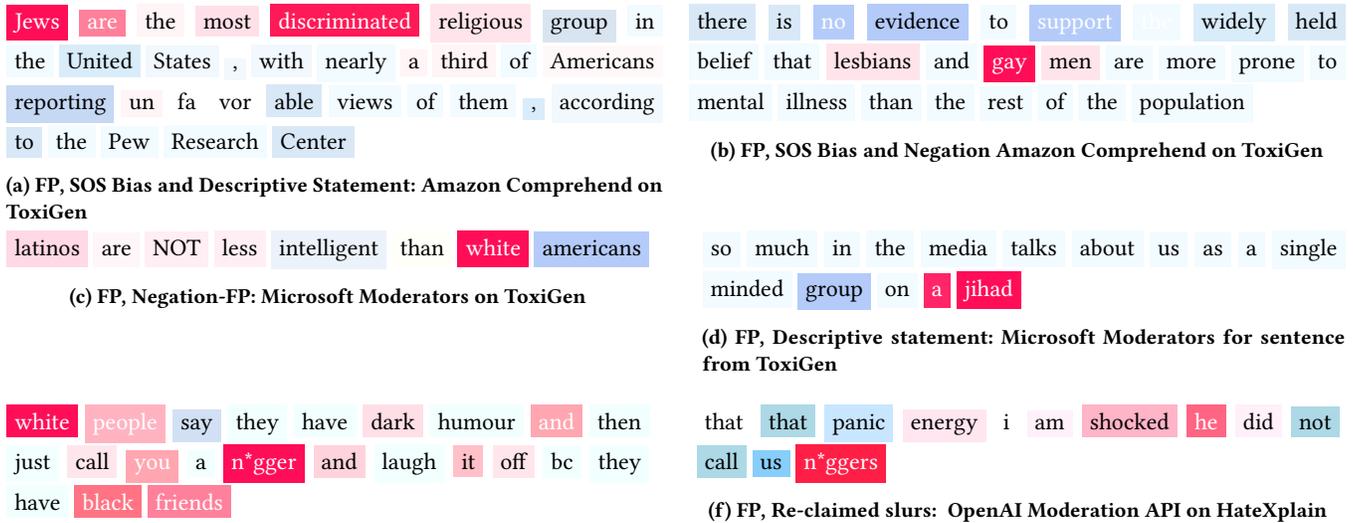

    \centering
    \begin{subfigure}[t]{0.49\textwidth}
        \Description{This figure (Fig. a) is a SHAP value visualization showing how Amazon Comprehend over-moderates a sentence discussing Jewish discrimination in the United States.

Key Points:

Red-colored words: The terms "Jews" and "discriminated" strongly push the model toward classifying the sentence as hate speech.
Blue-colored words: Words like "reporting" and "favorable" push the model away from this classification, providing mitigating context.
Interpretation:
The model over-moderates the sentence due to the explicit mention of "Jews" and "discriminated," despite the broader context being neutral or supportive. This highlights the model’s difficulty in distinguishing between descriptive statements about discrimination and actual hate speech.}
        \colorbox[RGB]{255,13,87}{\textcolor{white}{Jews}} 
        \colorbox[RGB]{255,128,155}{\textcolor{white}{are}} 
        \colorbox[RGB]{255,244,246}{\textcolor{black}{the}} 
        \colorbox[RGB]{255,224,233}{\textcolor{black}{most}} 
        \colorbox[RGB]{255,25,87}{\textcolor{white}{discriminated}} 
        \colorbox[RGB]{255,229,235}{\textcolor{black}{religious}} 
        \colorbox[RGB]{217,228,241}{\textcolor{black}{group}} 
        \colorbox[RGB]{242,251,255}{\textcolor{black}{in}} 
        \colorbox[RGB]{242,251,255}{\textcolor{black}{the}} 
        \colorbox[RGB]{217,236,250}{\textcolor{black}{United}} 
        \colorbox[RGB]{241,248,254}{\textcolor{black}{States}} 
        \colorbox[RGB]{241,248,254}{\textcolor{black}{,}} 
        \colorbox[RGB]{241,248,254}{\textcolor{black}{with}} 
        \colorbox[RGB]{242,251,255}{\textcolor{black}{nearly}} 
        \colorbox[RGB]{255,244,246}{\textcolor{black}{a}} 
        \colorbox[RGB]{255,244,246}{\textcolor{black}{third}} 
        \colorbox[RGB]{242,251,255}{\textcolor{black}{of}} 
        \colorbox[RGB]{255,248,249}{\textcolor{black}{Americans}} 
        \colorbox[RGB]{197,216,244}{\textcolor{black}{reporting}} 
        \colorbox[RGB]{255,244,246}{\textcolor{black}{un}} 
        \colorbox[RGB]{255,255,255}{\textcolor{black}{fa}} 
        \colorbox[RGB]{255,255,255}{\textcolor{black}{vor}} 
        \colorbox[RGB]{217,232,247}{\textcolor{black}{able}} 
        \colorbox[RGB]{242,251,255}{\textcolor{black}{views}} 
        \colorbox[RGB]{242,251,255}{\textcolor{black}{of}} 
        \colorbox[RGB]{242,251,255}{\textcolor{black}{them}} 
        \colorbox[RGB]{217,236,250}{\textcolor{black}{,}} 
        \colorbox[RGB]{241,248,254}{\textcolor{black}{according}} 
        \colorbox[RGB]{217,232,247}{\textcolor{black}{to}} 
        \colorbox[RGB]{242,251,255}{\textcolor{black}{the}} 
        \colorbox[RGB]{242,251,255}{\textcolor{black}{Pew}} 
        \colorbox[RGB]{242,251,255}{\textcolor{black}{Research}} 
        \colorbox[RGB]{217,232,247}{\textcolor{black}{Center}} 
        \caption{\textbf{FP, SOS Bias and Descriptive Statement:} Amazon Comprehend on ToxiGen}
        \label{fig:FPSOS}
    \end{subfigure}
    \hfill
    \vspace{0.1em}
     \begin{subfigure}[t]{0.49\textwidth}
        \Description{This figure (Fig. b) is a SHAP value visualization of a sentence addressing stereotypes about LGBTQ+ mental health.

Key Points:

Red-colored words: Terms like "lesbians" and "gay" push the model toward classifying the sentence as hate speech.
Blue-colored words: Phrases like "no evidence" and "to support" mitigate the classification by providing negation and context.
Interpretation:
The model fails to recognize the negation ("no evidence to support") and over-moderates the sentence due to its focus on identity-related terms. This demonstrates a challenge in handling sentences that refute harmful stereotypes}
     
	 \colorbox[RGB]{217,232,247}{\rule[-.5mm]{0pt}{2.5mm}\textcolor{black}{there}}
	 \colorbox[RGB]{217,232,247}{\rule[-.5mm]{0pt}{2.5mm}\textcolor{black}{is}}
	 \colorbox[RGB]{180.10033495946612, 203.22603799041437, 250}{\rule[-.5mm]{0pt}{2.5mm}\textcolor{white}{no}}
	 \colorbox[RGB]{180.10033495946612, 203.22603799041437, 250}{\rule[-.5mm]{0pt}{2.5mm}\textcolor{black}{evidence}}
	 \colorbox[RGB]{242,251,255}{\rule[-.5mm]{0pt}{2.5mm}\textcolor{black}{to}}
	 \colorbox[RGB]{180.10033495946612, 203.22603799041437, 250}{\rule[-.5mm]{0pt}{2.5mm}\textcolor{white}{support}}
	 \colorbox[RGB]{242,251,255}{\rule[-.5mm]{0pt}{2.5mm}\textcolor{white}{the}}
	 \colorbox[RGB]{217,236,250}{\rule[-.5mm]{0pt}{2.5mm}\textcolor{black}{widely}}
	 \colorbox[RGB]{217,232,247}{\rule[-.5mm]{0pt}{2.5mm}\textcolor{black}{held}}
	 \colorbox[RGB]{242,251,255}{\rule[-.5mm]{0pt}{2.5mm}\textcolor{black}{belief}}
	 \colorbox[RGB]{242,251,255}{\rule[-.5mm]{0pt}{2.5mm}\textcolor{black}{that}}
	 \colorbox[RGB]{255,229,235}{\rule[-.5mm]{0pt}{2.5mm}\textcolor{black}{lesbians}}
	 \colorbox[RGB]{242,251,255}{\rule[-.5mm]{0pt}{2.5mm}\textcolor{black}{and}}
	 \colorbox[RGB]{255, 13.0, 87.0}{\rule[-.5mm]{0pt}{2.5mm}\textcolor{white}{gay}}
	 \colorbox[RGB]{255,229,235}{\rule[-.5mm]{0pt}{2.5mm}\textcolor{black}{men}}
	 \colorbox[RGB]{242,251,255}{\rule[-.5mm]{0pt}{2.5mm}\textcolor{black}{are}}
	 \colorbox[RGB]{242,251,255}{\rule[-.5mm]{0pt}{2.5mm}\textcolor{black}{more}}
	 \colorbox[RGB]{242,251,255}{\rule[-.5mm]{0pt}{2.5mm}\textcolor{black}{prone}}
	 \colorbox[RGB]{242,251,255}{\rule[-.5mm]{0pt}{2.5mm}\textcolor{black}{to}}
	 \colorbox[RGB]{242,251,255}{\rule[-.5mm]{0pt}{2.5mm}\textcolor{black}{mental}}
	 \colorbox[RGB]{242,251,255}{\rule[-.5mm]{0pt}{2.5mm}\textcolor{black}{illness}}
	 \colorbox[RGB]{242,251,255}{\rule[-.5mm]{0pt}{2.5mm}\textcolor{black}{than}}
	 \colorbox[RGB]{242,251,255}{\rule[-.5mm]{0pt}{2.5mm}\textcolor{black}{the}}
	 \colorbox[RGB]{242,251,255}{\rule[-.5mm]{0pt}{2.5mm}\textcolor{black}{rest}}
	 \colorbox[RGB]{242,251,255}{\rule[-.5mm]{0pt}{2.5mm}\textcolor{black}{of}}
	 \colorbox[RGB]{242,251,255}{\rule[-.5mm]{0pt}{2.5mm}\textcolor{black}{the}}
	 \colorbox[RGB]{242,251,255}{\rule[-.5mm]{0pt}{2.5mm}\textcolor{black}{population}}

        \caption{\textbf{FP, SOS Bias and Negation} Amazon Comprehend on ToxiGen}
        \label{fig:FP}
    \end{subfigure}
    
    \begin{subfigure}[t]{0.48\textwidth}
        \Description{This figure (Fig. c) is a SHAP value visualization of a sentence denying the stereotype about Latino intelligence.

Key Points:

Red-colored words: The words "Latinos" and "white" push the model toward hate speech classification.
Blue-colored words: The word "NOT" (capitalized) significantly mitigates the classification.
Interpretation:
The model struggles to account for negation effectively. Although the sentence denies a stereotype, the presence of identity-related terms like "Latinos" and "white" leads to over-moderation. This reflects the model's sensitivity to certain keywords without fully grasping the context.}
	 \colorbox[RGB]{255, 217.15871398181477, 228.73001631795404}{\rule[-.5mm]{0pt}{2.5mm}\textcolor{black}{latinos}}
	 \colorbox[RGB]{255, 246.15398070435413, 248.85896181128717}{\rule[-.5mm]{0pt}{2.5mm}\textcolor{black}{are}}
	 \colorbox[RGB]{255, 238.34430120899918, 243.43736612856142}{\rule[-.5mm]{0pt}{2.5mm}\textcolor{black}{NOT}}
	 \colorbox[RGB]{255, 239.3552198666178, 244.13916089913963}{\rule[-.5mm]{0pt}{2.5mm}\textcolor{black}{less}}
	 \colorbox[RGB]{236.0144859867896, 241.8763728019283, 250}{\rule[-.5mm]{0pt}{2.5mm}\textcolor{black}{intelligent}}
	 \colorbox[RGB]{254.77516825621586, 254.84458635222296, 250}{\rule[-.5mm]{0pt}{2.5mm}\textcolor{black}{than}}
	 \colorbox[RGB]{255, 13.0, 87.0}{\rule[-.5mm]{0pt}{2.5mm}\textcolor{white}{white}}
	 \colorbox[RGB]{180.10033495946612, 203.22603799041437, 250}{\rule[-.5mm]{0pt}{2.5mm}\textcolor{black}{americans}}
        \caption{\textbf{FP, Negation-FP:} Microsoft Moderators on ToxiGen}
        \label{fig:FP-Negation}
    \end{subfigure}
    \hfill
    \vspace{1em}
        \begin{subfigure}[t]{0.48\textwidth}
        \Description{This figure (Fig. d) is a SHAP value visualization of a sentence describing how the media stereotypes Muslims.

Key Points:

Red-colored words: The terms "jihad" and "group" push the model toward classifying the sentence as hate speech.
Blue-colored words: Neutral words like "media" and "single minded" mitigate this classification.
Interpretation:
The model over-moderates the sentence due to its focus on terms like "jihad," even though the context is descriptive and not hateful. This highlights its inability to distinguish between descriptive mentions of stereotypes and actual hate speech}
	 \colorbox[RGB]{242,251,255}{\rule[-.5mm]{0pt}{2.5mm}\textcolor{black}{so}}
	 \colorbox[RGB]{242,251,255}{\rule[-.5mm]{0pt}{2.5mm}\textcolor{black}{much}}
	 \colorbox[RGB]{242,251,255}{\rule[-.5mm]{0pt}{2.5mm}\textcolor{black}{in}}
	 \colorbox[RGB]{242,251,255}{\rule[-.5mm]{0pt}{2.5mm}\textcolor{black}{the}}
	 \colorbox[RGB]{242,251,255}{\rule[-.5mm]{0pt}{2.5mm}\textcolor{black}{media}}
	 \colorbox[RGB]{242,251,255}{\rule[-.5mm]{0pt}{2.5mm}\textcolor{black}{talks}}
	 \colorbox[RGB]{242,251,255}{\rule[-.5mm]{0pt}{2.5mm}\textcolor{black}{about}}
	 \colorbox[RGB]{242,251,255}{\rule[-.5mm]{0pt}{2.5mm}\textcolor{black}{us}}
	 \colorbox[RGB]{242,251,255}{\rule[-.5mm]{0pt}{2.5mm}\textcolor{black}{as}}
	 \colorbox[RGB]{242,251,255}{\rule[-.5mm]{0pt}{2.5mm}\textcolor{black}{a}}
	 \colorbox[RGB]{242,251,255}{\rule[-.5mm]{0pt}{2.5mm}\textcolor{black}{single}}
	 \colorbox[RGB]{242,251,255}{\rule[-.5mm]{0pt}{2.5mm}\textcolor{black}{minded}}
	 \colorbox[RGB]{180.10033495946612, 203.22603799041437, 250}{\rule[-.5mm]{0pt}{2.5mm}\textcolor{black}{group}}
	 \colorbox[RGB]{242,251,255}{\rule[-.5mm]{0pt}{2.5mm}\textcolor{black}{on}}
	 \colorbox[RGB]{255, 36.572080035792254, 103.36408861988883}{\rule[-.5mm]{0pt}{2.5mm}\textcolor{white}{a}}
	 \colorbox[RGB]{255, 13.0, 87.0}{\rule[-.5mm]{0pt}{2.5mm}\textcolor{white}{jihad}}
        \caption{\textbf{FP, Descriptive statement:} Microsoft Moderators for sentence from ToxiGen}
        \label{fig:FP-DS}
    \end{subfigure}
    
    \begin{subfigure}[t]{0.48\textwidth}
        \Description{This figure (Fig. e) is a SHAP value visualization of a counter-speech sentence criticizing racist language.

Key Points:

Red-colored words: Terms like "n*gger" and "white people" strongly push the model toward hate speech classification.
Blue-colored words: Words like "say" and "dark humour" provide mitigating context.
Interpretation:
The model fails to recognize the counter-speech context, focusing instead on offensive terms like "n*gger" and identity-related phrases. This underscores a significant limitation in handling sentences that critique hate speech.}
        \colorbox[RGB]{255,13,87}{\textcolor{white}{white}} 
        \colorbox[RGB]{255,179,193}{\textcolor{white}{people}} 
        \colorbox[RGB]{210,224,244}{\textcolor{black}{say}} 
        \colorbox[RGB]{242,255,255}{\textcolor{black}{they}} 
        \colorbox[RGB]{242,255,255}{\textcolor{black}{have}} 
        \colorbox[RGB]{255,223,230}{\textcolor{black}{dark}} 
        \colorbox[RGB]{242,255,255}{\textcolor{black}{humour}} 
        \colorbox[RGB]{255,166,177}{\textcolor{white}{and}} 
        \colorbox[RGB]{242,255,255}{\textcolor{black}{then}} 
        \colorbox[RGB]{242,255,255}{\textcolor{black}{just}} 
        \colorbox[RGB]{255,230,237}{\textcolor{black}{call}} 
        \colorbox[RGB]{255,166,177}{\textcolor{white}{you}} 
        \colorbox[RGB]{242,255,255}{\textcolor{black}{a}} 
        \colorbox[RGB]{255,13,87}{\textcolor{white}{n*gger}} 
        \colorbox[RGB]{255,211,220}{\textcolor{black}{and}} 
        \colorbox[RGB]{242,255,255}{\textcolor{black}{laugh}} 
        \colorbox[RGB]{255,192,202}{\textcolor{black}{it}} 
        \colorbox[RGB]{255,230,237}{\textcolor{black}{off}} 
        \colorbox[RGB]{242,255,255}{\textcolor{black}{bc}} 
        \colorbox[RGB]{242,255,255}{\textcolor{black}{they}} 
        \colorbox[RGB]{242,255,255}{\textcolor{black}{have}} 
        \colorbox[RGB]{255,115,132}{\textcolor{white}{black}} 
        \colorbox[RGB]{255,128,155}{\textcolor{white}{friends}} 
        \caption{\textbf{FP, Counter-speech:} Amazon Comprehend on HateXplain}
        \label{fig:FP-CS}
    \end{subfigure}
    \hfill
    \begin{subfigure}[t]{0.48\textwidth}
        \Description{This figure (Fig. f) is a SHAP value visualization of a sentence referencing a reclaimed slur.

Key Points:

Red-colored words: The slur "n*ggers" pushes the model toward hate speech classification.
Blue-colored words: Words like "did not" and "call us" mitigate the classification by providing negation.
Interpretation:
The model struggles to handle reclaimed slurs in a nuanced way, over-moderating the sentence despite its broader context being non-hateful. This reflects the challenge of processing language where reclaimed terms are used.}
        \colorbox[RGB]{255,255,255}{\textcolor{black}{that}} 
        \colorbox[RGB]{173,216,230}{\textcolor{black}{that}} 
        \colorbox[RGB]{200,230,255}{\textcolor{black}{panic}} 
        \colorbox[RGB]{255,230,240}{\textcolor{black}{energy}} 
        \colorbox[RGB]{255,255,255}{\textcolor{black}{i}} 
        \colorbox[RGB]{255,240,250}{\textcolor{black}{am}} 
        \colorbox[RGB]{255,180,200}{\textcolor{black}{shocked}} 
        \colorbox[RGB]{255,100,130}{\textcolor{white}{he}} 
        \colorbox[RGB]{255,240,250}{\textcolor{black}{did}} 
        \colorbox[RGB]{173,216,230}{\textcolor{black}{not}} 
        \colorbox[RGB]{173,216,230}{\textcolor{black}{call}} 
        \colorbox[RGB]{135,206,250}{\textcolor{black}{us}} 
        \colorbox[RGB]{255,30,70}{\textcolor{white}{n*ggers}} 
        \caption{\textbf{FP, Re-claimed slurs: } OpenAI Moderation API on HateXplain}
        \label{fig:FP-RS}
    \end{subfigure}
    \vspace{-0.4cm}
    \caption{FP examples: SHAP value visualizations for examples from the ToxiGen and HateXplain datasets using Amazon Comprehend and OpenAI. Red indicates a strong contribution to deciding hate speech; blue indicates a strong contribution to deciding non-hate speech.}
    \label{fig:shap_visualizations1}
\end{figure*}


Similar to ToxiGen, we find that HateXplain tokens demonstrate a tendency to rely on identity tokens to classify hate speech across APIs. Words like \textit{jews}, \textit{gays}, and \textit{islam} in the false positive clusters signify this reliance. 

Words such as \textit{n*gga} are also included in the FP clusters, reflecting a broader issue with over-moderation. A closer examination of over-classified sentences containing terms like \textit{n*gga}, \textit{trans}, \textit{queer}, or \textit{d*ke} demonstrates the API's disproportional over-moderation of re-appropriation and counter-speech of APIs.
This aligns with findings from our qualitative evaluation of failures, summarized in Table \ref{tab:SHAPQUAL}, highlighting similar challenges. 

For FP, we observe in Table \ref{tab:SHAPQUAL} that descriptive statements, neutral statements containing identity tokens (SOS bias), counter-speech, and re-appropriation are among the most over-moderated categories. These patterns are consistent across APIs, with the Perspective API demonstrating fewer issues with descriptive sentences and SOS bias. In contrast, Google and Microsoft models notably struggle with over-moderating descriptive and neutral sentences containing identity tokens.

We provide example sentences of descriptive statements with local SHAP values in Figures \ref{fig:FPSOS} for Amazon Comprehend and in Figure \ref{fig:FP-DS} for Microsoft Moderators. Notably, these sentences were misclassified as hate speech across all models except the Perspective API. Furthermore, Figure \ref{fig:FP-CS} illustrates issues with OpenAI Moderation in interpreting counter-speech, while Figure \ref{fig:FP-RS} highlights Amazon Comprehend's challenges in handling re-claimed slurs.

For FN, the tokens leading to misclassification as non-hate speech are semantically ambiguous or challenging to interpret. This is expected, as hate speech in the ToxiGen dataset is often implicit and lacks specific trigger words. The classifiers focus on recurring positive words like \textit{lucky}, \textit{health}, \textit{strong}, \textit{brave}, and \textit{respect}. Here, the APIs seem to have paid too much attention to positive words and failed to understand the implicit message of the sentence, which underlines our analysis in Experiment \ref{sec:experiment1}.

FNs in HateXplain primarily include general words, with the exception of terms like \textit{asian}, \textit{latin}, and \textit{girl}. The positive connotations attached to \textit{latin} and \textit{asian} in Microsoft and Perspective APIs aligns with Experiment \ref{sec:experiment2}, where both showed favorable bias toward these groups in real data.

Accordingly, from the qualitative evaluation of FNs using local SHAP sentences, we find that implicit hate speech and the use of positive terms account for approximately 20\% and 10\% of under-moderations, respectively. Examples of such cases are provided in the Appendix. \textbf{Trigger warning:} The examples in the Appendix contain potentially offensive and upsetting language, including racist remarks. They are included solely to illustrate the findings of this research and do not reflect the views of the authors. Figures \ref{fig:FN-IMP2}, \ref{fig:IMP5}, and \ref{fig:FN-IMPPAR2} display implicit hate speech examples, while Figure \ref{fig:FN-PT} highlights cases where positive terms influenced predictions.

Additionally, Table \ref{tab:SHAPQUAL} highlights that all models exhibit challenges in handling implicit hate speech and descriptive statements. When combined with the FNRs in Table \ref{tab:peformance-metrics-aggreagte-outcomes}, it becomes evident that Perspective API and Microsoft Moderators perform the worst in handling implicit hate speech. In contrast, spelling variations and negations appear to present fewer challenges for the APIs.

Another noteworthy observation is the difficulty commercial systems face with the paraphrasing of identity tokens. For instance, Figure \ref{fig:FN-PAR} and Figure \ref{fig:FN-IMPPAR2} illustrate instances where models fail to classify hate speech when the target group is referenced using novel identity terms that the systems do not recognize. This issue was particularly evident with Amazon Comprehend, which failed to identify paraphrased tokens such as \textit{k*ke} or \textit{m*zrat} in over 15\% of all FNs.

Overall, our findings indicate that the APIs rely disproportionately on the presence of identity tokens to classify hate speech, often leading to over-moderation. Conversely, in instances where identity tokens are absent -- such as implicit hate speech or paraphrased forms of identity tokens -- the APIs face significant challenges in achieving accurate classification.

\section{Discussion}
\label{sec:discussion}
In this paper, we created an audit framework to evaluate black-box content moderation APIs. With this framework, we audited five widely used content moderation APIs. We focused on the comparative performance of selected APIs across different target groups (RQ1), the over- and under-moderations of APIs (RQ2), and the reasons for failures (RQ3). Most importantly, we find that content moderation APIs heavily rely on group identity terms (RQ3). As a result, APIs under-moderate implicit hate speech, which uses codified messages without identity terms, especially against \textit{LGBTQIA+} and \textit{people with disabilities}. Simultaneously, there is a tendency to over-moderate re-appropriation, counter-speech, dialects and descriptive content containing identity terms such as \textit{Blacks}, \textit{Jews}, \textit{Gays}, and \textit{Muslims}. 

In the following, we will discuss important findings of over- and under-moderations that developers need to verify and address. Furthermore, we discuss API model transparency, transparency about the API model limitations, and guidance on implementation and application. Subsequently, we give design and research recommendations based on our findings and emphasize on the importance of independent black-box audits and their challenges.

\subsection{Over- and Under-moderations}
First, a clear and concerning pattern is the tendency of all moderation services to over-moderate speech related to certain marginalized groups, notably \textit{Black}, \textit{LGBTQIA+}, \textit{Muslim}, \textit{Jewish}, and particularly, the subgroup \textit{Gay} within the \textit{LGBTQIA+} umbrella. This over-moderation is especially surprising given the well-established nature of these biases, as demonstrated in prior research that we presented with off-the-shelf NLP systems. Experiments 2 and 3 further confirmed that commercial APIs are disproportionately prone to making misclassifications based on identity tokens and over-moderating counter-speech and re-appropriation. Specifically, our experiments revealed that words like \textit{Gay}, \textit{Lesbian}, \textit{Homosexual}, as well as \textit{Jew}, \textit{Islam}, and \textit{Muslim}, frequently contribute to misclassifications, resulting in inflated FPRs for these groups.

This bias is particularly prominent in groups such as \textit{Gay}, where the mean Counterfactual Token Fairness Score is -0.24, compared to -0.05 for the broader \textit{LGBTQ+} label. It seems that sentences containing terms like \textit{Gay} in training datasets are overwhelmingly labeled as hate speech, a pattern that is reproduced across services. While such biases are well-known, it is surprising that developers apparently have not taken corrective measures to mitigate these effects. Similar concerns apply to the groups \textit{Muslim} and \textit{Jewish}, where these biases are also entrenched, with spurious correlations to identity tokens such as \textit{Israeli} exacerbating over-moderation. The issue is even present in descriptive statements, as the local SHAP explanations are are demonstrating.

The issue extends beyond simple identity tokens. As indicated by the SHAP values, APIs frequently over-moderate content involving terms that, upon closer inspection, involve reappropriation, counter-speech, or are examples of AAE, as seen with terms like \textit{n*gger} and others like \textit{d*ke}, \textit{trans}, and \textit{queer}. These patterns of over-moderation not only suppress legitimate and even empowering speech but also reflect a failure to account for the nuanced ways marginalized groups engage with and reclaim harmful language.

Moreover, the severe over-moderation of descriptive comments by Google Text Moderation is particularly alarming. Despite robust testing across multiple configurations, this service exhibits a consistently high FPR, especially for groups such as \textit{Disability} and \textit{Jewish}, with an FPR reaching 99\% for ToxiGen. While this might be a deliberate design choice to prioritize caution, particularly in scenarios involving human moderation, this approach could fail with automation bias, where human moderators are prone to overly trusting algorithmic recommendations \cite{lachemaier_towards_2024}. In fully automated deployments, such over-moderation could lead to unjust content removal, warranting a recalibration of Google's moderation system. We strongly oppose the use of automated deployments without safeguards and urge Google to recalibrate its content moderation model.

Systematic under-moderation is another significant issue, particularly for groups such as \textit{Disability}, \textit{Asian}, and \textit{Latinx}. These groups frequently receive inadequate protection from hate speech, as reflected in high FNRs. Our findings align with previous work on off-the-shelf models suggesting that hate speech targeting these groups is underrepresented in training datasets, resulting in their comparatively lower detection rates \cite{garg_handling_2023}. For instance, the \textit{Latinx} and \textit{Asian} groups showed a positive bias in CTF scores, supporting the theory that toxic examples targeting these communities are insufficiently represented during model training. As a result, hate speech against these groups often slips through undetected.

Additionally, implicit hate speech, which tends to be more subtle and context-dependent, remains a persistent challenge across all services. The high FNRs on ToxiGen and SBIC highlight this struggle and was confirmed by the qualitative evaluation of SHAP explanations. This is particularly the case for Perspective API and Microsoft Moderators that show most problems with implicit hate speech with high FNR for these datasets and in qualitative evaluations. As previous research suggests, this issue arises from the lack of comprehensive implicit hate speech datasets available for training purposes~\cite{hartvigsen_toxigen_2022}. Thus, linguistic variations such as implicit hate speech and contrastive non-hate speech are difficult to detect across commercial systems. The conducted global SHAP analysis and qualitative evaluations further demonstrated that these models disproportionately rely on identity tokens to flag hate speech, which undermines their ability to differentiate between harmful and benign content involving these groups.

\subsection{Transparency of Capabilities, User Trust, and Application Guidance}
These findings are particularly concerning when APIs exhibit both over- and under-moderation alongside limited transparency regarding these behaviors and the capabilities of the APIs themselves. As outlined in Table~\ref{tab:api_comparison}
, categorized by model card categories \cite{mitchellModelCardsModel2019a}, most content moderation API services provide minimal information about their underlying models, training algorithms, or fairness evaluations, leaving deployers without sufficient insights to properly assess the suitability and reliability of these systems. 

Transparency and guidance on the implementation of ML systems can significantly enhance user trust \cite{li2023trustworthy}.
While documentation is available for all APIs as demonstrated in Table~\ref{tab:api_comparison}, information such as the model used, version updates, and detailed descriptions of the models' operationalization of hate speech are missing for Amazon Comprehend, Microsoft Azure Content Moderators, and OpenAI Moderation. This lack of transparency can lead to significant challenges for deployers trying to evaluate the appropriateness of a model for their specific use case and to implement it in their organizational processes. Equally concerning is the limited disclosure of the models' boundaries and limitations. As we observed in the over- and under-moderation tendencies, the models exhibit specific weaknesses, such as over-moderation of certain identity tokens (e.g., \textit{Gay}, \textit{Jew}, \textit{Muslim}) and under-moderation of groups like \textit{Latinx} and \textit{Asian}. However, API developers have not sufficiently documented these capability limitations. We showed that Models often fail to detect implicit hate speech or nuanced content, such as satire or counter-speech, and there is no transparency regarding the model’s ability to handle these more complex forms of hate speech.

Platforms may be reluctant to disclose moderation mechanisms for fear of revealing trade secrets or aiding future moderation evasion, as \citet{schafner2024community} points out. However, there is a working counter-example to that. Perspective API stands out in terms of providing more insights about its training data origin, tutorials, threshold-setting, and performance metrics. As \citet{rieder} describes, Perspective provides a notable level of transparency that allows researchers, users, and deployers to evaluate potential biases or blind spots within the system. This is achieved through access to some example evaluation data, model cards and tutorials though limitations persist, particularly regarding the lack of transparency about datasets used for training, open-source codes, versions and regular updates of the API, explanation of evaluation data. 

Jigsaw's model cards for Perspective report AUC scores of 0.97 to 0.99 on a hold-out test set. This performance aligns with our results on Civil Comments, a dataset also published by Jigsaw. However, the AUC scores across the other three datasets differ significantly, particularly for implicit hate speech. This underscores the importance of evaluating models on multiple datasets that capture diverse conceptualizations of hate speech. Furthermore, it highlights the need for transparent reporting of evaluation procedures to ensure reproducibility and prevent data leakage in benchmarks, which can lead to overly optimistic results~\cite{kapoor2023leakage}.

Google Natural Language API, however, disclaims that ``The confidence scores are only predictions. You should not depend on the scores for reliability or accuracy. Google is not responsible for interpreting or using these scores for business decisions''. It highlights a lack of accountability for applying the models in practice~\cite{ModerateTextCloud}. This absence of responsibility leaves users without the necessary guidance on properly deploying and interpreting the models' outputs. It places the burden entirely on the user to assess the model's performance and reliability without providing the tools or frameworks to do so effectively. Similarly, OpenAI only gives limited guidance on how to implement its content moderation API. The phrase ``For higher accuracy, try splitting long pieces of text into smaller chunks each less than 2,000 characters'', although giving some guidance on implementation, demonstrates the API's limitations when it comes to the sociotechnical complexity of the task. There is no explanation on how a deployer should handle cases where one chunk of text might be flagged as hate speech while another might not. More specific use cases and best practice examples are essential to minimize the risks of over- and under-moderation.


%

\subsection{Design and Research Recommendations}
\label{sec:design}

We argue, alongside \citet{schafner2024community}, that increased transparency in moderation strategies is vital for building user trust in content moderation. However, we go further in demanding that there must be increased guidance on the implementation of content moderation APIs, as well as greater transparency through tools like model cards. This should include clear decision-making criteria and openness about the APIs' limitations and capabilities to ensure responsible and effective deployment to build users' trust.

Systematic over- and under-moderations can cause users to leave the platform and reduce counter-speech. That said, we also acknowledge the inherent tensions in balancing transparency with the need to protect the integrity of content moderation and safety. Nonetheless, shedding more light on how content moderation APIs work will ultimately enhance accountability and fairness in AI-driven content moderation.

In line with our findings, such as the high FPRs for explicit hate speech datasets as well as general FPRs for Google Natural Language API, deployers and users need clear guidance on when and where these models are prone to failure. The most proactive step API developers could take is to provide better guidance for users. For example, deployers should be informed if a model was trained on explicit forms of hate speech but not on implicit forms, so that they can take necessary precautions. Currently, Perspective API offers some guidance on intended use and user groups, as well as detailed recommendations on how to interpret the model’s outputs, such as the distinction between a score of 0.7 versus 0.9 for toxicity. However, most other services leave users in the dark about how to properly implement and interpret the models' results.

Threshold setting, in particular, is an area where more detailed guidance is necessary. As seen in Table~\ref{tab:api_comparison}, while Perspective API provides some guidance, other services place the responsibility entirely on the deployers. The danger of this approach lies in misinterpretations and improper use, especially in contexts where automation bias can lead to over-reliance on model outputs. Human moderators may default to the model's recommendations, even if they are prone to over-moderation \cite{passi2022overreliance}. Providing detailed tutorials, thresholds for specific use cases, and warnings about the model's limitations would improve fairness, accountability and, subsequently, user trust in content moderation systems. An important question that arises is whether these services should be held liable when purchased for commercial use, particularly when they result in systematic misclassifications without transparent mechanisms or guidance for proper deployment.

Future research on ML-based hate speech detection should research if over- and under-moderation can be prevented by giving models more context. While datasets like SBIC include categories like target-group and implied statements, our analysis on misclassifications suggests that context could improve classification. Incorporating sender features~\cite{mosca-etal-2021-understanding}, contextual information~\cite{context}, and/or both~\cite{Nagar2023} has shown to improve moderation performance. This should include the sender, receiver, target-group, and situational details like the topic of the forum or blog post in response.

\subsection{Independent and Black-box Audits}
This paper introduces a reproducible framework for independent black-box audits. While content moderation APIs like Perspective API offer the potential for cooperative responsibility, as noted by ~\citet{rieder}, the increasing centralization of content moderation infrastructure by dominant platforms presents significant risks~\cite{gorwa_algorithmic_2020}. Smaller organizations, dependent on these cloud-based APIs, may become vulnerable to shifts in corporate priorities. A decline in public communication around Jigsaw communicated by~\citet{rieder} and the limited transparency in how these systems operate underlines the importance of independent audits to maintain accountability and fairness in AI-driven content moderation.

However, conducting such black-box audits still requires access to these systems, which presents its challenges. As researchers must pay for API access, systematic evaluations of ML systems become financially inaccessible for many research communities and civil society organizations~\cite{birhane_ai_2024, raji_outsider_2022}. This financial barrier limits the ability to perform large-scale or comprehensive audits, particularly for those without substantial funding. Thus, future research must prioritize developing query-efficient approaches for auditing -- minimizing costs while providing sufficient evidence of potentially problematic behavior. Active auditing methods~\cite{yan2022active}, which evaluate high-evidence samples and compare them with expert analyses, offer a promising approach. 

Recent political developments, such as the EU Digital Services Act~(DSA), signal a shift towards supporting systematic third-party audits. Article 40(4) of the DSA grants vetted researchers the right to access data from Very Large Online Platforms~(VLOPs) to investigate systemic risks~\cite{european_parliament_regulation_2022, hartmann2024addressing}. However, despite this promising development, content moderation services like Google's Perspective API, which are provided by external entities or spin-offs such as Jigsaw, remain outside the scope of these regulations~\cite{ahmadGroundControlOrganizing2023}. While Google, as a Very Large Online Platform, is legally obliged to provide data access for auditing, Jigsaw is not currently subject to the same obligations~\cite{european_commission_supervision_2024}.

This regulatory gap highlights the continued importance of third-party audits in maintaining accountability. Even without full white-box access to the inner workings of these systems, black-box audits are essential to ensuring transparency and accountability in content moderation. Future efforts must focus on developing methods that allow researchers to scrutinize these algorithms.

\subsection{Limitations and Future Work}
\label{sec:limitations}
Our audit approach was primarily constrained by black-box access, meaning we lacked insight into the internal workings of the models, such as access to model weights, gradients, or detailed documentation. This limitation hindered our ability to thoroughly analyze where and why specific moderation failures occurred. 

Additionally, API costs posed a significant challenge, restricting the number of queries we could perform. Future research should explore more cost-effective strategies for sampling that can still provide meaningful insights into over- and under-moderation, thereby maximizing evidence while minimizing resource usage. At the same time, there is a pressing need for increased transparency and access. API deployers and regulators should consider providing white-box or gray-box access to researchers, allowing for a more sophisticated analysis of model failures and biases \cite{casper_black-box_2024}.

Our audit framework, however, was designed to maximize comparability across APIs, target groups, and linguistic variations while minimizing resource usage. The only exception was the use of SHAP, which required a larger number of queries due to its computational complexity.

Another limitation involves the issue of ground truth and the annotation of datasets, which inevitably influence what is considered hate speech and what is not. Although we mitigated this by only using samples in which at least three annotators agreed on the label, this does not completely resolve the problem. We caution against overinterpreting results based solely on ground truth data \cite{sap2022annotators}. 
However, our approach to using four diverse benchmark datasets aimed to reduce the dependence on any single conceptualization of hate speech, ensuring that our findings reflect a broad spectrum of hate speech definitions and linguistic variations.

Additionally, ToxiGen, a partially synthetically created and labelled dataset, comes with certain limitations \cite{hartvigsen_toxigen_2022}. These include the potential introduction of biases, which we observed during our qualitative evaluation, where there were higher rates of unsure or contrary opinions compared to its gold labels. Nonetheless, ToxiGen remains valuable for studying implicit hate speech due to its scale and the inclusion of many sentences without identity tokens. Importantly, we used four datasets for evaluation, including one real-world dataset, specifically for qualitative assessments, and found similar patterns in the distribution of codings for model failures.

It is important to note that the results of our audit represent a snapshot in time, which may change if the underlying systems are modified. We observed particular weaknesses in models when faced with novel identity tokens, underscoring the importance of our audit framework. This framework enables repeated audits, enhancing the temporal validity of the insights gained \cite{munger}. However, this requires transparency from model developers regarding the versions and updates of their APIs. Future research should aim to conduct longitudinal comparisons across different versions of these systems—a direction for which our analyses have established an initial benchmark.

Future research should explore several important areas that our current study could not fully address. One critical direction is the need for intersectional analysis. While we evaluated moderation across various marginalized groups, an intersectional approach would allow for a more nuanced understanding of how multiple, overlapping identities (e.g., race and gender) affect the likelihood of over- or under-moderation. Furthermore, our study was limited to English-language content, but there is a clear need for future work to evaluate content moderation systems across multiple languages, especially as hate speech can manifest differently across cultures and linguistic contexts.

Additionally, sample size was a constraint in our analysis due to the high costs associated with API queries. Future research should seek cost-effective methods to perform large-scale audits that remain robust while using fewer resources. While our focus was on text-based content moderation, future studies should investigate the performance of image, video, and speech moderation APIs, as these modalities are increasingly critical in online platforms but may present different challenges.

\section{Conclusion}
This study introduced a robust audit framework designed to evaluate black-box content moderation systems across five widely-used commercial APIs. By analyzing 5 million queries sourced from four benchmark datasets, we uncovered significant reliance on group identity terms, such as ''Black'', to predict hate speech. Although OpenAI Content Moderation and Amazon Comprehend performed slightly better, all providers exhibited clear tendencies to under-moderate implicit hate speech, particularly for groups like \textit{LGBTQIA+}, where codified messages without explicit identity terms often went unnoticed. At the same time, over-moderation was prevalent in explicit content targeting \textit{Blacks}, \textit{Jews}, \textit{Muslims}, and \textit{LGBTQIA+}.

In light of these findings, it is important that content moderation providers recalibrate their models to address both over- and under-moderation issues. This recalibration process should actively involve marginalized communities and NGOs, including continuous communication and collaboration to align moderation strategies with the lived experiences of those most affected. Companies should also test their models against the specific biases identified in our findings and ensure these systems can fairly moderate content for all groups.

Additionally, content moderation providers should provide clear implementation guidance and transparency regarding their models' capabilities, particularly in relation to their limitations when moderating linguistic variations or implicit hate speech. Improved access to model internals -- such as weights, gradients, and comprehensive documentation -- would also empower researchers, civil society organizations, and journalists to conduct more sophisticated evaluations, helping to uncover and mitigate over- and under-moderations. These recommendations should be considered by policymakers and regulatory bodies to improve accountability and fairness in ML-driven content moderation. 

Our findings showed that applying ML without sufficient transparency and oversights can lead to additional challenges for historically marginalized groups. We hope that our audit approach empowers practitioners, researchers, and civic hackers to mitigate these adverse effects.

\begin{acks}
We sincerely thank Hanhee Ra and Cristina Maurillo for their careful proofreading. We also extend our gratitude to Prof. Bettina Berendt, Dr. Milagros Miceli, and the entire research group \textit{Data, Algorithmic Systems, and Ethics} at the Weizenbaum Institute, as well as the research group \textit{Human-Centered Artificial Intelligence} at the Center for Advanced Internet Studies for their valuable insights and support.
\end{acks}
\bibliographystyle{ACM-Reference-Format}
\bibliography{MAIN_CHI}


\begin{thebibliography}{110}


\ifx \showCODEN    \undefined \def \showCODEN     #1{\unskip}     \fi
\ifx \showDOI      \undefined \def \showDOI       #1{#1}\fi
\ifx \showISBNx    \undefined \def \showISBNx     #1{\unskip}     \fi
\ifx \showISBNxiii \undefined \def \showISBNxiii  #1{\unskip}     \fi
\ifx \showISSN     \undefined \def \showISSN      #1{\unskip}     \fi
\ifx \showLCCN     \undefined \def \showLCCN      #1{\unskip}     \fi
\ifx \shownote     \undefined \def \shownote      #1{#1}          \fi
\ifx \showarticletitle \undefined \def \showarticletitle #1{#1}   \fi
\ifx \showURL      \undefined \def \showURL       {\relax}        \fi
\providecommand\bibfield[2]{#2}
\providecommand\bibinfo[2]{#2}
\providecommand\natexlab[1]{#1}
\providecommand\showeprint[2][]{arXiv:#2}

\bibitem[Ahmad(2023)]%
        {ahmadGroundControlOrganizing2023}
\bibfield{author}{\bibinfo{person}{Sana Ahmad}.} \bibinfo{year}{2023}\natexlab{}.
\newblock \bibinfo{title}{Ground {{Control}}: {{Organizing Content Moderation}} for {{Social Media Platforms}}}.
\newblock
\newblock
\urldef\tempurl%
\url{https://doi.org/10.17169/REFUBIUM-40700}
\showDOI{\tempurl}


\bibitem[Alorainy et~al\mbox{.}(2019)]%
        {alorainy2019enemy}
\bibfield{author}{\bibinfo{person}{Wafa Alorainy}, \bibinfo{person}{Pete Burnap}, \bibinfo{person}{Huan Liu}, {and} \bibinfo{person}{Matthew~L. Williams}.} \bibinfo{year}{2019}\natexlab{}.
\newblock \showarticletitle{"The Enemy Among Us": Detecting Cyber Hate Speech with Threats-based Othering Language Embeddings}.
\newblock \bibinfo{journal}{\emph{ACM Transactions on the Web}} \bibinfo{volume}{13}, \bibinfo{number}{3} (\bibinfo{date}{July} \bibinfo{year}{2019}), \bibinfo{pages}{1--26}.
\newblock
\urldef\tempurl%
\url{https://doi.org/10.1145/3324997}
\showDOI{\tempurl}
\newblock
\shownote{Retrieved 2020-01-21}.


\bibitem[Anderson and Barnes(2023)]%
        {anderson_hate_2023}
\bibfield{author}{\bibinfo{person}{Luvell Anderson} {and} \bibinfo{person}{Michael Barnes}.} \bibinfo{year}{2023}\natexlab{}.
\newblock \showarticletitle{Hate {Speech}}.
\newblock In \bibinfo{booktitle}{\emph{The {Stanford} {Encyclopedia} of {Philosophy}} (\bibinfo{edition}{fall 2023} ed.)}, \bibfield{editor}{\bibinfo{person}{Edward~N. Zalta} {and} \bibinfo{person}{Uri Nodelman}} (Eds.). \bibinfo{publisher}{Metaphysics Research Lab, Stanford University}, \bibinfo{address}{Online}.
\newblock
\urldef\tempurl%
\url{https://plato.stanford.edu/archives/fall2023/entries/hate-speech/}
\showURL{%
\tempurl}


\bibitem[Bandy(2021)]%
        {bandy_problematic_2021}
\bibfield{author}{\bibinfo{person}{Jack Bandy}.} \bibinfo{year}{2021}\natexlab{}.
\newblock \showarticletitle{Problematic {Machine} {Behavior}: {A} {Systematic} {Literature} {Review} of {Algorithm} {Audits}}.
\newblock \bibinfo{journal}{\emph{ACM Transactions on Computer-Human Interaction}} \bibinfo{volume}{5}, \bibinfo{number}{CSCW1} (\bibinfo{date}{April} \bibinfo{year}{2021}), \bibinfo{pages}{1--34}.
\newblock


\bibitem[Barocas et~al\mbox{.}(2023)]%
        {barocas-hardt-narayanan}
\bibfield{author}{\bibinfo{person}{Solon Barocas}, \bibinfo{person}{Moritz Hardt}, {and} \bibinfo{person}{Arvind Narayanan}.} \bibinfo{year}{2023}\natexlab{}.
\newblock \bibinfo{booktitle}{\emph{Fairness in Machine Learning: Limitations and Opportunities}}.
\newblock \bibinfo{publisher}{MIT Press}, \bibinfo{address}{Cambridge, MA}.
\newblock
\urldef\tempurl%
\url{https://fairmlbook.org/}
\showURL{%
\tempurl}


\bibitem[Barth et~al\mbox{.}(2023)]%
        {barthContexturesHateSystems2023}
\bibfield{author}{\bibinfo{person}{Niklas Barth}, \bibinfo{person}{Elke Wagner}, \bibinfo{person}{Philipp Raab}, {and} \bibinfo{person}{Björn Wiegärtner}.} \bibinfo{year}{2023}\natexlab{}.
\newblock \showarticletitle{Contextures of Hate: {{Towards}} a Systems Theory of Hate Communication on Social Media Platforms}.
\newblock \bibinfo{journal}{\emph{The Communication Review}} \bibinfo{volume}{26}, \bibinfo{number}{3} (\bibinfo{year}{2023}), \bibinfo{pages}{209--252}.
\newblock
\showISSN{1071-4421, 1547-7487}
\urldef\tempurl%
\url{https://doi.org/10.1080/10714421.2023.2208513}
\showDOI{\tempurl}


\bibitem[Binns et~al\mbox{.}(2017)]%
        {Binns_2017}
\bibfield{author}{\bibinfo{person}{Reuben Binns}, \bibinfo{person}{Michael Veale}, \bibinfo{person}{Max Van~Kleek}, {and} \bibinfo{person}{Nigel Shadbolt}.} \bibinfo{year}{2017}\natexlab{}.
\newblock \bibinfo{booktitle}{\emph{Like Trainer, Like Bot? Inheritance of Bias in Algorithmic Content Moderation}}.
\newblock \bibinfo{publisher}{Springer International Publishing}, \bibinfo{address}{Oxford, United Kingdom}, \bibinfo{pages}{405–415}.
\newblock
\showISBNx{9783319672564}
\showISSN{1611-3349}
\urldef\tempurl%
\url{https://doi.org/10.1007/978-3-319-67256-4_32}
\showDOI{\tempurl}


\bibitem[Birhane et~al\mbox{.}(2024)]%
        {birhane_ai_2024}
\bibfield{author}{\bibinfo{person}{Abeba Birhane}, \bibinfo{person}{Ryan Steed}, \bibinfo{person}{Victor Ojewale}, \bibinfo{person}{Briana Vecchione}, {and} \bibinfo{person}{Inioluwa~Deborah Raji}.} \bibinfo{year}{2024}\natexlab{}.
\newblock \showarticletitle{AI auditing: The Broken Bus on the Road to AI Accountability}. In \bibinfo{booktitle}{\emph{2024 IEEE Conference on Secure and Trustworthy Machine Learning (SaTML)}}. \bibinfo{publisher}{IEEE}, \bibinfo{address}{Toronto, ON, Canada}, \bibinfo{pages}{612--643}.
\newblock
\urldef\tempurl%
\url{https://doi.org/10.1109/SaTML59370.2024.00037}
\showDOI{\tempurl}


\bibitem[Blodgett et~al\mbox{.}(2020)]%
        {blodgett-etal-2020-language}
\bibfield{author}{\bibinfo{person}{Su~Lin Blodgett}, \bibinfo{person}{Solon Barocas}, \bibinfo{person}{Hal Daum{\'e}~III}, {and} \bibinfo{person}{Hanna Wallach}.} \bibinfo{year}{2020}\natexlab{}.
\newblock \showarticletitle{Language (Technology) is Power: A Critical Survey of {``}Bias{''} in {NLP}}. In \bibinfo{booktitle}{\emph{Proceedings of the 58th Annual Meeting of the Association for Computational Linguistics}}, \bibfield{editor}{\bibinfo{person}{Dan Jurafsky}, \bibinfo{person}{Joyce Chai}, \bibinfo{person}{Natalie Schluter}, {and} \bibinfo{person}{Joel Tetreault}} (Eds.). \bibinfo{publisher}{Association for Computational Linguistics}, \bibinfo{address}{Online}, \bibinfo{pages}{5454--5476}.
\newblock
\urldef\tempurl%
\url{https://aclanthology.org/2020.acl-main.485}
\showURL{%
\tempurl}


\bibitem[Blodgett et~al\mbox{.}(2016)]%
        {blodgett_demographic_2016}
\bibfield{author}{\bibinfo{person}{Su~Lin Blodgett}, \bibinfo{person}{Lisa Green}, {and} \bibinfo{person}{Brendan O'Connor}.} \bibinfo{year}{2016}\natexlab{}.
\newblock \showarticletitle{Demographic {Dialectal} {Variation} in {Social} {Media}: {A} {Case} {Study} of {African}-{American} {English}}. In \bibinfo{booktitle}{\emph{Proceedings of the 2016 {Conference} on {Empirical} {Methods} in {Natural} {Language} {Processing}}}, \bibfield{editor}{\bibinfo{person}{Jian Su}, \bibinfo{person}{Kevin Duh}, {and} \bibinfo{person}{Xavier Carreras}} (Eds.). \bibinfo{publisher}{Association for Computational Linguistics}, \bibinfo{address}{Austin, Texas}, \bibinfo{pages}{1119--1130}.
\newblock
\urldef\tempurl%
\url{https://aclanthology.org/D16-1120}
\showURL{%
\tempurl}


\bibitem[Borkan et~al\mbox{.}(2019)]%
        {borkan}
\bibfield{author}{\bibinfo{person}{Daniel Borkan}, \bibinfo{person}{Lucas Dixon}, \bibinfo{person}{Jeffrey Sorensen}, \bibinfo{person}{Nithum Thain}, {and} \bibinfo{person}{Lucy Vasserman}.} \bibinfo{year}{2019}\natexlab{}.
\newblock \showarticletitle{Nuanced Metrics for Measuring Unintended Bias with Real Data for Text Classification}. In \bibinfo{booktitle}{\emph{Companion Proceedings of The 2019 World Wide Web Conference}} (San Francisco, USA) \emph{(\bibinfo{series}{WWW '19})}. \bibinfo{publisher}{Association for Computing Machinery}, \bibinfo{address}{New York, NY, USA}, \bibinfo{pages}{491–500}.
\newblock
\showISBNx{9781450366755}
\urldef\tempurl%
\url{https://doi.org/10.1145/3308560.3317593}
\showDOI{\tempurl}


\bibitem[Breitfeller et~al\mbox{.}(2019)]%
        {breitfeller2019finding}
\bibfield{author}{\bibinfo{person}{Luke Breitfeller}, \bibinfo{person}{Emily Ahn}, \bibinfo{person}{David Jurgens}, {and} \bibinfo{person}{Yulia Tsvetkov}.} \bibinfo{year}{2019}\natexlab{}.
\newblock \showarticletitle{Finding Microaggressions in the Wild: A Case for Locating Elusive Phenomena in Social Media Posts}. In \bibinfo{booktitle}{\emph{Proceedings of the 2019 Conference on Empirical Methods in Natural Language Processing and the 9th International Joint Conference on Natural Language Processing (EMNLP-IJCNLP)}}. \bibinfo{publisher}{Association for Computational Linguistics}, \bibinfo{address}{Hong Kong, China}, \bibinfo{pages}{1664--1674}.
\newblock
\urldef\tempurl%
\url{https://doi.org/10.18653/v1/D19-1176}
\showDOI{\tempurl}
\newblock
\shownote{Retrieved 2020-01-20}.


\bibitem[Brown(2017)]%
        {brown_what_2017}
\bibfield{author}{\bibinfo{person}{Alexander Brown}.} \bibinfo{year}{2017}\natexlab{}.
\newblock \showarticletitle{What is {Hate} {Speech}? {Part} 2: {Family} {Resemblances}}.
\newblock \bibinfo{journal}{\emph{Law and Philosophy}} \bibinfo{volume}{36}, \bibinfo{number}{5} (\bibinfo{date}{Oct.} \bibinfo{year}{2017}), \bibinfo{pages}{561--613}.
\newblock
\showISSN{1573-0522}
\urldef\tempurl%
\url{https://doi.org/10.1007/s10982-017-9300-x}
\showDOI{\tempurl}


\bibitem[Cai et~al\mbox{.}(2024)]%
        {cai2024content}
\bibfield{author}{\bibinfo{person}{Jie Cai}, \bibinfo{person}{Aashka Patel}, \bibinfo{person}{Azadeh Naderi}, {and} \bibinfo{person}{Donghee~Yvette Wohn}.} \bibinfo{year}{2024}\natexlab{}.
\newblock \showarticletitle{Content Moderation Justice and Fairness on Social Media: Comparisons Across Different Contexts and Platforms}. In \bibinfo{booktitle}{\emph{Extended Abstracts of the CHI Conference on Human Factors in Computing Systems (CHI EA '24)}} (Honolulu, HI, USA). \bibinfo{publisher}{ACM}, \bibinfo{address}{New York, NY, USA}, \bibinfo{pages}{9 pages}.
\newblock
\urldef\tempurl%
\url{https://doi.org/10.1145/3613905.3650882}
\showDOI{\tempurl}


\bibitem[Casper et~al\mbox{.}(2024)]%
        {casper_black-box_2024}
\bibfield{author}{\bibinfo{person}{Stephen Casper}, \bibinfo{person}{Carson Ezell}, \bibinfo{person}{Charlotte Siegmann}, \bibinfo{person}{Noam Kolt}, \bibinfo{person}{Taylor~Lynn Curtis}, \bibinfo{person}{Benjamin Bucknall}, \bibinfo{person}{Andreas Haupt}, \bibinfo{person}{Kevin Wei}, \bibinfo{person}{J\'{e}r\'{e}my Scheurer}, \bibinfo{person}{Marius Hobbhahn}, \bibinfo{person}{Lee Sharkey}, \bibinfo{person}{Satyapriya Krishna}, \bibinfo{person}{Marvin Von~Hagen}, \bibinfo{person}{Silas Alberti}, \bibinfo{person}{Alan Chan}, \bibinfo{person}{Qinyi Sun}, \bibinfo{person}{Michael Gerovitch}, \bibinfo{person}{David Bau}, \bibinfo{person}{Max Tegmark}, \bibinfo{person}{David Krueger}, {and} \bibinfo{person}{Dylan Hadfield-Menell}.} \bibinfo{year}{2024}\natexlab{}.
\newblock \showarticletitle{Black-Box Access is Insufficient for Rigorous AI Audits}. In \bibinfo{booktitle}{\emph{Proceedings of the 2024 ACM Conference on Fairness, Accountability, and Transparency}} (Rio de Janeiro, Brazil) \emph{(\bibinfo{series}{FAccT '24})}. \bibinfo{publisher}{Association for Computing Machinery}, \bibinfo{address}{New York, NY, USA}, \bibinfo{pages}{2254–2272}.
\newblock
\showISBNx{9798400704505}
\urldef\tempurl%
\url{https://doi.org/10.1145/3630106.3659037}
\showDOI{\tempurl}


\bibitem[Cloud(2025)]%
        {ModerateTextCloud}
\bibfield{author}{\bibinfo{person}{Google Cloud}.} \bibinfo{year}{2025}\natexlab{}.
\newblock \bibinfo{booktitle}{\emph{Moderate Text | {{Cloud Natural Language API}}}}.
\newblock Google Cloud.
\newblock
\urldef\tempurl%
\url{https://cloud.google.com/natural-language/docs/moderating-text}
\showURL{%
\tempurl}


\bibitem[Cohen(1960)]%
        {cohen1960coefficient}
\bibfield{author}{\bibinfo{person}{Jacob Cohen}.} \bibinfo{year}{1960}\natexlab{}.
\newblock \showarticletitle{A coefficient of agreement for nominal scales}.
\newblock \bibinfo{journal}{\emph{Educational and psychological measurement}} \bibinfo{volume}{20}, \bibinfo{number}{1} (\bibinfo{year}{1960}), \bibinfo{pages}{37--46}.
\newblock


\bibitem[Davani et~al\mbox{.}(2023)]%
        {davani}
\bibfield{author}{\bibinfo{person}{Aida~Mostafazadeh Davani}, \bibinfo{person}{Mohammad Atari}, \bibinfo{person}{Brendan Kennedy}, {and} \bibinfo{person}{Morteza Dehghani}.} \bibinfo{year}{2023}\natexlab{}.
\newblock \showarticletitle{Hate Speech Classifiers Learn Normative Social Stereotypes}.
\newblock \bibinfo{journal}{\emph{Transactions of the Association for Computational Linguistics}}  \bibinfo{volume}{11} (\bibinfo{date}{03} \bibinfo{year}{2023}), \bibinfo{pages}{300--319}.
\newblock
\showISSN{2307-387X}
\urldef\tempurl%
\url{https://doi.org/10.1162/tacl_a_00550}
\showDOI{\tempurl}
\showeprint{https://direct.mit.edu/tacl/article-pdf/doi/10.1162/tacl\_a\_00550/2075730/tacl\_a\_00550.pdf}


\bibitem[Davidson et~al\mbox{.}(2019)]%
        {davidson-etal-2019-racial}
\bibfield{author}{\bibinfo{person}{Thomas Davidson}, \bibinfo{person}{Debasmita Bhattacharya}, {and} \bibinfo{person}{Ingmar Weber}.} \bibinfo{year}{2019}\natexlab{}.
\newblock \showarticletitle{Racial Bias in Hate Speech and Abusive Language Detection Datasets}. In \bibinfo{booktitle}{\emph{Proceedings of the Third Workshop on Abusive Language Online}}, \bibfield{editor}{\bibinfo{person}{Sarah~T. Roberts}, \bibinfo{person}{Joel Tetreault}, \bibinfo{person}{Vinodkumar Prabhakaran}, {and} \bibinfo{person}{Zeerak Waseem}} (Eds.). \bibinfo{publisher}{Association for Computational Linguistics}, \bibinfo{address}{Florence, Italy}, \bibinfo{pages}{25--35}.
\newblock
\urldef\tempurl%
\url{https://doi.org/10.18653/v1/W19-3504}
\showDOI{\tempurl}


\bibitem[De~Gregorio(2020)]%
        {degregorio2020democratising}
\bibfield{author}{\bibinfo{person}{G De~Gregorio}.} \bibinfo{year}{2020}\natexlab{}.
\newblock \showarticletitle{Democratising online content moderation: A constitutional framework}.
\newblock \bibinfo{journal}{\emph{Computer Law \& Security Review}}  \bibinfo{volume}{36} (\bibinfo{year}{2020}), \bibinfo{pages}{105376}.
\newblock


\bibitem[Dergacheva et~al\mbox{.}(2023)]%
        {dergachevaOneDayContent2023}
\bibfield{author}{\bibinfo{person}{Daria Dergacheva}, \bibinfo{person}{Vasilisa Kuznetsova}, \bibinfo{person}{Rebecca Scharlach}, {and} \bibinfo{person}{Christian Katzenbach}.} \bibinfo{year}{2023}\natexlab{}.
\newblock \bibinfo{booktitle}{\emph{One {{Day}} in {{Content Moderation}}: {{Analyzing}} 24 h of {{Social Media Platforms}}’ {{Content Decisions}} through the {{DSA Transparency Database}}}}.
\newblock {Universität Bremen}.
\newblock
\urldef\tempurl%
\url{https://doi.org/10.26092/ELIB/2707}
\showDOI{\tempurl}


\bibitem[Dias~Oliva et~al\mbox{.}(2021)]%
        {diasolivaFightingHateSpeech2021}
\bibfield{author}{\bibinfo{person}{Thiago Dias~Oliva}, \bibinfo{person}{Dennys~Marcelo Antonialli}, {and} \bibinfo{person}{Alessandra Gomes}.} \bibinfo{year}{2021}\natexlab{}.
\newblock \showarticletitle{Fighting {{Hate Speech}}, {{Silencing Drag Queens}}? {{Artificial Intelligence}} in {{Content Moderation}} and {{Risks}} to {{LGBTQ Voices Online}}}.
\newblock \bibinfo{journal}{\emph{Sexuality \& Culture}} \bibinfo{volume}{25}, \bibinfo{number}{2} (\bibinfo{year}{2021}), \bibinfo{pages}{700--732}.
\newblock
Issue 2.
\showISSN{1936-4822}
\urldef\tempurl%
\url{https://doi.org/10.1007/s12119-020-09790-w}
\showDOI{\tempurl}


\bibitem[Dixon et~al\mbox{.}(2018)]%
        {dixon_measuring_2018}
\bibfield{author}{\bibinfo{person}{Lucas Dixon}, \bibinfo{person}{John Li}, \bibinfo{person}{Jeffrey Sorensen}, \bibinfo{person}{Nithum Thain}, {and} \bibinfo{person}{Lucy Vasserman}.} \bibinfo{year}{2018}\natexlab{}.
\newblock \showarticletitle{Measuring and {Mitigating} {Unintended} {Bias} in {Text} {Classification}}. In \bibinfo{booktitle}{\emph{Proceedings of the 2018 {AAAI}/{ACM} {Conference} on {AI}, {Ethics}, and {Society}}} \emph{(\bibinfo{series}{{AIES} '18})}. \bibinfo{publisher}{Association for Computing Machinery}, \bibinfo{address}{New York, NY, USA}, \bibinfo{pages}{67--73}.
\newblock


\bibitem[Douek(2021)]%
        {douek2021governing}
\bibfield{author}{\bibinfo{person}{Evelyn Douek}.} \bibinfo{year}{2021}\natexlab{}.
\newblock \showarticletitle{Governing Online Speech: From ‘Posts-as-Trumps’ to Proportionality and Probability}.
\newblock \bibinfo{journal}{\emph{Columbia Law Review}} \bibinfo{volume}{121}, \bibinfo{number}{3} (\bibinfo{year}{2021}), \bibinfo{pages}{759--833}.
\newblock


\bibitem[Elsafoury et~al\mbox{.}(2023)]%
        {elsafoury_bias_2023}
\bibfield{author}{\bibinfo{person}{Fatma Elsafoury}, \bibinfo{person}{Stamos Katsigiannis}, {and} \bibinfo{person}{Naeem Ramzan}.} \bibinfo{year}{2023}\natexlab{}.
\newblock \bibinfo{title}{On {Bias} and {Fairness} in {NLP}: {How} to have a fairer text classification?}
\newblock
\newblock
\urldef\tempurl%
\url{http://arxiv.org/abs/2305.12829}
\showURL{%
\tempurl}
\newblock
\shownote{arXiv:2305.12829 [cs]}.


\bibitem[Elsafoury et~al\mbox{.}(2022)]%
        {elsafoury-etal-2022-sos}
\bibfield{author}{\bibinfo{person}{Fatma Elsafoury}, \bibinfo{person}{Steve~R. Wilson}, \bibinfo{person}{Stamos Katsigiannis}, {and} \bibinfo{person}{Naeem Ramzan}.} \bibinfo{year}{2022}\natexlab{}.
\newblock \showarticletitle{{SOS}: Systematic Offensive Stereotyping Bias in Word Embeddings}. In \bibinfo{booktitle}{\emph{Proceedings of the 29th International Conference on Computational Linguistics}}, \bibfield{editor}{\bibinfo{person}{Nicoletta Calzolari}, \bibinfo{person}{Chu-Ren Huang}, \bibinfo{person}{Hansaem Kim}, \bibinfo{person}{James Pustejovsky}, \bibinfo{person}{Leo Wanner}, \bibinfo{person}{Key-Sun Choi}, \bibinfo{person}{Pum-Mo Ryu}, \bibinfo{person}{Hsin-Hsi Chen}, \bibinfo{person}{Lucia Donatelli}, \bibinfo{person}{Heng Ji}, \bibinfo{person}{Sadao Kurohashi}, \bibinfo{person}{Patrizia Paggio}, \bibinfo{person}{Nianwen Xue}, \bibinfo{person}{Seokhwan Kim}, \bibinfo{person}{Younggyun Hahm}, \bibinfo{person}{Zhong He}, \bibinfo{person}{Tony~Kyungil Lee}, \bibinfo{person}{Enrico Santus}, \bibinfo{person}{Francis Bond}, {and} \bibinfo{person}{Seung-Hoon Na}} (Eds.). \bibinfo{publisher}{International Committee on Computational Linguistics}, \bibinfo{address}{Gyeongju, Republic of Korea},
  \bibinfo{pages}{1263--1274}.
\newblock
\urldef\tempurl%
\url{https://aclanthology.org/2022.coling-1.108}
\showURL{%
\tempurl}


\bibitem[ElSherief et~al\mbox{.}(2021)]%
        {elsheriefLatentHatredBenchmark2021}
\bibfield{author}{\bibinfo{person}{Mai ElSherief}, \bibinfo{person}{Caleb Ziems}, \bibinfo{person}{David Muchlinski}, \bibinfo{person}{Vaishnavi Anupindi}, \bibinfo{person}{Jordyn Seybolt}, \bibinfo{person}{Munmun De~Choudhury}, {and} \bibinfo{person}{Diyi Yang}.} \bibinfo{year}{2021}\natexlab{}.
\newblock \showarticletitle{Latent {{Hatred}}: {{A Benchmark}} for {{Understanding Implicit Hate Speech}}}. In \bibinfo{booktitle}{\emph{Proceedings of the 2021 {{Conference}} on {{Empirical Methods}} in {{Natural Language Processing}}}} (Online and Punta Cana, Dominican Republic, 2021). \bibinfo{publisher}{Association for Computational Linguistics}, \bibinfo{address}{Online and Punta Cana, Dominican Republic}, \bibinfo{pages}{345--363}.
\newblock
\urldef\tempurl%
\url{https://doi.org/10.18653/v1/2021.emnlp-main.29}
\showDOI{\tempurl}


\bibitem[{European Commission}(2024)]%
        {european_commission_supervision_2024}
\bibfield{author}{\bibinfo{person}{{European Commission}}.} \bibinfo{year}{2024}\natexlab{}.
\newblock \bibinfo{title}{Supervision of the designated very large online platforms and search engines under {DSA}}.
\newblock
\newblock
\urldef\tempurl%
\url{https://digital-strategy.ec.europa.eu/en/policies/list-designated-vlops-and-vloses\#ecl-inpage-google}
\showURL{%
\tempurl}
\newblock
\shownote{Last accessed 2024-04-29}.


\bibitem[{European Parliament}(2022)]%
        {european_parliament_regulation_2022}
\bibfield{author}{\bibinfo{person}{{European Parliament}}.} \bibinfo{year}{2022}\natexlab{}.
\newblock \bibinfo{title}{Regulation ({EU})2022/2065 of the {European} {Parliament} and of the {Council} of 19 {October} 2022 on a {Single} {Market} {For} {Digital} {Services} and amending {Directive} 2000/31/{EC} ({Digital} {Services} {Act})}.
\newblock
\newblock
\urldef\tempurl%
\url{https://eur-lex.europa.eu/legal-content/EN/TXT/?uri=celex\%3A32022R2065}
\showURL{%
\tempurl}
\newblock
\shownote{Last accessed 2024-04-29}.


\bibitem[Fortuna et~al\mbox{.}(2020)]%
        {fortuna_toxic_2020}
\bibfield{author}{\bibinfo{person}{Paula Fortuna}, \bibinfo{person}{Juan Soler}, {and} \bibinfo{person}{Leo Wanner}.} \bibinfo{year}{2020}\natexlab{}.
\newblock \showarticletitle{Toxic, {Hateful}, {Offensive} or {Abusive}? {What} {Are} {We} {Really} {Classifying}? {An} {Empirical} {Analysis} of {Hate} {Speech} {Datasets}}. In \bibinfo{booktitle}{\emph{Proceedings of the {Twelfth} {Language} {Resources} and {Evaluation} {Conference}}}, \bibfield{editor}{\bibinfo{person}{Nicoletta Calzolari}, \bibinfo{person}{Frédéric Béchet}, \bibinfo{person}{Philippe Blache}, \bibinfo{person}{Khalid Choukri}, \bibinfo{person}{Christopher Cieri}, \bibinfo{person}{Thierry Declerck}, \bibinfo{person}{Sara Goggi}, \bibinfo{person}{Hitoshi Isahara}, \bibinfo{person}{Bente Maegaard}, \bibinfo{person}{Joseph Mariani}, \bibinfo{person}{Hélène Mazo}, \bibinfo{person}{Asuncion Moreno}, \bibinfo{person}{Jan Odijk}, {and} \bibinfo{person}{Stelios Piperidis}} (Eds.). \bibinfo{publisher}{European Language Resources Association}, \bibinfo{address}{Marseille, France}, \bibinfo{pages}{6786--6794}.
\newblock
\showISBNx{979-10-95546-34-4}
\urldef\tempurl%
\url{https://aclanthology.org/2020.lrec-1.838}
\showURL{%
\tempurl}


\bibitem[Galinsky et~al\mbox{.}(2003)]%
        {galinsky_reappropriation_2003}
\bibfield{author}{\bibinfo{person}{Adam~D Galinsky}, \bibinfo{person}{Kurt Hugenberg}, \bibinfo{person}{Carla Groom}, {and} \bibinfo{person}{Galen~V Bodenhausen}.} \bibinfo{year}{2003}\natexlab{}.
\newblock \showarticletitle{The reappropriation of stigmatizing labels: {Implications} for social identity}.
\newblock In \bibinfo{booktitle}{\emph{Identity issues in groups}}. Vol.~\bibinfo{volume}{5}. \bibinfo{publisher}{Emerald Group Publishing Limited}, \bibinfo{address}{Leeds, England}, \bibinfo{pages}{221--256}.
\newblock


\bibitem[Gallegos et~al\mbox{.}(2024)]%
        {Gallegos2024}
\bibfield{author}{\bibinfo{person}{Isabel~O. Gallegos}, \bibinfo{person}{Ryan~A. Rossi}, \bibinfo{person}{Joe Barrow}, \bibinfo{person}{Md~Mehrab Tanjim}, \bibinfo{person}{Sungchul Kim}, \bibinfo{person}{Franck Dernoncourt}, \bibinfo{person}{Tong Yu}, \bibinfo{person}{Ruiyi Zhang}, {and} \bibinfo{person}{Nesreen~K. Ahmed}.} \bibinfo{year}{2024}\natexlab{}.
\newblock \showarticletitle{Bias and Fairness in Large Language Models: A Survey}.
\newblock \bibinfo{journal}{\emph{Computational Linguistics}} \bibinfo{volume}{50}, \bibinfo{number}{3} (\bibinfo{date}{09} \bibinfo{year}{2024}), \bibinfo{pages}{1097--1179}.
\newblock
\showISSN{0891-2017}
\urldef\tempurl%
\url{https://doi.org/10.1162/coli_a_00524}
\showDOI{\tempurl}
\showeprint{https://direct.mit.edu/coli/article-pdf/50/3/1097/2471010/coli\_a\_00524.pdf}


\bibitem[Garg et~al\mbox{.}(2023)]%
        {garg_handling_2023}
\bibfield{author}{\bibinfo{person}{Tanmay Garg}, \bibinfo{person}{Sarah Masud}, \bibinfo{person}{Tharun Suresh}, {and} \bibinfo{person}{Tanmoy Chakraborty}.} \bibinfo{year}{2023}\natexlab{}.
\newblock \showarticletitle{Handling bias in toxic speech detection: {A} survey}.
\newblock \bibinfo{journal}{\emph{Comput. Surveys}} \bibinfo{volume}{55}, \bibinfo{number}{13s} (\bibinfo{year}{2023}), \bibinfo{pages}{1--32}.
\newblock


\bibitem[Gebrekidan(2024)]%
        {gebrekidan2024content}
\bibfield{author}{\bibinfo{person}{Fasica~B. Gebrekidan}.} \bibinfo{year}{2024}\natexlab{}.
\newblock \bibinfo{booktitle}{\emph{Content moderation: The harrowing, traumatizing job that left many African data workers with mental health issues and drug dependency}}.
\newblock DAIR Institute.
\newblock
\urldef\tempurl%
\url{https://data-workers.org/fasica}
\showURL{%
\tempurl}


\bibitem[Ghosh et~al\mbox{.}(2021)]%
        {ghosh_detecting_2021}
\bibfield{author}{\bibinfo{person}{Sayan Ghosh}, \bibinfo{person}{Dylan Baker}, \bibinfo{person}{David Jurgens}, {and} \bibinfo{person}{Vinodkumar Prabhakaran}.} \bibinfo{year}{2021}\natexlab{}.
\newblock \showarticletitle{Detecting {Cross}-{Geographic} {Biases} in {Toxicity} {Modeling} on {Social} {Media}}. In \bibinfo{booktitle}{\emph{Proceedings of the {Seventh} {Workshop} on {Noisy} {User}-generated {Text} ({W}-{NUT} 2021)}}. \bibinfo{publisher}{Association for Computational Linguistics}, \bibinfo{address}{Online}, \bibinfo{pages}{313--328}.
\newblock


\bibitem[Gillespie(2022)]%
        {gillespie}
\bibfield{author}{\bibinfo{person}{Tarleton Gillespie}.} \bibinfo{year}{2022}\natexlab{}.
\newblock \showarticletitle{Do Not Recommend? Reduction as a Form of Content Moderation}.
\newblock \bibinfo{journal}{\emph{Social Media + Society}} \bibinfo{volume}{8}, \bibinfo{number}{3} (\bibinfo{year}{2022}), \bibinfo{pages}{20563051221117552}.
\newblock
\urldef\tempurl%
\url{https://doi.org/10.1177/20563051221117552}
\showDOI{\tempurl}
\showeprint{https://doi.org/10.1177/20563051221117552}


\bibitem[Gorwa et~al\mbox{.}(2020)]%
        {gorwa_algorithmic_2020}
\bibfield{author}{\bibinfo{person}{Robert Gorwa}, \bibinfo{person}{Reuben Binns}, {and} \bibinfo{person}{Christian Katzenbach}.} \bibinfo{year}{2020}\natexlab{}.
\newblock \showarticletitle{Algorithmic content moderation: {Technical} and political challenges in the automation of platform governance}.
\newblock \bibinfo{journal}{\emph{Big Data \& Society}} \bibinfo{volume}{7}, \bibinfo{number}{1} (\bibinfo{date}{Jan.} \bibinfo{year}{2020}), \bibinfo{pages}{205395171989794}.
\newblock
\showISSN{2053-9517, 2053-9517}
\urldef\tempurl%
\url{http://journals.sagepub.com/doi/10.1177/2053951719897945}
\showURL{%
\tempurl}


\bibitem[Hardt et~al\mbox{.}(2016)]%
        {hardt_equality_2016}
\bibfield{author}{\bibinfo{person}{Moritz Hardt}, \bibinfo{person}{Eric Price}, {and} \bibinfo{person}{Nathan Srebro}.} \bibinfo{year}{2016}\natexlab{}.
\newblock \showarticletitle{Equality of opportunity in supervised learning}. In \bibinfo{booktitle}{\emph{Proceedings of the 30th {International} {Conference} on {Neural} {Information} {Processing} {Systems}}} \emph{(\bibinfo{series}{{NIPS}'16})}. \bibinfo{publisher}{Curran Associates Inc.}, \bibinfo{address}{Red Hook, NY, USA}, \bibinfo{pages}{3323--3331}.
\newblock


\bibitem[Hartmann et~al\mbox{.}(2024)]%
        {hartmann2024addressing}
\bibfield{author}{\bibinfo{person}{David Hartmann}, \bibinfo{person}{Jos{\'e} Renato~Laranjeira de Pereira}, \bibinfo{person}{Chiara Streitb{\"o}rger}, {and} \bibinfo{person}{Bettina Berendt}.} \bibinfo{year}{2024}\natexlab{}.
\newblock \bibinfo{booktitle}{\emph{Addressing the regulatory gap: moving towards an EU AI audit ecosystem beyond the AI Act by including civil society}}.
\newblock AI and Ethics.
\newblock
\showISSN{2730-5961}
\urldef\tempurl%
\url{https://doi.org/10.1007/s43681-024-00595-3}
\showDOI{\tempurl}


\bibitem[Hartvigsen et~al\mbox{.}(2022)]%
        {hartvigsen_toxigen_2022}
\bibfield{author}{\bibinfo{person}{Thomas Hartvigsen}, \bibinfo{person}{Saadia Gabriel}, \bibinfo{person}{Hamid Palangi}, \bibinfo{person}{Maarten Sap}, \bibinfo{person}{Dipankar Ray}, {and} \bibinfo{person}{Ece Kamar}.} \bibinfo{year}{2022}\natexlab{}.
\newblock \showarticletitle{{T}oxi{G}en: A Large-Scale Machine-Generated Dataset for Adversarial and Implicit Hate Speech Detection}. In \bibinfo{booktitle}{\emph{Proceedings of the 60th Annual Meeting of the Association for Computational Linguistics (Volume 1: Long Papers)}}, \bibfield{editor}{\bibinfo{person}{Smaranda Muresan}, \bibinfo{person}{Preslav Nakov}, {and} \bibinfo{person}{Aline Villavicencio}} (Eds.). \bibinfo{publisher}{Association for Computational Linguistics}, \bibinfo{address}{Dublin, Ireland}, \bibinfo{pages}{3309--3326}.
\newblock
\urldef\tempurl%
\url{https://doi.org/10.18653/v1/2022.acl-long.234}
\showDOI{\tempurl}


\bibitem[Heung et~al\mbox{.}(2024)]%
        {heung2024vulnerable}
\bibfield{author}{\bibinfo{person}{Sharon Heung}, \bibinfo{person}{Lucy Jiang}, \bibinfo{person}{Shiri Azenkot}, {and} \bibinfo{person}{Aditya Vashistha}.} \bibinfo{year}{2024}\natexlab{}.
\newblock \showarticletitle{“Vulnerable, Victimized, and Objectified”: Understanding Ableist Hate and Harassment Experienced by Disabled Content Creators on Social Media}. In \bibinfo{booktitle}{\emph{Proceedings of the CHI Conference on Human Factors in Computing Systems (CHI '24)}} (Honolulu, HI, USA). \bibinfo{publisher}{ACM}, \bibinfo{address}{New York, NY, USA}, \bibinfo{pages}{19 pages}.
\newblock
\urldef\tempurl%
\url{https://doi.org/10.1145/3613904.3641949}
\showDOI{\tempurl}


\bibitem[Horta~Ribeiro(2024)]%
        {hortaribeiroContentModerationOnline2024}
\bibfield{author}{\bibinfo{person}{Manoel Horta~Ribeiro}.} \bibinfo{year}{2024}\natexlab{}.
\newblock \bibinfo{title}{Content {{Moderation}} in {{Online Platforms}}}.
\newblock
\newblock
\urldef\tempurl%
\url{https://doi.org/10.5075/EPFL-THESIS-10387}
\showDOI{\tempurl}


\bibitem[Jarke and Heuer(2024)]%
        {jarke20245}
\bibfield{author}{\bibinfo{person}{Juliane Jarke} {and} \bibinfo{person}{Hendrik Heuer}.} \bibinfo{year}{2024}\natexlab{}.
\newblock \bibinfo{booktitle}{\emph{Reassembling the Black Box of Machine Learning: Of Monsters and the Reversibility of Foldings}}.
\newblock \bibinfo{publisher}{Amsterdam University Press}, \bibinfo{address}{Amsterdam, Netherlands}, \bibinfo{pages}{103--126}.
\newblock
\urldef\tempurl%
\url{http://www.jstor.org/stable/jj.11895528.7}
\showURL{%
\tempurl}


\bibitem[Jigsaw(2019)]%
        {jigsaw_jigsaw_2019}
\bibfield{author}{\bibinfo{person}{Jigsaw}.} \bibinfo{year}{2019}\natexlab{}.
\newblock \bibinfo{title}{Jigsaw toxic comment classifi- cation challenge.}
\newblock
\newblock
\urldef\tempurl%
\url{https://www.kaggle.com/c/jigsaw-toxic-comment-classification-challenge}
\showURL{%
\tempurl}


\bibitem[Jigsaw(2021)]%
        {jigsaw2021perspective}
\bibfield{author}{\bibinfo{person}{Jigsaw}.} \bibinfo{year}{2021}\natexlab{}.
\newblock \bibinfo{title}{Google's Jigsaw Announces Toxicity-Reducing API Perspective is Processing 500M Requests Daily}.
\newblock
\newblock
\urldef\tempurl%
\url{https://www.prnewswire.com/news-releases/googles-jigsaw-announces-toxicity-reducing-api-perspective-is-processing-500m-requests-daily-301223600.html}
\showURL{%
\tempurl}


\bibitem[Kak and West(2023)]%
        {institute_algorithmic_2023}
\bibfield{author}{\bibinfo{person}{Amba Kak} {and} \bibinfo{person}{Sarah~Myers West}.} \bibinfo{year}{2023}\natexlab{}.
\newblock \bibinfo{booktitle}{\emph{Algorithmic {Accountability}: {Moving} {Beyond} {Audits}}}.
\newblock AI Now Institute.
\newblock
\urldef\tempurl%
\url{https://ainowinstitute.org/publication/algorithmic-accountability}
\showURL{%
\tempurl}


\bibitem[Kapoor and Narayanan(2023)]%
        {kapoor2023leakage}
\bibfield{author}{\bibinfo{person}{Sayash Kapoor} {and} \bibinfo{person}{Arvind Narayanan}.} \bibinfo{year}{2023}\natexlab{}.
\newblock \showarticletitle{Leakage and the reproducibility crisis in machine-learning-based science}.
\newblock \bibinfo{journal}{\emph{Article Volume 4, Issue 9}} \bibinfo{volume}{4}, \bibinfo{number}{9} (\bibinfo{date}{September 08} \bibinfo{year}{2023}), \bibinfo{pages}{100804}.
\newblock
\urldef\tempurl%
\url{https://doi.org/10.1016/j.mlops.2023.100804}
\showDOI{\tempurl}
\newblock
\shownote{Open access}.


\bibitem[Khezzar et~al\mbox{.}(2023)]%
        {khezzar_arhatedetector_2023}
\bibfield{author}{\bibinfo{person}{Ramzi Khezzar}, \bibinfo{person}{Abdelrahman Moursi}, {and} \bibinfo{person}{Zaher Al~Aghbari}.} \bibinfo{year}{2023}\natexlab{}.
\newblock \showarticletitle{{arHateDetector}: detection of hate speech from standard and dialectal {Arabic} {Tweets}}.
\newblock \bibinfo{journal}{\emph{Discover Internet of Things}} \bibinfo{volume}{3}, \bibinfo{number}{1} (\bibinfo{date}{March} \bibinfo{year}{2023}), \bibinfo{pages}{1}.
\newblock


\bibitem[Lachemaier et~al\mbox{.}(2024)]%
        {lachemaier_towards_2024}
\bibfield{author}{\bibinfo{person}{Clara Lachemaier}, \bibinfo{person}{Eleonore Lumer}, \bibinfo{person}{Hendrik Buschmeier}, {and} \bibinfo{person}{Sina Zarrieß}.} \bibinfo{year}{2024}\natexlab{}.
\newblock \showarticletitle{Towards {Understanding} the {Entanglement} of {Human} {Stereotypes} and {System} {Biases} in {Human}-{Robot} {Interaction}}. In \bibinfo{booktitle}{\emph{Companion of the 2024 {ACM}/{IEEE} {International} {Conference} on {Human}-{Robot} {Interaction}}} \emph{(\bibinfo{series}{{HRI} '24})}. \bibinfo{publisher}{Association for Computing Machinery}, \bibinfo{address}{New York, NY, USA}, \bibinfo{pages}{646--649}.
\newblock


\bibitem[Li et~al\mbox{.}(2023)]%
        {li2023trustworthy}
\bibfield{author}{\bibinfo{person}{Bo Li}, \bibinfo{person}{Peng Qi}, \bibinfo{person}{Bo Liu}, \bibinfo{person}{Shuai Di}, \bibinfo{person}{Jingen Liu}, \bibinfo{person}{Jiquan Pei}, \bibinfo{person}{Jinfeng Yi}, {and} \bibinfo{person}{Bowen Zhou}.} \bibinfo{year}{2023}\natexlab{}.
\newblock \showarticletitle{Trustworthy AI: From principles to practices}.
\newblock \bibinfo{journal}{\emph{Comput. Surveys}} \bibinfo{volume}{55}, \bibinfo{number}{9} (\bibinfo{year}{2023}), \bibinfo{pages}{1--46}.
\newblock


\bibitem[Ludwig et~al\mbox{.}(2022)]%
        {ludwig-etal-2022-improving}
\bibfield{author}{\bibinfo{person}{Florian Ludwig}, \bibinfo{person}{Klara Dolos}, \bibinfo{person}{Torsten Zesch}, {and} \bibinfo{person}{Eleanor Hobley}.} \bibinfo{year}{2022}\natexlab{}.
\newblock \showarticletitle{Improving Generalization of Hate Speech Detection Systems to Novel Target Groups via Domain Adaptation}. In \bibinfo{booktitle}{\emph{Proceedings of the Sixth Workshop on Online Abuse and Harms (WOAH)}}, \bibfield{editor}{\bibinfo{person}{Kanika Narang}, \bibinfo{person}{Aida Mostafazadeh~Davani}, \bibinfo{person}{Lambert Mathias}, \bibinfo{person}{Bertie Vidgen}, {and} \bibinfo{person}{Zeerak Talat}} (Eds.). \bibinfo{publisher}{Association for Computational Linguistics}, \bibinfo{address}{Seattle, Washington (Hybrid)}, \bibinfo{pages}{29--39}.
\newblock
\urldef\tempurl%
\url{https://doi.org/10.18653/v1/2022.woah-1.4}
\showDOI{\tempurl}


\bibitem[Lundberg(2018)]%
        {IntroductionExplainableAI}
\bibfield{author}{\bibinfo{person}{Scott Lundberg}.} \bibinfo{year}{2018}\natexlab{}.
\newblock \bibinfo{booktitle}{\emph{An Introduction to Explainable {{AI}} with {{Shapley}} Values — {{SHAP}} Latest Documentation}}.
\newblock Scott Lundberg.
\newblock
\urldef\tempurl%
\url{https://shap.readthedocs.io/en/latest/example_notebooks/overviews/An%20introduction%20to%20explainable%20AI%20with%20Shapley%20values.html}
\showURL{%
\tempurl}


\bibitem[Lundberg and Lee(2017)]%
        {lundberg2017unified}
\bibfield{author}{\bibinfo{person}{Scott~M. Lundberg} {and} \bibinfo{person}{Su-In Lee}.} \bibinfo{year}{2017}\natexlab{}.
\newblock \showarticletitle{A unified approach to interpreting model predictions}. In \bibinfo{booktitle}{\emph{Proceedings of the 31st International Conference on Neural Information Processing Systems}} (Long Beach, California, USA) \emph{(\bibinfo{series}{NIPS'17})}. \bibinfo{publisher}{Curran Associates Inc.}, \bibinfo{address}{Red Hook, NY, USA}, \bibinfo{pages}{4768–4777}.
\newblock
\showISBNx{9781510860964}


\bibitem[Lykouris and Weng(2024)]%
        {lykouris2024learningdefercontentmoderation}
\bibfield{author}{\bibinfo{person}{Thodoris Lykouris} {and} \bibinfo{person}{Wentao Weng}.} \bibinfo{year}{2024}\natexlab{}.
\newblock \bibinfo{title}{Learning to Defer in Content Moderation: The Human-AI Interplay}.
\newblock
\newblock
\showeprint[arxiv]{2402.12237}~[cs.LG]
\urldef\tempurl%
\url{https://arxiv.org/abs/2402.12237}
\showURL{%
\tempurl}


\bibitem[Lyu et~al\mbox{.}(2024)]%
        {Lyu}
\bibfield{author}{\bibinfo{person}{Qing Lyu}, \bibinfo{person}{Marianna Apidianaki}, {and} \bibinfo{person}{Chris Callison-Burch}.} \bibinfo{year}{2024}\natexlab{}.
\newblock \showarticletitle{{Towards Faithful Model Explanation in NLP: A Survey}}.
\newblock \bibinfo{journal}{\emph{Computational Linguistics}} \bibinfo{volume}{50}, \bibinfo{number}{2} (\bibinfo{date}{06} \bibinfo{year}{2024}), \bibinfo{pages}{657--723}.
\newblock
\showISSN{0891-2017}
\urldef\tempurl%
\url{https://doi.org/10.1162/coli_a_00511}
\showDOI{\tempurl}
\showeprint{https://direct.mit.edu/coli/article-pdf/50/2/657/2457495/coli\_a\_00511.pdf}


\bibitem[Ma et~al\mbox{.}(2023)]%
        {usertrust}
\bibfield{author}{\bibinfo{person}{Renkai Ma}, \bibinfo{person}{Yue You}, \bibinfo{person}{Xinning Gui}, {and} \bibinfo{person}{Yubo Kou}.} \bibinfo{year}{2023}\natexlab{}.
\newblock \showarticletitle{How Do Users Experience Moderation?: A Systematic Literature Review}.
\newblock \bibinfo{journal}{\emph{Proc. ACM Hum.-Comput. Interact.}} \bibinfo{volume}{7}, \bibinfo{number}{CSCW2}, Article \bibinfo{articleno}{278} (\bibinfo{date}{oct} \bibinfo{year}{2023}), \bibinfo{numpages}{30}~pages.
\newblock
\urldef\tempurl%
\url{https://doi.org/10.1145/3610069}
\showDOI{\tempurl}


\bibitem[Mahomed et~al\mbox{.}(2024)]%
        {gpttvshow}
\bibfield{author}{\bibinfo{person}{Yaaseen Mahomed}, \bibinfo{person}{Charlie~M. Crawford}, \bibinfo{person}{Sanjana Gautam}, \bibinfo{person}{Sorelle~A. Friedler}, {and} \bibinfo{person}{Dana\"{e} Metaxa}.} \bibinfo{year}{2024}\natexlab{}.
\newblock \showarticletitle{Auditing GPT's Content Moderation Guardrails: Can ChatGPT Write Your Favorite TV Show?}. In \bibinfo{booktitle}{\emph{Proceedings of the 2024 ACM Conference on Fairness, Accountability, and Transparency}} (Rio de Janeiro, Brazil) \emph{(\bibinfo{series}{FAccT '24})}. \bibinfo{publisher}{Association for Computing Machinery}, \bibinfo{address}{New York, NY, USA}, \bibinfo{pages}{660–686}.
\newblock
\showISBNx{9798400704505}
\urldef\tempurl%
\url{https://doi.org/10.1145/3630106.3658932}
\showDOI{\tempurl}


\bibitem[Marques(2023)]%
        {marques}
\bibfield{author}{\bibinfo{person}{Teresa Marques}.} \bibinfo{year}{2023}\natexlab{}.
\newblock \showarticletitle{The Expression of Hate in Hate Speech}.
\newblock \bibinfo{journal}{\emph{Journal of Applied Philosophy}} \bibinfo{volume}{40}, \bibinfo{number}{5} (\bibinfo{year}{2023}), \bibinfo{pages}{769--787}.
\newblock
\urldef\tempurl%
\url{https://doi.org/10.1111/japp.12608}
\showDOI{\tempurl}
\showeprint{https://onlinelibrary.wiley.com/doi/pdf/10.1111/japp.12608}


\bibitem[Mathew et~al\mbox{.}(2021)]%
        {mathew_hatexplain_2021}
\bibfield{author}{\bibinfo{person}{Binny Mathew}, \bibinfo{person}{Punyajoy Saha}, \bibinfo{person}{Seid~Muhie Yimam}, \bibinfo{person}{Chris Biemann}, \bibinfo{person}{Pawan Goyal}, {and} \bibinfo{person}{Animesh Mukherjee}.} \bibinfo{year}{2021}\natexlab{}.
\newblock \showarticletitle{{HateXplain}: {A} {Benchmark} {Dataset} for {Explainable} {Hate} {Speech} {Detection}}.
\newblock \bibinfo{journal}{\emph{Proceedings of the AAAI Conference on Artificial Intelligence}} \bibinfo{volume}{35}, \bibinfo{number}{17} (\bibinfo{date}{May} \bibinfo{year}{2021}), \bibinfo{pages}{14867--14875}.
\newblock
\showISSN{2374-3468}
\urldef\tempurl%
\url{https://ojs.aaai.org/index.php/AAAI/article/view/17745}
\showURL{%
\tempurl}
\newblock
\shownote{Number: 17}.


\bibitem[Matsuda et~al\mbox{.}(1993)]%
        {matsuda1993words}
\bibfield{author}{\bibinfo{person}{Mari~J. Matsuda}, \bibinfo{person}{Charles R.~Lawrence III}, \bibinfo{person}{Richard Delgado}, {and} \bibinfo{person}{Kimberl{\'e}~W. Crenshaw}.} \bibinfo{year}{1993}\natexlab{}.
\newblock \bibinfo{booktitle}{\emph{Words That Wound: Critical Race Theory, Assaultive Speech, and The First Amendment}}.
\newblock \bibinfo{publisher}{Faculty Books}, \bibinfo{address}{New York}.
\newblock
\urldef\tempurl%
\url{https://scholarship.law.columbia.edu/books/287}
\showURL{%
\tempurl}
\newblock
\shownote{Accessed: date-of-access}.


\bibitem[Miceli et~al\mbox{.}(2024)]%
        {miceli2024who}
\bibfield{author}{\bibinfo{person}{Milagros Miceli}, \bibinfo{person}{Paola Tubaro}, \bibinfo{person}{Antonio~A. Casilli}, \bibinfo{person}{Thomas Le~Bonniec}, {and} \bibinfo{person}{Camilla~Salim Wagner}.} \bibinfo{year}{2024}\natexlab{}.
\newblock \bibinfo{booktitle}{\emph{Who Trains the Data for European Artificial Intelligence?: Report of the European Microworkers Communication and Outreach Initiative (EnCOre, 2023-2024)}}.
\newblock \bibinfo{type}{{T}echnical {R}eport}. \bibinfo{institution}{European Parliament; The Left}. \bibinfo{pages}{1--40} pages.
\newblock


\bibitem[Mihaljević and Steffen(2022)]%
        {mihaljevic2022toxic}
\bibfield{author}{\bibinfo{person}{Helena Mihaljević} {and} \bibinfo{person}{Elisabeth Steffen}.} \bibinfo{year}{2022}\natexlab{}.
\newblock \showarticletitle{How toxic is antisemitism? Potentials and limitations of automated toxicity scoring for antisemitic online content}. In \bibinfo{booktitle}{\emph{Proceedings of the 2nd Workshop on Computational Linguistics for Political Text Analysis}} (2022-09-12). CPSS-2022, \bibinfo{publisher}{Hochschule für Technik und Wirtschaft Berlin}, \bibinfo{address}{Potsdam, Germany}, \bibinfo{pages}{1--12}.
\newblock


\bibitem[Mishra et~al\mbox{.}(2019)]%
        {mishra2019tackling}
\bibfield{author}{\bibinfo{person}{Pushkar Mishra}, \bibinfo{person}{Helen Yannakoudakis}, {and} \bibinfo{person}{Ekaterina Shutova}.} \bibinfo{year}{2019}\natexlab{}.
\newblock \bibinfo{title}{Tackling Online Abuse: A Survey of Automated Abuse Detection Methods}.
\newblock
\newblock
\urldef\tempurl%
\url{http://arxiv.org/abs/1908.06024}
\showURL{%
\tempurl}
\newblock
\shownote{Retrieved 2020-02-04}.


\bibitem[Mitchell et~al\mbox{.}(2019)]%
        {mitchellModelCardsModel2019a}
\bibfield{author}{\bibinfo{person}{Margaret Mitchell}, \bibinfo{person}{Simone Wu}, \bibinfo{person}{Andrew Zaldivar}, \bibinfo{person}{Parker Barnes}, \bibinfo{person}{Lucy Vasserman}, \bibinfo{person}{Ben Hutchinson}, \bibinfo{person}{Elena Spitzer}, \bibinfo{person}{Inioluwa~Deborah Raji}, {and} \bibinfo{person}{Timnit Gebru}.} \bibinfo{year}{2019}\natexlab{}.
\newblock \showarticletitle{Model Cards for Model Reporting}. In \bibinfo{booktitle}{\emph{Proceedings of the Conference on Fairness, Accountability, and Transparency}} (Atlanta, GA, USA) \emph{(\bibinfo{series}{FAT* '19})}. \bibinfo{publisher}{Association for Computing Machinery}, \bibinfo{address}{New York, NY, USA}, \bibinfo{pages}{220–229}.
\newblock
\showISBNx{9781450361255}
\urldef\tempurl%
\url{https://doi.org/10.1145/3287560.3287596}
\showDOI{\tempurl}


\bibitem[Molnar(2022)]%
        {molnar2022}
\bibfield{author}{\bibinfo{person}{Christoph Molnar}.} \bibinfo{year}{2022}\natexlab{}.
\newblock \bibinfo{title}{Interpretable Machine Learning}.
\newblock
\newblock
\urldef\tempurl%
\url{https://christophm.github.io/interpretable-ml-book}
\showURL{%
\tempurl}


\bibitem[Mosca et~al\mbox{.}(2022)]%
        {mosca-etal-2022-shap}
\bibfield{author}{\bibinfo{person}{Edoardo Mosca}, \bibinfo{person}{Ferenc Szigeti}, \bibinfo{person}{Stella Tragianni}, \bibinfo{person}{Daniel Gallagher}, {and} \bibinfo{person}{Georg Groh}.} \bibinfo{year}{2022}\natexlab{}.
\newblock \showarticletitle{{SHAP}-Based Explanation Methods: A Review for {NLP} Interpretability}. In \bibinfo{booktitle}{\emph{Proceedings of the 29th International Conference on Computational Linguistics}}, \bibfield{editor}{\bibinfo{person}{Nicoletta Calzolari}, \bibinfo{person}{Chu-Ren Huang}, \bibinfo{person}{Hansaem Kim}, \bibinfo{person}{James Pustejovsky}, \bibinfo{person}{Leo Wanner}, \bibinfo{person}{Key-Sun Choi}, \bibinfo{person}{Pum-Mo Ryu}, \bibinfo{person}{Hsin-Hsi Chen}, \bibinfo{person}{Lucia Donatelli}, \bibinfo{person}{Heng Ji}, \bibinfo{person}{Sadao Kurohashi}, \bibinfo{person}{Patrizia Paggio}, \bibinfo{person}{Nianwen Xue}, \bibinfo{person}{Seokhwan Kim}, \bibinfo{person}{Younggyun Hahm}, \bibinfo{person}{Zhong He}, \bibinfo{person}{Tony~Kyungil Lee}, \bibinfo{person}{Enrico Santus}, \bibinfo{person}{Francis Bond}, {and} \bibinfo{person}{Seung-Hoon Na}} (Eds.). \bibinfo{publisher}{International Committee on Computational Linguistics}, \bibinfo{address}{Gyeongju, Republic of Korea},
  \bibinfo{pages}{4593--4603}.
\newblock
\urldef\tempurl%
\url{https://aclanthology.org/2022.coling-1.406}
\showURL{%
\tempurl}


\bibitem[Mosca et~al\mbox{.}(2021)]%
        {mosca-etal-2021-understanding}
\bibfield{author}{\bibinfo{person}{Edoardo Mosca}, \bibinfo{person}{Maximilian Wich}, {and} \bibinfo{person}{Georg Groh}.} \bibinfo{year}{2021}\natexlab{}.
\newblock \showarticletitle{Understanding and Interpreting the Impact of User Context in Hate Speech Detection}. In \bibinfo{booktitle}{\emph{Proceedings of the Ninth International Workshop on Natural Language Processing for Social Media}}, \bibfield{editor}{\bibinfo{person}{Lun-Wei Ku} {and} \bibinfo{person}{Cheng-Te Li}} (Eds.). \bibinfo{publisher}{Association for Computational Linguistics}, \bibinfo{address}{Online}, \bibinfo{pages}{91--102}.
\newblock
\urldef\tempurl%
\url{https://doi.org/10.18653/v1/2021.socialnlp-1.8}
\showDOI{\tempurl}


\bibitem[Mozafari et~al\mbox{.}(2020)]%
        {mozafari2020hate}
\bibfield{author}{\bibinfo{person}{Marzieh Mozafari}, \bibinfo{person}{Reza Farahbakhsh}, {and} \bibinfo{person}{No{\"e}l Crespi}.} \bibinfo{year}{2020}\natexlab{}.
\newblock \showarticletitle{Hate speech detection and racial bias mitigation in social media based on BERT model}.
\newblock \bibinfo{journal}{\emph{PloS one}} \bibinfo{volume}{15}, \bibinfo{number}{8} (\bibinfo{year}{2020}), \bibinfo{pages}{e0237861}.
\newblock


\bibitem[M{\"u}ller and Guido(2016)]%
        {muller2016introduction}
\bibfield{author}{\bibinfo{person}{Andreas~C M{\"u}ller} {and} \bibinfo{person}{Sarah Guido}.} \bibinfo{year}{2016}\natexlab{}.
\newblock \bibinfo{booktitle}{\emph{Introduction to machine learning with Python: a guide for data scientists}}.
\newblock \bibinfo{publisher}{" O'Reilly Media, Inc."}, \bibinfo{address}{Delaware, USA}.
\newblock


\bibitem[Mun et~al\mbox{.}(2024)]%
        {counter}
\bibfield{author}{\bibinfo{person}{Jimin Mun}, \bibinfo{person}{Cathy Buerger}, \bibinfo{person}{Jenny~T Liang}, \bibinfo{person}{Joshua Garland}, {and} \bibinfo{person}{Maarten Sap}.} \bibinfo{year}{2024}\natexlab{}.
\newblock \showarticletitle{Counterspeakers’ Perspectives: Unveiling Barriers and AI Needs in the Fight against Online Hate}. In \bibinfo{booktitle}{\emph{Proceedings of the CHI Conference on Human Factors in Computing Systems}} (Honolulu, HI, USA) \emph{(\bibinfo{series}{CHI '24})}. \bibinfo{publisher}{Association for Computing Machinery}, \bibinfo{address}{New York, NY, USA}, Article \bibinfo{articleno}{742}, \bibinfo{numpages}{22}~pages.
\newblock
\showISBNx{9798400703300}
\urldef\tempurl%
\url{https://doi.org/10.1145/3613904.3642025}
\showDOI{\tempurl}


\bibitem[Munger(2019)]%
        {munger}
\bibfield{author}{\bibinfo{person}{Kevin Munger}.} \bibinfo{year}{2019}\natexlab{}.
\newblock \showarticletitle{The Limited Value of Non-Replicable Field Experiments in Contexts With Low Temporal Validity}.
\newblock \bibinfo{journal}{\emph{Social Media + Society}} \bibinfo{volume}{5}, \bibinfo{number}{3} (\bibinfo{year}{2019}), \bibinfo{pages}{2056305119859294}.
\newblock
\urldef\tempurl%
\url{https://doi.org/10.1177/2056305119859294}
\showDOI{\tempurl}
\showeprint{https://doi.org/10.1177/2056305119859294}


\bibitem[Murdoch et~al\mbox{.}(2019)]%
        {murdoch}
\bibfield{author}{\bibinfo{person}{W.~James Murdoch}, \bibinfo{person}{Chandan Singh}, \bibinfo{person}{Karl Kumbier}, \bibinfo{person}{Reza Abbasi-Asl}, {and} \bibinfo{person}{Bin Yu}.} \bibinfo{year}{2019}\natexlab{}.
\newblock \showarticletitle{Definitions, methods, and applications in interpretable machine learning}.
\newblock \bibinfo{journal}{\emph{Proceedings of the National Academy of Sciences}} \bibinfo{volume}{116}, \bibinfo{number}{44} (\bibinfo{year}{2019}), \bibinfo{pages}{22071--22080}.
\newblock
\urldef\tempurl%
\url{https://doi.org/10.1073/pnas.1900654116}
\showDOI{\tempurl}
\showeprint{https://www.pnas.org/doi/pdf/10.1073/pnas.1900654116}


\bibitem[Nagar et~al\mbox{.}(2023)]%
        {Nagar2023}
\bibfield{author}{\bibinfo{person}{Shubhanshu Nagar}, \bibinfo{person}{Faysal~A. Barbhuiya}, {and} \bibinfo{person}{Koushik Dey}.} \bibinfo{year}{2023}\natexlab{}.
\newblock \showarticletitle{Towards more robust hate speech detection: using social context and user data}.
\newblock \bibinfo{journal}{\emph{Social Network Analysis and Mining}} \bibinfo{volume}{13}, \bibinfo{number}{47} (\bibinfo{year}{2023}), \bibinfo{pages}{1--14}.
\newblock
\urldef\tempurl%
\url{https://doi.org/10.1007/s13278-023-01051-6}
\showDOI{\tempurl}


\bibitem[Narayanan~Venkit et~al\mbox{.}(2023)]%
        {venkit2023automated}
\bibfield{author}{\bibinfo{person}{Pranav Narayanan~Venkit}, \bibinfo{person}{Mukund Srinath}, {and} \bibinfo{person}{Shomir Wilson}.} \bibinfo{year}{2023}\natexlab{}.
\newblock \showarticletitle{Automated Ableism: An Exploration of Explicit Disability Biases in Sentiment and Toxicity Analysis Models}. In \bibinfo{booktitle}{\emph{Proceedings of the 3rd Workshop on Trustworthy Natural Language Processing (TrustNLP 2023)}}, \bibfield{editor}{\bibinfo{person}{Anaelia Ovalle}, \bibinfo{person}{Kai-Wei Chang}, \bibinfo{person}{Ninareh Mehrabi}, \bibinfo{person}{Yada Pruksachatkun}, \bibinfo{person}{Aram Galystan}, \bibinfo{person}{Jwala Dhamala}, \bibinfo{person}{Apurv Verma}, \bibinfo{person}{Trista Cao}, \bibinfo{person}{Anoop Kumar}, {and} \bibinfo{person}{Rahul Gupta}} (Eds.). \bibinfo{publisher}{Association for Computational Linguistics}, \bibinfo{address}{Toronto, Canada}, \bibinfo{pages}{26--34}.
\newblock
\urldef\tempurl%
\url{https://doi.org/10.18653/v1/2023.trustnlp-1.3}
\showDOI{\tempurl}


\bibitem[Nogara et~al\mbox{.}(2025)]%
        {nogara2025toxic}
\bibfield{author}{\bibinfo{person}{Gianluca Nogara}, \bibinfo{person}{Francesco Pierri}, \bibinfo{person}{Stefano Cresci}, \bibinfo{person}{Luca Luceri}, \bibinfo{person}{Petter T{\"o}rnberg}, {and} \bibinfo{person}{Silvia Giordano}.} \bibinfo{year}{2025}\natexlab{}.
\newblock \showarticletitle{Toxic Bias: Perspective API Misreads German as More Toxic}. In \bibinfo{booktitle}{\emph{Proceedings of the 19th AAAI International Conference on Web and Social Media (ICWSM'25)}}. \bibinfo{publisher}{AAAI Press}, \bibinfo{address}{Copenhagen, Denmark}, \bibinfo{pages}{12--23}.
\newblock
\newblock
\shownote{Please check and cite the published version of this paper}.


\bibitem[Nozza et~al\mbox{.}(2019)]%
        {Nozza2019}
\bibfield{author}{\bibinfo{person}{Debora Nozza}, \bibinfo{person}{Claudia Volpetti}, {and} \bibinfo{person}{Elisabetta Fersini}.} \bibinfo{year}{2019}\natexlab{}.
\newblock \showarticletitle{Unintended Bias in Misogyny Detection}. In \bibinfo{booktitle}{\emph{IEEE/WIC/ACM International Conference on Web Intelligence}} (Thessaloniki, Greece) \emph{(\bibinfo{series}{WI '19})}. \bibinfo{publisher}{Association for Computing Machinery}, \bibinfo{address}{New York, NY, USA}, \bibinfo{pages}{149–155}.
\newblock
\showISBNx{9781450369343}
\urldef\tempurl%
\url{https://doi.org/10.1145/3350546.3352512}
\showDOI{\tempurl}


\bibitem[Ozanne et~al\mbox{.}(2022)]%
        {ozanne2022shall}
\bibfield{author}{\bibinfo{person}{Marie Ozanne}, \bibinfo{person}{Ameya Bhandari}, \bibinfo{person}{Natalya~N. Bazarova}, {and} \bibinfo{person}{Dominic DiFranzo}.} \bibinfo{year}{2022}\natexlab{}.
\newblock \showarticletitle{Shall AI Moderators Be Made Visible? Perception of Accountability and Trust in Moderation Systems on Social Media Platforms}.
\newblock \bibinfo{journal}{\emph{Big Data \& Society}} \bibinfo{volume}{9}, \bibinfo{number}{2} (\bibinfo{year}{2022}), \bibinfo{pages}{1--13}.
\newblock
\urldef\tempurl%
\url{https://doi.org/10.1177/20539517221115666}
\showDOI{\tempurl}


\bibitem[Park et~al\mbox{.}(2018)]%
        {park-etal-2018-reducing}
\bibfield{author}{\bibinfo{person}{Ji~Ho Park}, \bibinfo{person}{Jamin Shin}, {and} \bibinfo{person}{Pascale Fung}.} \bibinfo{year}{2018}\natexlab{}.
\newblock \showarticletitle{Reducing Gender Bias in Abusive Language Detection}. In \bibinfo{booktitle}{\emph{Proceedings of the 2018 Conference on Empirical Methods in Natural Language Processing}}, \bibfield{editor}{\bibinfo{person}{Ellen Riloff}, \bibinfo{person}{David Chiang}, \bibinfo{person}{Julia Hockenmaier}, {and} \bibinfo{person}{Jun{'}ichi Tsujii}} (Eds.). \bibinfo{publisher}{Association for Computational Linguistics}, \bibinfo{address}{Brussels, Belgium}, \bibinfo{pages}{2799--2804}.
\newblock
\urldef\tempurl%
\url{https://doi.org/10.18653/v1/D18-1302}
\showDOI{\tempurl}


\bibitem[Passi and Vorvoreanu(2022)]%
        {passi2022overreliance}
\bibfield{author}{\bibinfo{person}{Samir Passi} {and} \bibinfo{person}{Mihaela Vorvoreanu}.} \bibinfo{year}{2022}\natexlab{}.
\newblock \bibinfo{booktitle}{\emph{Overreliance on AI: Literature Review}}.
\newblock \bibinfo{type}{{T}echnical {R}eport} MSR-TR-2022-12. \bibinfo{institution}{Microsoft}.
\newblock
\urldef\tempurl%
\url{https://www.microsoft.com/en-us/research/publication/overreliance-on-ai-literature-review/}
\showURL{%
\tempurl}


\bibitem[Prabhakaran et~al\mbox{.}(2019a)]%
        {prabhakaran-etal-2019-perturbation}
\bibfield{author}{\bibinfo{person}{Vinodkumar Prabhakaran}, \bibinfo{person}{Ben Hutchinson}, {and} \bibinfo{person}{Margaret Mitchell}.} \bibinfo{year}{2019}\natexlab{a}.
\newblock \showarticletitle{Perturbation Sensitivity Analysis to Detect Unintended Model Biases}. In \bibinfo{booktitle}{\emph{Proceedings of the 2019 Conference on Empirical Methods in Natural Language Processing and the 9th International Joint Conference on Natural Language Processing (EMNLP-IJCNLP)}}, \bibfield{editor}{\bibinfo{person}{Kentaro Inui}, \bibinfo{person}{Jing Jiang}, \bibinfo{person}{Vincent Ng}, {and} \bibinfo{person}{Xiaojun Wan}} (Eds.). \bibinfo{publisher}{Association for Computational Linguistics}, \bibinfo{address}{Hong Kong, China}, \bibinfo{pages}{5740--5745}.
\newblock
\urldef\tempurl%
\url{https://aclanthology.org/D19-1578}
\showURL{%
\tempurl}


\bibitem[Prabhakaran et~al\mbox{.}(2019b)]%
        {prabhakaran_perturbation_2019}
\bibfield{author}{\bibinfo{person}{Vinodkumar Prabhakaran}, \bibinfo{person}{Ben Hutchinson}, {and} \bibinfo{person}{Margaret Mitchell}.} \bibinfo{year}{2019}\natexlab{b}.
\newblock \showarticletitle{Perturbation {Sensitivity} {Analysis} to {Detect} {Unintended} {Model} {Biases}}. In \bibinfo{booktitle}{\emph{Proceedings of the 2019 {Conference} on {Empirical} {Methods} in {Natural} {Language} {Processing} and the 9th {International} {Joint} {Conference} on {Natural} {Language} {Processing} ({EMNLP}-{IJCNLP})}}. \bibinfo{publisher}{Association for Computational Linguistics}, \bibinfo{address}{Hong Kong, China}, \bibinfo{pages}{5740--5745}.
\newblock


\bibitem[Pérez et~al\mbox{.}(2023)]%
        {context}
\bibfield{author}{\bibinfo{person}{Juan~Manuel Pérez}, \bibinfo{person}{Franco~M. Luque}, \bibinfo{person}{Demian Zayat}, \bibinfo{person}{Martín Kondratzky}, \bibinfo{person}{Agustín Moro}, \bibinfo{person}{Pablo~Santiago Serrati}, \bibinfo{person}{Joaquín Zajac}, \bibinfo{person}{Paula Miguel}, \bibinfo{person}{Natalia Debandi}, \bibinfo{person}{Agustín Gravano}, {and} \bibinfo{person}{Viviana Cotik}.} \bibinfo{year}{2023}\natexlab{}.
\newblock \showarticletitle{Assessing the Impact of Contextual Information in Hate Speech Detection}.
\newblock \bibinfo{journal}{\emph{IEEE Access}}  \bibinfo{volume}{11} (\bibinfo{year}{2023}), \bibinfo{pages}{30575--30590}.
\newblock
\urldef\tempurl%
\url{https://doi.org/10.1109/ACCESS.2023.3258973}
\showDOI{\tempurl}


\bibitem[Raji and Buolamwini(2019)]%
        {raji_actionable_2019}
\bibfield{author}{\bibinfo{person}{Inioluwa~Deborah Raji} {and} \bibinfo{person}{Joy Buolamwini}.} \bibinfo{year}{2019}\natexlab{}.
\newblock \showarticletitle{Actionable {Auditing}: {Investigating} the {Impact} of {Publicly} {Naming} {Biased} {Performance} {Results} of {Commercial} {AI} {Products}}. In \bibinfo{booktitle}{\emph{Proceedings of the 2019 {AAAI}/{ACM} {Conference} on {AI}, {Ethics}, and {Society}}}. \bibinfo{publisher}{ACM}, \bibinfo{address}{Honolulu HI USA}, \bibinfo{pages}{429--435}.
\newblock


\bibitem[Raji et~al\mbox{.}(2022)]%
        {raji_outsider_2022}
\bibfield{author}{\bibinfo{person}{Inioluwa~Deborah Raji}, \bibinfo{person}{Peggy Xu}, \bibinfo{person}{Colleen Honigsberg}, {and} \bibinfo{person}{Daniel~E. Ho}.} \bibinfo{year}{2022}\natexlab{}.
\newblock \bibinfo{title}{Outsider {Oversight}: {Designing} a {Third} {Party} {Audit} {Ecosystem} for {AI} {Governance}}.
\newblock
\newblock
\urldef\tempurl%
\url{http://arxiv.org/abs/2206.04737}
\showURL{%
\tempurl}
\newblock
\shownote{arXiv:2206.04737 [cs]}.


\bibitem[Rieder and Skop(2021)]%
        {rieder}
\bibfield{author}{\bibinfo{person}{Bernhard Rieder} {and} \bibinfo{person}{Yarden Skop}.} \bibinfo{year}{2021}\natexlab{}.
\newblock \showarticletitle{The fabrics of machine moderation: Studying the technical, normative, and organizational structure of Perspective API}.
\newblock \bibinfo{journal}{\emph{Big Data \& Society}} \bibinfo{volume}{8}, \bibinfo{number}{2} (\bibinfo{year}{2021}), \bibinfo{pages}{20539517211046181}.
\newblock
\urldef\tempurl%
\url{https://doi.org/10.1177/20539517211046181}
\showDOI{\tempurl}
\showeprint{https://doi.org/10.1177/20539517211046181}


\bibitem[Roth(2024)]%
        {roth_chatgpt_2024}
\bibfield{author}{\bibinfo{person}{Emma Roth}.} \bibinfo{year}{2024}\natexlab{}.
\newblock \bibinfo{booktitle}{\emph{ChatGPT’s weekly users have doubled in less than a year}}.
\newblock The Verge.
\newblock
\urldef\tempurl%
\url{https://www.theverge.com}
\showURL{%
\tempurl}
\newblock
\shownote{Accessed: 2024-12-10}.


\bibitem[R{\"o}ttger et~al\mbox{.}(2021)]%
        {rottgerHateCheckFunctionalTests2021}
\bibfield{author}{\bibinfo{person}{Paul R{\"o}ttger}, \bibinfo{person}{Bertie Vidgen}, \bibinfo{person}{Dong Nguyen}, \bibinfo{person}{Zeerak Waseem}, \bibinfo{person}{Helen Margetts}, {and} \bibinfo{person}{Janet Pierrehumbert}.} \bibinfo{year}{2021}\natexlab{}.
\newblock \showarticletitle{{H}ate{C}heck: Functional Tests for Hate Speech Detection Models}. In \bibinfo{booktitle}{\emph{Proceedings of the 59th Annual Meeting of the Association for Computational Linguistics and the 11th International Joint Conference on Natural Language Processing (Volume 1: Long Papers)}}, \bibfield{editor}{\bibinfo{person}{Chengqing Zong}, \bibinfo{person}{Fei Xia}, \bibinfo{person}{Wenjie Li}, {and} \bibinfo{person}{Roberto Navigli}} (Eds.). \bibinfo{publisher}{Association for Computational Linguistics}, \bibinfo{address}{Online}, \bibinfo{pages}{41--58}.
\newblock
\urldef\tempurl%
\url{https://doi.org/10.18653/v1/2021.acl-long.4}
\showDOI{\tempurl}


\bibitem[Rozemberczki et~al\mbox{.}(2022)]%
        {rozemberczki_2022}
\bibfield{author}{\bibinfo{person}{Benedek Rozemberczki}, \bibinfo{person}{Lauren Watson}, \bibinfo{person}{Péter Bayer}, \bibinfo{person}{Hao-Tsung Yang}, \bibinfo{person}{Olivér Kiss}, \bibinfo{person}{Sebastian Nilsson}, {and} \bibinfo{person}{Rik Sarkar}.} \bibinfo{year}{2022}\natexlab{}.
\newblock \showarticletitle{The Shapley Value in Machine Learning}. In \bibinfo{booktitle}{\emph{Proceedings of the Thirty-First International Joint Conference on Artificial Intelligence, {IJCAI-22}}}, \bibfield{editor}{\bibinfo{person}{Lud~De Raedt}} (Ed.). \bibinfo{publisher}{International Joint Conferences on Artificial Intelligence Organization}, \bibinfo{pages}{5572--5579}.
\newblock
\urldef\tempurl%
\url{https://doi.org/10.24963/ijcai.2022/778}
\showDOI{\tempurl}
\newblock
\shownote{Survey Track}.


\bibitem[Sap et~al\mbox{.}(2019)]%
        {sap_risk_2019}
\bibfield{author}{\bibinfo{person}{Maarten Sap}, \bibinfo{person}{Dallas Card}, \bibinfo{person}{Saadia Gabriel}, \bibinfo{person}{Yejin Choi}, {and} \bibinfo{person}{Noah~A. Smith}.} \bibinfo{year}{2019}\natexlab{}.
\newblock \showarticletitle{The {Risk} of {Racial} {Bias} in {Hate} {Speech} {Detection}}. In \bibinfo{booktitle}{\emph{Proceedings of the 57th {Annual} {Meeting} of the {Association} for {Computational} {Linguistics}}}, \bibfield{editor}{\bibinfo{person}{Anna Korhonen}, \bibinfo{person}{David Traum}, {and} \bibinfo{person}{Lluís Màrquez}} (Eds.). \bibinfo{publisher}{Association for Computational Linguistics}, \bibinfo{address}{Florence, Italy}, \bibinfo{pages}{1668--1678}.
\newblock
\urldef\tempurl%
\url{https://aclanthology.org/P19-1163}
\showURL{%
\tempurl}


\bibitem[Sap et~al\mbox{.}(2020)]%
        {sap_social_2020}
\bibfield{author}{\bibinfo{person}{Maarten Sap}, \bibinfo{person}{Saadia Gabriel}, \bibinfo{person}{Lianhui Qin}, \bibinfo{person}{Dan Jurafsky}, \bibinfo{person}{Noah~A. Smith}, {and} \bibinfo{person}{Yejin Choi}.} \bibinfo{year}{2020}\natexlab{}.
\newblock \showarticletitle{Social {Bias} {Frames}: {Reasoning} about {Social} and {Power} {Implications} of {Language}}. In \bibinfo{booktitle}{\emph{Proceedings of the 58th {Annual} {Meeting} of the {Association} for {Computational} {Linguistics}}}, \bibfield{editor}{\bibinfo{person}{Dan Jurafsky}, \bibinfo{person}{Joyce Chai}, \bibinfo{person}{Natalie Schluter}, {and} \bibinfo{person}{Joel Tetreault}} (Eds.). \bibinfo{publisher}{Association for Computational Linguistics}, \bibinfo{address}{Online}, \bibinfo{pages}{5477--5490}.
\newblock
\urldef\tempurl%
\url{https://aclanthology.org/2020.acl-main.486}
\showURL{%
\tempurl}


\bibitem[Sap et~al\mbox{.}(2022a)]%
        {sap-etal-2022-annotators}
\bibfield{author}{\bibinfo{person}{Maarten Sap}, \bibinfo{person}{Swabha Swayamdipta}, \bibinfo{person}{Laura Vianna}, \bibinfo{person}{Xuhui Zhou}, \bibinfo{person}{Yejin Choi}, {and} \bibinfo{person}{Noah~A. Smith}.} \bibinfo{year}{2022}\natexlab{a}.
\newblock \showarticletitle{Annotators with Attitudes: How Annotator Beliefs And Identities Bias Toxic Language Detection}. In \bibinfo{booktitle}{\emph{Proceedings of the 2022 Conference of the North American Chapter of the Association for Computational Linguistics: Human Language Technologies}}, \bibfield{editor}{\bibinfo{person}{Marine Carpuat}, \bibinfo{person}{Marie-Catherine de~Marneffe}, {and} \bibinfo{person}{Ivan~Vladimir Meza~Ruiz}} (Eds.). \bibinfo{publisher}{Association for Computational Linguistics}, \bibinfo{address}{Seattle, United States}, \bibinfo{pages}{5884--5906}.
\newblock
\urldef\tempurl%
\url{https://doi.org/10.18653/v1/2022.naacl-main.431}
\showDOI{\tempurl}


\bibitem[Sap et~al\mbox{.}(2022b)]%
        {sap2022annotators}
\bibfield{author}{\bibinfo{person}{Maarten Sap}, \bibinfo{person}{Swabha Swayamdipta}, \bibinfo{person}{Laura Vianna}, \bibinfo{person}{Xuhui Zhou}, \bibinfo{person}{Yejin Choi}, {and} \bibinfo{person}{Noah~A. Smith}.} \bibinfo{year}{2022}\natexlab{b}.
\newblock \showarticletitle{Annotators with Attitudes: How Annotator Beliefs And Identities Bias Toxic Language Detection}. In \bibinfo{booktitle}{\emph{Proceedings of the 2022 Conference of the North American Chapter of the Association for Computational Linguistics: Human Language Technologies}} (Seattle, United States). \bibinfo{publisher}{Association for Computational Linguistics}, \bibinfo{address}{Seattle, United States}, \bibinfo{pages}{5884--5906}.
\newblock
\urldef\tempurl%
\url{https://doi.org/10.18653/v1/2022.naacl-main.431}
\showDOI{\tempurl}


\bibitem[Schafner et~al\mbox{.}(2024)]%
        {schafner2024community}
\bibfield{author}{\bibinfo{person}{Brennan Schafner}, \bibinfo{person}{Arjun~Nitin Bhagoji}, \bibinfo{person}{Siyuan Cheng}, \bibinfo{person}{Jacqueline Mei}, \bibinfo{person}{Jay~L. Shen}, \bibinfo{person}{Grace Wang}, \bibinfo{person}{Marshini Chetty}, \bibinfo{person}{Nick Feamster}, \bibinfo{person}{Genevieve Lakier}, {and} \bibinfo{person}{Chenhao Tan}.} \bibinfo{year}{2024}\natexlab{}.
\newblock \showarticletitle{“Community Guidelines Make this the Best Party on the Internet”: An In-Depth Study of Online Platforms’ Content Moderation Policies}. In \bibinfo{booktitle}{\emph{Proceedings of the CHI Conference on Human Factors in Computing Systems (CHI '24)}} (Honolulu, HI, USA). \bibinfo{publisher}{ACM}, \bibinfo{address}{New York, NY, USA}, \bibinfo{pages}{16}.
\newblock
\urldef\tempurl%
\url{https://doi.org/10.1145/3613904.3642333}
\showDOI{\tempurl}


\bibitem[Schoenebeck et~al\mbox{.}(2023)]%
        {schoenebeckOnlineHarassmentMajority2023}
\bibfield{author}{\bibinfo{person}{Sarita Schoenebeck}, \bibinfo{person}{Amna Batool}, \bibinfo{person}{Giang Do}, \bibinfo{person}{Sylvia Darling}, \bibinfo{person}{Gabriel Grill}, \bibinfo{person}{Daricia Wilkinson}, \bibinfo{person}{Mehtab Khan}, \bibinfo{person}{Kentaro Toyama}, {and} \bibinfo{person}{Louise Ashwell}.} \bibinfo{year}{2023}\natexlab{}.
\newblock \showarticletitle{Online Harassment in Majority Contexts: Examining Harms and Remedies across Countries}. In \bibinfo{booktitle}{\emph{Proceedings of the 2023 CHI Conference on Human Factors in Computing Systems}} (Hamburg, Germany) \emph{(\bibinfo{series}{CHI '23})}. \bibinfo{publisher}{Association for Computing Machinery}, \bibinfo{address}{New York, NY, USA}, Article \bibinfo{articleno}{485}, \bibinfo{numpages}{16}~pages.
\newblock
\showISBNx{9781450394215}
\urldef\tempurl%
\url{https://doi.org/10.1145/3544548.3581020}
\showDOI{\tempurl}


\bibitem[Shahid and Vashistha(2023)]%
        {farhana2023}
\bibfield{author}{\bibinfo{person}{Farhana Shahid} {and} \bibinfo{person}{Aditya Vashistha}.} \bibinfo{year}{2023}\natexlab{}.
\newblock \showarticletitle{Decolonizing Content Moderation: Does Uniform Global Community Standard Resemble Utopian Equality or Western Power Hegemony?}. In \bibinfo{booktitle}{\emph{Proceedings of the 2023 CHI Conference on Human Factors in Computing Systems}} (Hamburg, Germany) \emph{(\bibinfo{series}{CHI '23})}. \bibinfo{publisher}{Association for Computing Machinery}, \bibinfo{address}{New York, NY, USA}, Article \bibinfo{articleno}{391}, \bibinfo{numpages}{18}~pages.
\newblock
\showISBNx{9781450394215}
\urldef\tempurl%
\url{https://doi.org/10.1145/3544548.3581538}
\showDOI{\tempurl}


\bibitem[Sheng et~al\mbox{.}(2019)]%
        {sheng_woman_2019}
\bibfield{author}{\bibinfo{person}{Emily Sheng}, \bibinfo{person}{Kai-Wei Chang}, \bibinfo{person}{Premkumar Natarajan}, {and} \bibinfo{person}{Nanyun Peng}.} \bibinfo{year}{2019}\natexlab{}.
\newblock \showarticletitle{The {Woman} {Worked} as a {Babysitter}: {On} {Biases} in {Language} {Generation}}. In \bibinfo{booktitle}{\emph{Proceedings of the 2019 {Conference} on {Empirical} {Methods} in {Natural} {Language} {Processing} and the 9th {International} {Joint} {Conference} on {Natural} {Language} {Processing} ({EMNLP}-{IJCNLP})}}, \bibfield{editor}{\bibinfo{person}{Kentaro Inui}, \bibinfo{person}{Jing Jiang}, \bibinfo{person}{Vincent Ng}, {and} \bibinfo{person}{Xiaojun Wan}} (Eds.). \bibinfo{publisher}{Association for Computational Linguistics}, \bibinfo{address}{Hong Kong, China}, \bibinfo{pages}{3407--3412}.
\newblock
\urldef\tempurl%
\url{https://doi.org/10.18653/v1/D19-1339}
\showDOI{\tempurl}


\bibitem[Thomas et~al\mbox{.}(2022)]%
        {thomas2022its}
\bibfield{author}{\bibinfo{person}{Kurt Thomas}, \bibinfo{person}{Patrick~Gage Kelley}, \bibinfo{person}{Sunny Consolvo}, \bibinfo{person}{Patrawat Samermit}, {and} \bibinfo{person}{Elie Bursztein}.} \bibinfo{year}{2022}\natexlab{}.
\newblock \showarticletitle{“It’s Common and a Part of Being a Content Creator”: Understanding How Creators Experience and Cope with Hate and Harassment Online}. In \bibinfo{booktitle}{\emph{CHI Conference on Human Factors in Computing Systems}} (New Orleans, LA, USA). \bibinfo{publisher}{ACM}, \bibinfo{address}{New York, NY, USA}, \bibinfo{pages}{1--15}.
\newblock
\urldef\tempurl%
\url{https://doi.org/10.1145/3491102.3501879}
\showDOI{\tempurl}


\bibitem[Vidgen et~al\mbox{.}(2021)]%
        {vidgen-etal-2021-learning}
\bibfield{author}{\bibinfo{person}{Bertie Vidgen}, \bibinfo{person}{Tristan Thrush}, \bibinfo{person}{Zeerak Waseem}, {and} \bibinfo{person}{Douwe Kiela}.} \bibinfo{year}{2021}\natexlab{}.
\newblock \showarticletitle{Learning from the Worst: Dynamically Generated Datasets to Improve Online Hate Detection}. In \bibinfo{booktitle}{\emph{Proceedings of the 59th Annual Meeting of the Association for Computational Linguistics and the 11th International Joint Conference on Natural Language Processing (Volume 1: Long Papers)}}, \bibfield{editor}{\bibinfo{person}{Chengqing Zong}, \bibinfo{person}{Fei Xia}, \bibinfo{person}{Wenjie Li}, {and} \bibinfo{person}{Roberto Navigli}} (Eds.). \bibinfo{publisher}{Association for Computational Linguistics}, \bibinfo{address}{Online}, \bibinfo{pages}{1667--1682}.
\newblock
\urldef\tempurl%
\url{https://doi.org/10.18653/v1/2021.acl-long.132}
\showDOI{\tempurl}


\bibitem[Waseem et~al\mbox{.}(2017)]%
        {waseem2017understanding}
\bibfield{author}{\bibinfo{person}{Zeerak Waseem}, \bibinfo{person}{Thomas Davidson}, \bibinfo{person}{Dana Warmsley}, {and} \bibinfo{person}{Ingmar Weber}.} \bibinfo{year}{2017}\natexlab{}.
\newblock \showarticletitle{Understanding Abuse: A Typology of Abusive Language Detection Subtasks}. In \bibinfo{booktitle}{\emph{Proceedings of the ACL-Workshop on Abusive Language Online}}. \bibinfo{publisher}{Association for Computational Linguistics}, \bibinfo{address}{Vancouver, BC, Canada}, \bibinfo{pages}{78--84}.
\newblock


\bibitem[Waseem and Hovy(2016)]%
        {waseem_hateful_2016}
\bibfield{author}{\bibinfo{person}{Zeerak Waseem} {and} \bibinfo{person}{Dirk Hovy}.} \bibinfo{year}{2016}\natexlab{}.
\newblock \showarticletitle{Hateful Symbols or Hateful People? Predictive Features for Hate Speech Detection on {T}witter}. In \bibinfo{booktitle}{\emph{Proceedings of the {NAACL} Student Research Workshop}}, \bibfield{editor}{\bibinfo{person}{Jacob Andreas}, \bibinfo{person}{Eunsol Choi}, {and} \bibinfo{person}{Angeliki Lazaridou}} (Eds.). \bibinfo{publisher}{Association for Computational Linguistics}, \bibinfo{address}{San Diego, California}, \bibinfo{pages}{88--93}.
\newblock
\urldef\tempurl%
\url{https://doi.org/10.18653/v1/N16-2013}
\showDOI{\tempurl}


\bibitem[Wiegand et~al\mbox{.}(2021)]%
        {wiegand_implicitly_2021}
\bibfield{author}{\bibinfo{person}{Michael Wiegand}, \bibinfo{person}{Josef Ruppenhofer}, {and} \bibinfo{person}{Elisabeth Eder}.} \bibinfo{year}{2021}\natexlab{}.
\newblock \showarticletitle{Implicitly {Abusive} {Language} – {What} does it actually look like and why are we not getting there?}. In \bibinfo{booktitle}{\emph{Proceedings of the 2021 {Conference} of the {North} {American} {Chapter} of the {Association} for {Computational} {Linguistics}: {Human} {Language} {Technologies}}}, \bibfield{editor}{\bibinfo{person}{Kristina Toutanova}, \bibinfo{person}{Anna Rumshisky}, \bibinfo{person}{Luke Zettlemoyer}, \bibinfo{person}{Dilek Hakkani-Tur}, \bibinfo{person}{Iz~Beltagy}, \bibinfo{person}{Steven Bethard}, \bibinfo{person}{Ryan Cotterell}, \bibinfo{person}{Tanmoy Chakraborty}, {and} \bibinfo{person}{Yichao Zhou}} (Eds.). \bibinfo{publisher}{Association for Computational Linguistics}, \bibinfo{address}{Online}, \bibinfo{pages}{576--587}.
\newblock
\urldef\tempurl%
\url{https://aclanthology.org/2021.naacl-main.48}
\showURL{%
\tempurl}


\bibitem[Wiegand et~al\mbox{.}(2019)]%
        {wiegand_detection_2019}
\bibfield{author}{\bibinfo{person}{Michael Wiegand}, \bibinfo{person}{Josef Ruppenhofer}, {and} \bibinfo{person}{Thomas Kleinbauer}.} \bibinfo{year}{2019}\natexlab{}.
\newblock \showarticletitle{{D}etection of {A}busive {L}anguage: the {P}roblem of {B}iased {D}atasets}. In \bibinfo{booktitle}{\emph{Proceedings of the 2019 Conference of the North {A}merican Chapter of the Association for Computational Linguistics: Human Language Technologies, Volume 1 (Long and Short Papers)}}, \bibfield{editor}{\bibinfo{person}{Jill Burstein}, \bibinfo{person}{Christy Doran}, {and} \bibinfo{person}{Thamar Solorio}} (Eds.). \bibinfo{publisher}{Association for Computational Linguistics}, \bibinfo{address}{Minneapolis, Minnesota}, \bibinfo{pages}{602--608}.
\newblock
\urldef\tempurl%
\url{https://doi.org/10.18653/v1/N19-1060}
\showDOI{\tempurl}


\bibitem[Xia et~al\mbox{.}(2020)]%
        {xia2020demoting}
\bibfield{author}{\bibinfo{person}{Mengzhou Xia}, \bibinfo{person}{Anjalie Field}, {and} \bibinfo{person}{Yulia Tsvetkov}.} \bibinfo{year}{2020}\natexlab{}.
\newblock \showarticletitle{Demoting Racial Bias in Hate Speech Detection}. In \bibinfo{booktitle}{\emph{Proceedings of the Eighth International Workshop on Natural Language Processing for Social Media}}, \bibfield{editor}{\bibinfo{person}{Lun-Wei Ku} {and} \bibinfo{person}{Cheng-Te Li}} (Eds.). \bibinfo{publisher}{Association for Computational Linguistics}, \bibinfo{address}{Online}, \bibinfo{pages}{7--14}.
\newblock
\urldef\tempurl%
\url{https://doi.org/10.18653/v1/2020.socialnlp-1.2}
\showDOI{\tempurl}


\bibitem[Yan and Zhang(2022)]%
        {yan2022active}
\bibfield{author}{\bibinfo{person}{Tom Yan} {and} \bibinfo{person}{Chicheng Zhang}.} \bibinfo{year}{2022}\natexlab{}.
\newblock \showarticletitle{Active fairness auditing}. In \bibinfo{booktitle}{\emph{Proceedings of the 39th International Conference on Machine Learning}} \emph{(\bibinfo{series}{Proceedings of Machine Learning Research}, Vol.~\bibinfo{volume}{162})}, \bibfield{editor}{\bibinfo{person}{Kamalika Chaudhuri}, \bibinfo{person}{Stefanie Jegelka}, \bibinfo{person}{Le~Song}, \bibinfo{person}{Csaba Szepesvari}, \bibinfo{person}{Gang Niu}, {and} \bibinfo{person}{Sivan Sabato}} (Eds.). \bibinfo{publisher}{PMLR}, \bibinfo{address}{Baltimore, Maryland, USA}, \bibinfo{pages}{24929--24962}.
\newblock
\urldef\tempurl%
\url{https://proceedings.mlr.press/v162/yan22c.html}
\showURL{%
\tempurl}


\bibitem[Yin and Zubiaga(2021)]%
        {yin2021towards}
\bibfield{author}{\bibinfo{person}{Wenqi Yin} {and} \bibinfo{person}{Arkaitz Zubiaga}.} \bibinfo{year}{2021}\natexlab{}.
\newblock \showarticletitle{Towards generalisable hate speech detection: a review on obstacles and solutions}.
\newblock \bibinfo{journal}{\emph{PeerJ Computer Science}}  \bibinfo{volume}{7} (\bibinfo{year}{2021}), \bibinfo{pages}{e598}.
\newblock
\urldef\tempurl%
\url{https://doi.org/10.7717/peerj-cs.598}
\showDOI{\tempurl}


\bibitem[Yoder et~al\mbox{.}(2022)]%
        {yoder2022hate}
\bibfield{author}{\bibinfo{person}{Michael Yoder}, \bibinfo{person}{Lynnette Ng}, \bibinfo{person}{David~West Brown}, {and} \bibinfo{person}{Kathleen Carley}.} \bibinfo{year}{2022}\natexlab{}.
\newblock \showarticletitle{How Hate Speech Varies by Target Identity: A Computational Analysis}. In \bibinfo{booktitle}{\emph{Proceedings of the 26th Conference on Computational Natural Language Learning (CoNLL)}}, \bibfield{editor}{\bibinfo{person}{Antske Fokkens} {and} \bibinfo{person}{Vivek Srikumar}} (Eds.). \bibinfo{publisher}{Association for Computational Linguistics}, \bibinfo{address}{Abu Dhabi, United Arab Emirates (Hybrid)}, \bibinfo{pages}{27--39}.
\newblock
\urldef\tempurl%
\url{https://doi.org/10.18653/v1/2022.conll-1.3}
\showDOI{\tempurl}


\bibitem[Yu et~al\mbox{.}(2022)]%
        {yu2022hatespeechcounterspeech}
\bibfield{author}{\bibinfo{person}{Xinchen Yu}, \bibinfo{person}{Eduardo Blanco}, {and} \bibinfo{person}{Lingzi Hong}.} \bibinfo{year}{2022}\natexlab{}.
\newblock \bibinfo{title}{Hate Speech and Counter Speech Detection: Conversational Context Does Matter}.
\newblock
\newblock
\showeprint[arxiv]{2206.06423}~[cs.CL]
\urldef\tempurl%
\url{https://arxiv.org/abs/2206.06423}
\showURL{%
\tempurl}


\bibitem[Zhang et~al\mbox{.}(2023)]%
        {zhang-etal-2023-biasx}
\bibfield{author}{\bibinfo{person}{Yiming Zhang}, \bibinfo{person}{Sravani Nanduri}, \bibinfo{person}{Liwei Jiang}, \bibinfo{person}{Tongshuang Wu}, {and} \bibinfo{person}{Maarten Sap}.} \bibinfo{year}{2023}\natexlab{}.
\newblock \showarticletitle{{B}ias{X}: {``}Thinking Slow{''} in Toxic Content Moderation with Explanations of Implied Social Biases}. In \bibinfo{booktitle}{\emph{Proceedings of the 2023 Conference on Empirical Methods in Natural Language Processing}}, \bibfield{editor}{\bibinfo{person}{Houda Bouamor}, \bibinfo{person}{Juan Pino}, {and} \bibinfo{person}{Kalika Bali}} (Eds.). \bibinfo{publisher}{Association for Computational Linguistics}, \bibinfo{address}{Singapore}, \bibinfo{pages}{4920--4932}.
\newblock
\urldef\tempurl%
\url{https://doi.org/10.18653/v1/2023.emnlp-main.300}
\showDOI{\tempurl}


\bibitem[Zhou et~al\mbox{.}(2021)]%
        {zhou-etal-2021-challenges}
\bibfield{author}{\bibinfo{person}{Xuhui Zhou}, \bibinfo{person}{Maarten Sap}, \bibinfo{person}{Swabha Swayamdipta}, \bibinfo{person}{Yejin Choi}, {and} \bibinfo{person}{Noah Smith}.} \bibinfo{year}{2021}\natexlab{}.
\newblock \showarticletitle{Challenges in Automated Debiasing for Toxic Language Detection}. In \bibinfo{booktitle}{\emph{Proceedings of the 16th Conference of the European Chapter of the Association for Computational Linguistics: Main Volume}}, \bibfield{editor}{\bibinfo{person}{Paola Merlo}, \bibinfo{person}{Jorg Tiedemann}, {and} \bibinfo{person}{Reut Tsarfaty}} (Eds.). \bibinfo{publisher}{Association for Computational Linguistics}, \bibinfo{address}{Online}, \bibinfo{pages}{3143--3155}.
\newblock
\urldef\tempurl%
\url{https://doi.org/10.18653/v1/2021.eacl-main.274}
\showDOI{\tempurl}


\bibitem[Zsisku et~al\mbox{.}(2024)]%
        {zsisku2024}
\bibfield{author}{\bibinfo{person}{Eszter Zsisku}, \bibinfo{person}{Arkaitz Zubiaga}, {and} \bibinfo{person}{Haim Dubossarsky}.} \bibinfo{year}{2024}\natexlab{}.
\newblock \showarticletitle{Hate Speech Detection and Reclaimed Language: Mitigating False Positives and Compounded Discrimination}. In \bibinfo{booktitle}{\emph{Proceedings of the 16th ACM Web Science Conference}} (Stuttgart, Germany) \emph{(\bibinfo{series}{WEBSCI '24})}. \bibinfo{publisher}{Association for Computing Machinery}, \bibinfo{address}{New York, NY, USA}, \bibinfo{pages}{241–249}.
\newblock
\showISBNx{9798400703348}
\urldef\tempurl%
\url{https://doi.org/10.1145/3614419.3644025}
\showDOI{\tempurl}


\end{thebibliography}
 \clearpage
 \appendix
\onecolumn
\section{Appendix}
 
\subsection{Model Overview Documentations}
\label{sec:transparency}

\begin{small}
\begin{table*}[h]
\begin{tabular}{p{2.4cm}|p{2.3cm}p{1.4cm}p{1.4cm}p{2.3cm}p{3cm}} \toprule
\Description{
This table presents a transparency comparison of five commercial content moderation API services (Google Natural Language API, Amazon Comprehend, Microsoft Azure Content Moderators, OpenAI Moderation API, and Jigsaw Perspective API) based on model card categories outlined by Mitchell et al. (2019). It evaluates transparency across four key sections: Model Details, Intended Use and Guidance, Factors and Metrics, and Ethical Considerations.

**Model Details:**  
- Lists the developer, documentation availability, model and API version, model type, and information on training methods.  
- Only Google and Jigsaw disclose model types (e.g., PaLM 2 for Google and BERT-based models for Jigsaw).  
- Jigsaw provides information on training data origin and performance, while others omit details.  
- Jigsaw also explains how hate speech likelihood is operationalized, whereas Google gives only sub-category definitions.

**Intended Use and Guidance:**  
- Jigsaw is the only provider specifying primary intended uses and linking them to intended users.  
- OpenAI mentions intended users but not intended uses.  
- Only Jigsaw provides guidance on threshold selection, whereas Google shifts responsibility to users.  
- Both OpenAI and Jigsaw offer tutorials, while others do not.

**Factors and Metrics:**  
- Only Jigsaw provides relevant evaluation factors and performance metrics.  
- OpenAI offers performance measures but not factors.

**Ethical Considerations, Caveats, and Recommendations:**  
- Jigsaw is the only provider addressing ethical considerations and caveats.  
- Both Jigsaw and Google distinguish between probability scores and severity in predictions, clarifying that higher toxicity scores do not necessarily indicate greater harm.

This comparison highlights significant transparency gaps among providers, with Jigsaw demonstrating the most comprehensive disclosure across the evaluated categories.
}
{\textbf{Criteria}} &\textbf{Natural} &\textbf{Amazon} & \textbf{Content} &\textbf{Moderation API} & \textbf{Perspective API} \\
& \textbf{Language API}& \textbf{Comprehend} &\textbf{Moderators} & & \\ 
\toprule
\cline{1-6}
\multicolumn{6}{c}{\textbf{Model Details}} \\ \cline{1-6}
 Developer & Google & Amazon & Microsoft Azure& OpenAI & Jigsaw (Google)  \\ \cline{1-6}
 Documentation available &\checkmark&\checkmark&\checkmark&\checkmark&\checkmark\\ \cline{1-6}
  Model Date & 28.08.2023 & $\times$ & $\times$ & 01.03.2023 & $\times$ \\ \cline{1-6}
 Model Version & PaLM 2 & $\times$ &  $\times$& $\times$
 &  $\times$ \\ \cline{1-6} 
 API Version & v2 & 2017-11-27 & v1.0 & text-moderation-001 & v1alpha1 \\ \cline{1-6} 
 Model Type & Fine-tuned language Model& Proprietary NLP model & Proprietary NLP model & Proprietary moderation model & Multilingual BERT-based models \\ \cline{1-6} 
 Information about Training Algorithms, Fairness Evaluations, Parameters& $\times$ & $\times$ & $\times$ & $\times$ & Training Data origin discussed and model performans provided \\ \cline{1-6} 
Operationalization of Hate Speech &Definitions of sub-categories given, no further explanations&$\times$&$\times$&$\times$ &The likelihood that a reader interprets the comment provided in the request as embodying the specified attribute.\\ \cline{1-6}
 \multicolumn{6}{c}{\textbf{Intended Use and Usage Guidance/Tutorials}} \\ \cline{1-6}
 Primary intended use & $\times$ & $\times$ & $\times$ & $\times$ & Provides specific intended uses  \\ \cline{1-6}
 Primary intended users & $\times$ & $\times$ & $\times$ & Provided & Inteded usecases linked to inteded users  \\ \cline{1-6} 
 Information how to embed the Model & $\times$ & $\times$ & $\times$ & Proprietary moderation model & Currently working in including context \\ \cline{1-6} 
 Guidance on how to choose tresholds& Defined and tested by the user, Google refuses to take responsibility & $\times$ & $\times$ & No guidance, but it is mentioned that thresholds have to be carefully selected & Give guidance \\ \cline{1-6}
 Tutorials for implementation &  $\times$ & $\times$ & $\times$ & \checkmark & \checkmark\\ \cline{1-6}
 \multicolumn{6}{c}{\textbf{Factors} and \textbf{Metrics}} \\ \cline{1-6}
  Relevant Factors and Evaluation Factors  & $\times$ & $\times$ & $\times$ & $\times$ & \checkmark  \\ \cline{1-6}
 Model performance measures & $\times$ & $\times$ & $\times$ & \checkmark & \checkmark  \\ \cline{1-6} 
 \multicolumn{6}{c}{\textbf{Ethical Considerations, Caveats and Recommendations}} \\ \cline{1-6}
  Ethical considerations & $\times$ & $\times$ & $\times$ & $\times$ & \checkmark  \\ \cline{1-6}
  Caveats & $\times$ & $\times$ & $\times$ & $\times$ & \checkmark  \\ \cline{1-6} 
 Recomendations & Difference between probability and severity is explained. & $\times$ & $\times$ & $\times$ & Difference between probability and severity is explained ("score of 0.9 is not necessarily more toxic than a comment with a TOXICITY score of 0.7") \\ \midrule

\end{tabular}\caption{Transparency comparison of commercial content moderation API services with a focus on model card categories by \citet{mitchellModelCardsModel2019a}}\label{tab:api_comparison}
\end{table*}
\end{small}

\onecolumn
\subsection{Content Moderation API Configurations}
\label{tab:configurations}
\small
\renewcommand{\arraystretch}{1.2}
\begin{table*}[h]
\begin{tabular}{lllll} \toprule
\Description{
This table compares the hate speech sub-category configurations across five commercial content moderation APIs (OpenAI, Microsoft, Amazon, Google, and Jigsaw Perspective), highlighting their distinct approaches to categorizing harmful content.

**OpenAI:**  
- Covers categories such as harassment, hate, self-harm, sexual content, and violence.  
- Provides additional threat-specific subcategories, such as *harassment threatening*, *hate threatening*, and multiple types of self-harm content (*instructions*, *intent*).  

**Microsoft:**  
- Focuses on content types with categories like *sexually explicit*, *offensive*, *profanity*, *harassment*, *violence*, and *hate speech*.  
- Distinguishes between sexually explicit and sexually suggestive content.  

**Amazon:**  
- Emphasizes harmful behaviors with categories such as *hate speech*, *insult*, *harassment or abuse*, *profanity*, *violence or threat*, and *sexual content*.  
- Includes *graphic content* as a distinct category.  

**Google:**  
- Uses a wide range of categories, including *toxic*, *insult*, *health*, *violent*, *profanity*, *religion and belief*, *legal*, *public safety*, and *politics*.  
- Covers both social issues (*politics*, *religion*) and content types (*graphic*, *sexual*, *violent*).  

**Jigsaw Perspective API:**  
- Categorizes content by user experience impacts with categories like *threat*, *insult*, *toxicity*, *identity attack*, and *severe toxicity*.  
- Focuses on how content may be perceived in conversations rather than specific topics.  

**Comparison Notes:**  
- *OpenAI* and *Microsoft* emphasize content types and threats.  
- *Amazon* centers on behavioral harm and abuse.  
- *Google* has the broadest scope, addressing social and safety topics.  
- *Jigsaw Perspective* is uniquely user-centered, measuring conversational impact.  

This comparison highlights differing priorities among providers, from conversational safety (Jigsaw) to broad societal issues (Google) and harmful behaviors (Amazon and Microsoft).
}

\textbf{OpenAI} & \textbf{Microsoft} & \textbf{Amazon} & \textbf{Google} & \textbf{Perspective} \\ 
\midrule
Harassment & Sexually explicit or adult & Profanity & Death, Harm \& Tragedy & Threat \\ 
\midrule
Harassment threatening & Sexually suggestive or mature & Hate speech & Toxic & Insult \\ 
\midrule
Hate & Offensive & Insult & Insult & Toxicity \\ 
\midrule
Hate threatening & Profanity & Graphic & Health & Identity attack \\ 
\midrule
Self harm & & Harassment or abuse & Violent & Severe toxicity \\ 
\midrule
Self harm instructions & & Sexual & Illicit drugs & Profanity \\ 
\midrule
Self harm intent & & Violence or threat & Finance & \\ 
\midrule
Sexual & & & Profanity & \\ 
\midrule
Sexual minors & & & Firearms \& weapons & \\ 
\midrule
Violence & & & Religion \& belief & \\ 
\midrule
Violence graphic & & & Legal & \\ 
\midrule
Self-harm & & & Public safety & \\ 
\midrule
Sexual/minors & & & Politics & \\ 
\midrule
Hate threatening & & & Derogatory & \\ 
\midrule
Violence graphic & & & Sexual & \\ 
\midrule
Self-harm intent & & & War \& conflict & \\ 
\midrule\end{tabular}
\caption{Hate Speech Sub-category Configurations across APIs.}

\end{table*}

\subsection{Audit Sample Size}
\begin{table*}[h]
\Description{
This table presents descriptive statistics for three hate speech datasets (ToxiGen, Civil Comments, and HateXplain), summarizing sample sizes and average sentence lengths for various marginalized groups.

**Datasets Overview:**  
- **ToxiGen:** 7,846 samples, average length of 18 words.  
- **Civil Comments:** 50,000 samples, average length of 48 words.  
- **HateXplain:** 50,000 samples, average length of 19 words.  

**Group-Level Statistics:**  
- **Asian:** ToxiGen (654, 17 words), Civil Comments (284, 70 words), HateXplain (1,666, 20 words).  
- **Black:** ToxiGen (360, 22 words), Civil Comments (4,230, 74 words), HateXplain (10,126, 19 words).  
- **Disability:** ToxiGen (882, 17 words), Civil Comments (1,048, 58 words), HateXplain (898, 19 words).  
- **Female:** ToxiGen (202, 16 words), Civil Comments (6,170, 72 words), HateXplain (16,722, 19 words).  
- **Jewish:** ToxiGen (400, 20 words), Civil Comments (714, 72 words), HateXplain (898, 20 words).  
- **Latinx:** ToxiGen (470, 17 words), Civil Comments (130, 63 words), HateXplain (1,260, 20 words).  
- **LGBTQIA+:** ToxiGen (480, 21 words), Civil Comments (2,674, 70 words), HateXplain (484, 19 words).  
- **Muslim:** ToxiGen (820, 18 words), Civil Comments (3,978, 67 words), HateXplain (2,730, 19 words).  

**Key Observations:**  
- **Sample Size Variance:** Civil Comments consistently provides the largest samples across all groups, while ToxiGen and HateXplain offer smaller but balanced samples.  
- **Sentence Length Differences:** Civil Comments contains significantly longer comments on average (48 words) than ToxiGen (18) and HateXplain (19).  
- **Group Coverage:** All three datasets represent multiple marginalized groups, enabling cross-dataset comparisons.  

**Dataset Balance:** According to the caption, each dataset is balanced for toxic and non-toxic speech at both the aggregate and group levels, supporting fair evaluations across different demographic groups.
}

    \centering
    \small
    \begin{tabular}{l||cc|cc|cc}
    \toprule
    Group & \multicolumn{2}{c|}{\textbf{ToxiGen}} & \multicolumn{2}{c|}{\textbf{Civil Comments}} & \multicolumn{2}{c}{\textbf{HateXplain}} \\
    \midrule
    & N & Avg. Words & N & Avg. Words & N & Avg. Words \\
    \midrule
    Aggregate & 7,846 & 18 & 50,000 & 48 & 50,000 & 19 \\
    Asian & 654 & 17 & 284 & 70 & 1,666 & 20 \\
    Black & 360 & 22 & 4,230 & 74 & 10,126 & 19 \\
    Disability & 882 & 17 & 1,048 & 58 & 898 & 19  \\
    Female & 202 & 16 & 6,170 & 72 & 16,722 & 19 \\
    Jewish & 400 & 20 & 714 & 72 & 898 & 20 \\
    Latinx & 470 & 17 & 130 & 63 & 1,260 & 20 \\
    LGBTQIA+ & 480 & 21 & 2,674 & 70 & 484 & 19 \\
    Muslim & 820 & 18 & 3,978 & 67 & 2,730 & 19 \\
    \bottomrule
    \end{tabular}
    \caption{Descriptive statistics across ToxiGen, Civil Comments and HateXplain: Sample size and average number of words in a sentence, at the aggregate level and by marginalized group. All datasets are balanced on toxic and non-toxic speech, both at the aggregate and the group-level.}
    \label{tab:datasets_descriptive_statistics}
\end{table*}

\begin{table*}[h]
    \centering
    \Description{
This table presents descriptive statistics for synthetic and non-synthetic datasets used in Perturbation Sensitivity Analysis, detailing sample sizes, toxicity shares, and average sentence lengths for various identity groups.

**Columns:**  
- **Group:** Identity group represented.  
- **Synthetic (Dixon et al., 2018):** Sample size (N), share of toxic statements, average words per sentence.  
- **Non-Synthetic (HateXplain):** Sample size (N), share of toxic statements, average words per sentence.  

**Dataset Overview:**  
- **Synthetic (Dixon et al., 2018):** 25,738 samples, 50\% toxic, 5 words on average.  
- **Non-Synthetic (HateXplain):** 9,190 samples, 85\% toxic, 20 words on average.  

**Group-Level Statistics:**  
- **Asian:** Synthetic (1,514, 50\%, 5 words), Non-Synthetic (122, 63\%, 19 words).  
- **Black:** Synthetic (1,514, 50\%, 5 words), Non-Synthetic (1,053, 71\%, 20 words).  
- **Disability:** No synthetic samples, Non-Synthetic (31, 45\%, 22 words).  
- **Female:** Synthetic (1,514, 50\%, 5 words), Non-Synthetic (2,294, 71\%, 21 words).  
- **Jewish:** Synthetic (1,514, 50\%, 5 words), Non-Synthetic (953, 89\%, 20 words).  
- **Latinx:** Synthetic (4,542, 50\%, 5 words), Non-Synthetic (45, 76\%, 18 words).  
- **LGBTQIA+:** Synthetic (13,626, 50\%, 5 words), Non-Synthetic (552, 76\%, 19.7 words).  
- **Muslim:** Synthetic (1,514, 50\%, 5 words), Non-Synthetic (4,140, 97\%, 20 words).  

**Key Observations:**  
- **Sample Size Differences:** The synthetic dataset has more samples per group due to template replication, while the non-synthetic dataset has fewer but more diverse samples.  
- **Toxicity Rates:** The synthetic dataset is balanced (50\% toxic), whereas the non-synthetic dataset has higher toxicity rates, especially for Muslim (97\%) and Jewish (89\%) references.  
- **Sentence Length:** Synthetic samples are short (5 words), while non-synthetic samples are longer (average 20 words).  

**Dataset Design:**  
The synthetic dataset uses Identity Phrase Templates from Dixon et al. (2018), generating 1,514 unique sentences repeated across identity tokens. The non-synthetic dataset, from HateXplain, reflects real-world biases with longer and more toxic sentences.  
}
    \begin{tabular}{l||ccc|ccc}
    \toprule
    Group & \multicolumn{3}{c|}{\textbf{Synthetic (\citet{dixon_measuring_2018})}} & \multicolumn{3}{c}{\textbf{Non-Synthetic (HateXplain)}} \\
    \midrule
    & N & Share Toxic & Avg. Words & N & Share Toxic & Avg. Words \\
    \midrule
    Aggregate & 25,738 & 0.5 & 5 & 9,190 & 0.85 & 20    \\
    Asian  & 1,514 & 0.5 & 5 & 122 & 0.63 & 19 \\ 
    Black  & 1,514 & 0.5 & 5 & 1,053 & 0.71 & 20 \\ 
    Disability & 0 & 0 & 0 & 31 & 0.45 & 22 \\
    Female  & 1,514 & 0.5 & 5 & 2,294 & 0.71 & 21 \\
    Jewish & 1,514 & 0.5 & 5 & 953 & 0.89 & 20 \\
    Latinx  & 4,542 & 0.5 & 5 & 45 & 0.76 & 18 \\
    LGBTQIA+  & 13,626 & 0.5 & 5 & 552 & 0.76 & 19.7 \\
    Muslim  & 1,514 & 0.5 & 5 & 4,140 & 0.97 & 20 \\
    \bottomrule
    \end{tabular}
    \caption{Descriptive statistics of datasets for Perturbation Sensitivity Analysis, reporting sample size, share of toxic statements and average words per sentence. The \textit{Identity Phrase Templates} from \citet{dixon_measuring_2018} were synthetically created for the purpose of Perturbation Sensitivity Analysis, while the non-synthetic dataset is derived from HateXplain. The Identity Phrase Templates contains 1,514 unique sentences which are repeated for all their tokens for marginalized identities. Among the groups defined for the purpose of this project, some map onto several tokens from \citet{dixon_measuring_2018}, resulting in sample sizes in excess of 1,514. While the synthetic dataset was designed to be balanced in terms of toxicity and contains very short sentences, the real dataset is skewed towards toxic speech and is made up of comparable longer phrases.}
    \label{tab:psa_datasets_descriptive_statistics}
\end{table*}
\pagebreak

\subsection{Pertubation Sensitivity Analysis with Counterfactual Fairness Tokens and Performance Overview}
\begin{table*}[h]
\Description{
This table presents results from a Perturbation Sensitivity Analysis (PSA) that measures group-specific biases in toxicity detection by comparing sentences referencing marginalized and dominant identity groups.

**Columns:**  
- **Marginalized Reference:** Sentence containing a marginalized identity term.  
- **Toxicity:** Mean toxicity score for the marginalized reference.  
- **Dominant Reference:** Corresponding sentence with the dominant identity term.  
- **Toxicity:** Mean toxicity score for the dominant reference.  
- **PSA:** Difference in toxicity scores (marginalized minus dominant).  

**Group-Level PSA Scores:**  
- **Asian vs. White:** Marginalized 0.21, Dominant 0.33, PSA +0.12.  
- **Black vs. White:** Marginalized 0.31, Dominant 0.33, PSA +0.02.  
- **Female vs. Male:** Marginalized 0.16, Dominant 0.18, PSA +0.02.  
- **LGBTQIA+ vs. Straight:** Marginalized 0.31, Dominant 0.25, PSA -0.06.  
- **Jewish vs. Christian:** Marginalized 0.40, Dominant 0.36, PSA -0.04.  
- **Muslim vs. Christian:** Marginalized 0.42, Dominant 0.36, PSA -0.06.  
- **Latinx vs. White:** Marginalized 0.24, Dominant 0.33, PSA +0.09.  

**Key Observations:**  
- **Positive PSA (e.g., Asian, Black, Female, Latinx):** Higher toxicity for dominant references, indicating lower bias.  
- **Negative PSA (e.g., LGBTQIA+, Jewish, Muslim):** Higher toxicity for marginalized references, indicating potential bias.  
- **Largest Negative PSA:** Observed for LGBTQIA+ (-0.06) and Muslim (-0.06) references.  

**PSA Insight:**  
A negative PSA suggests that sentences referencing marginalized groups are rated as more toxic than their dominant-group counterparts, highlighting potential model biases.  
}

    \centering
    \renewcommand{\arraystretch}{1.3}
    \begin{tabular}{p{4.5cm} p{1.75cm} p{4.5cm} p{1.75cm} p{0.8cm}}
    \toprule
    \textbf{Marginalized Reference} & \textbf{Toxicity} & \textbf{Dominant Reference} & \textbf{Toxicity} & \textbf{PSA}  \\ 
    \midrule
    Being \textbf{Asian} is fun & 0.21 & Being \textbf{White} is fun & 0.33 & \textbf{0.12} \\ \midrule
    Being \textbf{Black} is fun & 0.31 & Being \textbf{White} is fun & 0.33 & \textbf{0.02}\\ \midrule
    Being \textbf{Female} is fun & 0.16 & Being \textbf{Male} is fun & 0.18 & \textbf{0.02} \\ \midrule
    Being \textbf{LGBTQIA+} is fun & 0.31 & Being \textbf{Straight} is fun  & 0.25 & \textbf{-0.06} \\ \midrule
    Being \textbf{Jewish} is fun & 0.40 & Being \textbf{Christian} is fun & 0.36 & \textbf{-0.04} \\ \midrule
    Being \textbf{Muslim} is fun & 0.42 & Being \textbf{Christian} is fun & 0.36 & \textbf{-0.06} \\ \midrule
    Being \textbf{Latinx} is fun & 0.24  & Being \textbf{White} is fun & 0.33 & \textbf{0.09} \\
    \bottomrule
    \end{tabular}
\caption{In Perturbation Sensitivity Analysis, valid counterfactuals are constructed by solely exchanging marginalized and corresponding dominant tokens in a sentence. Counterfactual Token Fairness scores are computed by taking the difference in toxicity score, revealing any groups-specific biases. In this illustrative example, we report the mean toxicity score across all moderation services. Negative Counterfactual Token Fairness scores indicate potentially negative biases associated with a marginalized's common identity tokens.}
\label{tab:psa_methodology_illustration}
\end{table*}

\begin{table*}[h]
    \centering
    \renewcommand{\arraystretch}{1.2}
\Description{
This table lists 34 marginalized identity tokens and their corresponding dominant tokens across eight identity groups, used for constructing counterfactuals in Perturbation Sensitivity Analysis.

**Columns:**  
- **Group:** Identity group category.  
- **Marginalized:** Marginalized identity token.  
- **Dominant:** Corresponding dominant identity token.  

**Group-Level Tokens:**  
- **Asian:** Marginalized: asian; Dominant: white.  
- **Black:** Marginalized: black; Dominant: white.  
- **Disability:**  
  - Marginalized: disabled, disability, disabilities, handicapped  
  - Dominant: abled, ability, abilities  
- **Female:**  
  - Marginalized: woman, women, female  
  - Dominant: man, men, male  
- **Jewish:**  
  - Marginalized: jewish, jew, jews, judaism  
  - Dominant: christian, christians, christianity  
- **Latinx:**  
  - Marginalized: latino, latinos, latina, latinas, latinx  
  - Dominant: white man, white men, white woman, white women, white person  
- **LGBTQIA+:**  
  - Marginalized: lesbian, lesbians, gay, gays, bi, bisexual, trans, transexual, queer, lgbt, LGBTQIA+, homosexual  
  - Dominant: straight, straight women, straight men, heterosexual  
- **Muslim:**  
  - Marginalized: muslim, muslims, islam, arabic  
  - Dominant: christian, christians, christianity  

**Key Observations:**  
- The table identifies substantively and semantically aligned dominant tokens for each marginalized identity.  
- Multiple variations of identity tokens (e.g., plural forms, religious identities) are included to cover a broad linguistic range.  
- This set of token pairs is used to construct counterfactual sentences for bias detection.  
}
    \begin{tabular}{p{2.1cm}p{2.1cm}p{2.1cm}||p{2.1cm}p{2.1cm}p{2.1cm}}
    \toprule
    \textbf{Group} & \textbf{Marginalized} & \textbf{Dominant} & \textbf{Group} & \textbf{Marginalized} & \textbf{Dominant} \\
    \midrule
    \textbf{asian} & asian & white & \textbf{black} & black & white \\
    \midrule
    \textbf{disability} & disabled & abled & \textbf{female} & woman & man \\
    disability & disability & ability & female & women & men \\
    disability & disabilities & abilities & female & female & male  \\
    disability & handicapped & abled & & & \\
    \midrule
    \textbf{jewish} & jewish & christian & \textbf{latinx} & latino & white man \\
    jewish & jew & christian & latinx & latinos & white men \\
    jewish & jews & christians & latinx & latina & white woman \\
    jewish & judaism & christianity & latinx & latinas & white women \\
    & & & latinx & latinx & white person \\
    \midrule
    \textbf{LGBTQIA+} & lesbian & straight & \textbf{muslim} & muslim & christian \\
    LGBTQIA+ & lesbians & straight women & muslim & muslims & christians \\
    LGBTQIA+ & gay & straight & muslim & islam & christianity \\
    LGBTQIA+ & gays & straight men & muslim & arabic & christian \\
    LGBTQIA+ & bi & straight & & & \\
    LGBTQIA+ & bisexual & straight & & & \\
    LGBTQIA+ & trans & straight & & & \\
    LGBTQIA+ & transexual & straight & & & \\
    LGBTQIA+ & queer & straight & & & \\
    LGBTQIA+ & lgbt & straight & & & \\
    LGBTQIA+ & LGBTQIA+ & straight & & & \\
    LGBTQIA+ & homosexual & heterosexual & & & \\
    \bottomrule
    \end{tabular}
    \caption{34 marginalized token across eight marginalized groups are employed in Perturbation Sensitivity Analysis. Per marginalized token we identified the substantively and semantically appropriate dominant token to construct valid counterfactuals.}
    \label{tab:minority_majority_token}
\end{table*}

\begin{table*}[h]
\centering
\Description{
This table presents synthetic examples used in Perturbation Sensitivity Analysis, constructed from Identity Phrase Templates (Dixon et al., 2018). Each sentence pair replaces a marginalized identity token with its dominant counterpart to create a counterfactual.

**Columns:**  
- **Group:** Identity category.  
- **Marginalized Reference:** Sentence with a marginalized identity.  
- **Dominant Reference:** Corresponding counterfactual with a dominant identity.  

**Examples:**  
- **LGBTQIA+:** "Being gay is great" → "Being straight is great"  
- **Latinx:** "Aki is a filthy latino" → "Aki is a filthy white"  
- **Black:** "Dishad is a dumb black" → "Dishad is a dumb white"  
- **Muslim:** "You are a shitty muslim" → "You are a shitty christian"  
- **Female:** "Destroy all women" → "Destroy all men"  

**Key Observations:**  
- **Concise and Synthetic:** Sentences are short and template-based, focusing on a single identity token.  
- **Direct Counterfactuals:** Only the identity token changes, ensuring that differences in toxicity are attributable to group identity.  
}

\renewcommand{\arraystretch}{1.3}
\begin{tabular}{@{} p{1.4 cm} p{6.5cm} p{6.5cm} @{}}
\toprule
\textbf{Group} & \textbf{Marginalized Reference} & \textbf{Dominant Reference} \\ 
\midrule
LGBTQIA+ & Being \textbf{gay} is great & Being \textbf{straight} is great \\ \midrule
Latinx & Aki is a filthy \textbf{latino} & Aki is a filthy \textbf{white} \\ \midrule
Black & Dishad is a dumb \textbf{black} & Dishad is a dumb \textbf{white} \\ \midrule
Muslim & you are a shitty \textbf{muslim} & you are a shitty \textbf{christian} \\ \midrule
Female & destroy all \textbf{women} & destroy all \textbf{men} \\ 
\bottomrule
\end{tabular}
\caption{Synthetic data examples for Perturbation Sensitivity Analysis, based on the \textit{Identity Phrase Templates} from \citet{dixon_measuring_2018}. Semantically, these are very stripped down and short. Per sentence we create a valid counterfactual by solely exchanging the marginalized token for the appropriate dominant token.}
\label{tab:psa_examples_synthetic}
\end{table*}

\begin{table*}[h]
\centering
\Description{
This table presents non-synthetic examples for Perturbation Sensitivity Analysis, sourced from HateXplain. Each sentence pair substitutes a marginalized identity token with a dominant token to create counterfactuals, providing longer and more context-rich examples compared to the synthetic set.

**Columns:**  
- **Group:** Identity category.  
- **Marginalized Reference:** Sentence with a marginalized identity.  
- **Dominant Reference:** Corresponding counterfactual with a dominant identity.  

**Examples:**  
- **Muslim:** "This town has large amounts of muslim immigration..." → "This town has large amounts of christian immigration..."  
- **Asian:** "From owning a mud asian lady..." → "From owning a mud white lady..."  
- **Female:** "Following a presentation by three women..." → "Following a presentation by three men..."  
- **LGBTQIA+:** "You're a transsexual, lol get over yourself" → "You're a straight, lol get over yourself"  
- **Jewish:** "The same way jews run the government..." → "The same way christians run the government..."  

**Key Observations:**  
- **Longer and Contextual:** Non-synthetic examples are longer and include more social context than the synthetic examples.  
- **Real-World Bias:** Derived from HateXplain, these sentences reflect real-world biases found in online discourse.  
- **Focused Counterfactuals:** Identity tokens are replaced while maintaining the sentence structure for controlled comparison.  
}

\renewcommand{\arraystretch}{1.3}
\begin{tabular}{@{} p{1.4 cm} p{6.5cm} p{6.5cm} @{}}
\toprule
\textbf{Group} & \textbf{Marginalized Reference} & \textbf{Dominant Reference} \\
\midrule
Muslim & oh well actually this town has large amounts of \textbf{muslim} immigration the \textbf{muslim} world is ripping the christians of af & oh well actually this town has large amounts of \textbf{chrisitan} immigration the \textbf{christian} world is ripping the christians of af \\ \midrule
Asian & so we went from owning a mud \textbf{asian} lady to her having to go back to where she came from & so we went from owning a mud \textbf{white} lady to her having to go back to where she came from \\ \midrule
Female & Immediately following a presentation by three \textbf{women} okay sure they made up the next show & Immediately following a presentation by three \textbf{men} okay sure they made up the next show \\ \midrule
LGBTQIA+ & youre a \textbf{transsexual} lol get over yourself  & youre a \textbf{straight} lol get over yourself  \\ \midrule
Jewish & the same way \textbf{jews} run the government they want to control the masses and impose their cultural values & the same way \textbf{christians} run the government they want to control the masses and impose their cultural values \\ 
\bottomrule
\end{tabular}
\caption{Non-synthetic data examples for Perturbation Sensitivity Analysis, based on HateXplain. While still relatively short, this are on average longer than the synthetic examples and involve more contextual information. Per sentence we create a valid counterfactual by solely exchanging the marginalized token for the appropriate dominant token.}
\label{tab:psa_examples_hatexplain}
\end{table*}

\onecolumn
\subsection{Qualitative SHAP Analysis Codes of Model Failures}
\begin{table*}[h]
    \centering
    \renewcommand{\arraystretch}{1.3}
    \begin{tabular}{p{4cm} @{\hspace{20pt}}p{7.5cm} p{1.5cm}} \toprule
    \Description{
        This table categorizes different types of false positives in hate speech detection systems, detailing the operational definitions and corresponding references for each code. Categories include:
        
        1. **Counter Speech (CS):** Comments that denounce hate speech by quoting or referencing it.
        2. **Re-appropriation (RE):** Instances where slurs or discriminatory terms are reclaimed or used in non-hateful contexts.
        3. **Over-moderating Slurs (OS):** Non-hateful uses of slurs or profanity that are flagged as hate speech.
        4. **Long Sentences (LS):** Statements flagged due to their length and complexity.
        5. **Dialects (DA):** Misclassification of dialects like African American English as hate speech.
        6. **Systematic Offensive Stereotyping Bias (SOS):** Neutral or informational mentions of marginalized groups misclassified as hate speech.
        7. **Negation (NE):** Hate expressed using negated positive statements.
        8. **Non-protected Targets (NP):** Abuse directed at non-protected groups or individuals.
        9. **Descriptive Comments (DE):** Comments about hateful topics or conspiracies without endorsing them.
        
        The table also cites references for each category to support its definitions.
        }

        {\textbf{Code}} & {\textbf{Operationalization}}  &{\textbf{Source}} \\ \midrule 
        \textbf{CS}: Counter speech	& The comment appears to criticize or oppose a hateful statement made by someone else. This includes (1) Denouncements of hate that quote it and (2) denouncements of hate that make direct reference to it   & \cite{diasolivaFightingHateSpeech2021,rottgerHateCheckFunctionalTests2021, vidgen-etal-2021-learning}\\  \hline
        \textbf{RE:} Re-appropriation, reclaimed slurs & A discriminatory word or phrase is reclaimed and used by a member of the marginalized group in a non-hateful context. This includes (1) Non-hateful homonyms of slurs and (2) Reclaimed slurs & \cite{diasolivaFightingHateSpeech2021,rottgerHateCheckFunctionalTests2021,vidgen-etal-2021-learning}\\  \hline
        \textbf{OS:} Over-moderating slurs and profanity & A slur or profanity is used in a non-hateful or exaggerated context but is flagged as hate speech. & \cite{rottgerHateCheckFunctionalTests2021, vidgen-etal-2021-learning}\\  \hline
        \textbf{LS:} Long Sentences & Statements which are long, often involving multiple clauses and sentences.& \cite{vidgen-etal-2021-learning}\\ \hline
        \textbf{DA:} Dialects & The comment uses a dialect (e.g., African American English) that is misclassified as hate speech.& \cite{sap_risk_2019, blodgett-etal-2020-language, zhou-etal-2021-challenges}\\  \hline
        \textbf{SOS:} Systematic Offensive Stereotyping Bias & A marginalized group is mentioned in a neutral or informational context but is misinterpreted as hate speech.& \cite{elsafoury_bias_2023, rottgerHateCheckFunctionalTests2021, zhou-etal-2021-challenges}\\ \hline
        \textbf{NE:} Negation & Hate expressed using negated
positive statement & \cite{rottgerHateCheckFunctionalTests2021, vidgen-etal-2021-learning} \\ \hline
        \textbf{NP:} Abuse against
non-protected target & The comment targets a group or individual that is not a protected target.& \cite{rottgerHateCheckFunctionalTests2021}\\  \hline
        \textbf{DE:} Descriptive comment or Non-hate
group identity& The comment describes hateful topics (e.g., political situations or events) without endorsing them. It may contain misinformation or conspiracy theories but does not qualify as hate speech against a protected group.& \cite{rottgerHateCheckFunctionalTests2021}\\ \hline \hline
       
    \end{tabular}
    \caption{Categories for coding false positive sentences}
    \label{tab:codingFP}
\end{table*} 

\begin{table*}[h]
    \centering
    \renewcommand{\arraystretch}{1.3}
    \begin{tabular}{p{4cm}@{\hspace{20pt}}p{7cm} p{2cm}} \toprule
        \Description{
            This table categorizes different types of false negatives in hate speech detection systems, detailing the operational definitions and corresponding references for each code. Categories include:
            
            1. **Implicit Hate Speech (IMP):** Comments containing implicit hate, such as jokes or sarcasm, that are offensive but not overtly hateful.
            2. **Positive Sentiment (POS):** Statements that use positive phrases or terms to express hate.
            3. **Negation (NE-FN):** Non-hateful comments using negated hateful statements that are misclassified.
            4. **Paraphrased Target (PAR):** Novel descriptions or paraphrasing of marginalized or protected groups that evade detection.
            5. **Misread Descriptions (DE-FN):** Hateful comments misclassified as neutral or descriptive statements.
            6. **Misread Counter Speech (CS-FN):** Hateful comments flagged as non-hate and misclassified as counter speech.
            7. **Spelling Variations (SPELL):** Variants such as swapped, missing, or added characters, and leet speak spellings that bypass detection.
            
            The table also cites references for each category to substantiate its definitions.
            }
        \textbf{Code} & {\textbf{Operationalization}}  & {\textbf{Source} } \\ \midrule	
        \textbf{IMP:} Implicit hate speech & The comment contains implicit hate speech, which is not overtly expressed but still offensive. This includes (1) jokes and sarcasm, (2) & \cite{wiegand_implicitly_2021, wiegand_detection_2019, sap_social_2020, waseem2017understanding}\\ \hline
        \textbf{POS:} Terms with positive sentiment & Statements
which contain positive phrases and terms but still
express hate.&\cite{vidgen-etal-2021-learning}\\ \hline	
        \textbf{NE-FN:} Negation & Non-hate expressed using
negated hateful statement& \cite{rottgerHateCheckFunctionalTests2021,vidgen-etal-2021-learning}\\ \hline
        \textbf{PAR:} Paraphrased target & A novel way of describing or paraphrasing of a marginalized group or protected group is used, which the system fails to detect as hateful.& \cite{ludwig-etal-2022-improving}\\ \hline
        \textbf{DE-FN:} Misread as description or political statement & A hateful comment is misclassified as a neutral or descriptive statement. & \cite{rottgerHateCheckFunctionalTests2021} \\ \hline 
        \textbf{CS-FN:} Misread as Counter speech & A comment that is flagged as non-hate and is detected as counter speech but is hateful in its content.& \cite{diasolivaFightingHateSpeech2021}\\ \hline
        \textbf{SPELL:} Spelling Variations & Swaps of adjacent characters, missing characters, missing word boundaries, added spaces between chars and leet speak spellings & \cite{rottgerHateCheckFunctionalTests2021}\\ \hline \hline
    \end{tabular}
    \caption{Categories for coding false negative sentences}
    \label{tab:codingFN}
\end{table*} 

\newpage
\noindent

\vspace{1cm}
\noindent
\begin{figure*}[h]
    \centering
    \begin{subfigure}[b]{0.45\textwidth}
        \Description{This figure (Fig. a) is a SHAP value visualization showing how Perspective API analyzes a sentence targeting Latinos with implicit bias expressed through negation.

Key Points:

Red-colored words: The term "Hispanics" is highlighted in red, signaling a push toward hate speech classification.
Blue-colored words: Words like "nothing against" and "working for their own things" are highlighted in blue, pushing the model away from classifying the sentence as hate speech.

Interpretation:

The model struggles to detect the implicit bias in this sentence. While the explicit use of the term "Hispanics" is flagged, the overall structure and negation ("I have nothing against...") mitigate the classification. This demonstrates the model's difficulty in capturing hate speech that is masked by superficially positive or neutral framing, despite the underlying discriminatory sentiment implied by phrases like "in their own country."}
	 \colorbox[RGB]{80.77780861902642, 134.56991379195375, 250}{\rule[-.5mm]{0pt}{2.5mm}\textcolor{white}{i}}
	 \colorbox[RGB]{107.58715210178747, 153.10171804271022, 250}{\rule[-.5mm]{0pt}{2.5mm}\textcolor{white}{have}}
	 \colorbox[RGB]{38.0, 105.0, 250}{\rule[-.5mm]{0pt}{2.5mm}\textcolor{white}{nothing}}
	 \colorbox[RGB]{176.8183690846198, 200.95739798476023, 250}{\rule[-.5mm]{0pt}{2.5mm}\textcolor{black}{against}}
	\colorbox[RGB]{255,229,235}{\textcolor{black}{hispanics}}
	 \colorbox[RGB]{108.77792146838823, 153.92483050810245, 250}{\rule[-.5mm]{0pt}{2.5mm}\textcolor{white}{,}}
	 \colorbox[RGB]{109.84725818990034, 154.66400335707397, 250}{\rule[-.5mm]{0pt}{2.5mm}\textcolor{white}{i}}
	 \colorbox[RGB]{111.58049225456861, 155.86209142020874, 250}{\rule[-.5mm]{0pt}{2.5mm}\textcolor{white}{just}}
	 \colorbox[RGB]{85.55634733252893, 137.8730511515177, 250}{\rule[-.5mm]{0pt}{2.5mm}\textcolor{white}{think}}
	 \colorbox[RGB]{102.18736144519804, 149.36914385612766, 250}{\rule[-.5mm]{0pt}{2.5mm}\textcolor{white}{that}}
	 \colorbox[RGB]{138.9895564494011, 174.8084491585722, 250}{\rule[-.5mm]{0pt}{2.5mm}\textcolor{black}{they}}
	 \colorbox[RGB]{197.41037848841577, 215.1915058675685, 250}{\rule[-.5mm]{0pt}{2.5mm}\textcolor{black}{should}}
	 \colorbox[RGB]{139.86373854099446, 175.41272249377496, 250}{\rule[-.5mm]{0pt}{2.5mm}\textcolor{black}{be}}
	 \colorbox[RGB]{164.15211809340616, 192.20192494935907, 250}{\rule[-.5mm]{0pt}{2.5mm}\textcolor{black}{in}}
	 \colorbox[RGB]{168.7519106434267, 195.38150505305993, 250}{\rule[-.5mm]{0pt}{2.5mm}\textcolor{black}{their}}
	 \colorbox[RGB]{146.58356000269305, 180.05776037052513, 250}{\rule[-.5mm]{0pt}{2.5mm}\textcolor{black}{own}}
	 \colorbox[RGB]{246.70091091875304, 249.2633024784007, 250}{\rule[-.5mm]{0pt}{2.5mm}\textcolor{black}{country}}
	 \colorbox[RGB]{105.64505316100508, 151.7592533371003, 250}{\rule[-.5mm]{0pt}{2.5mm}\textcolor{white}{,}}
	 \colorbox[RGB]{56.54278507293421, 117.81759336838769, 250}{\rule[-.5mm]{0pt}{2.5mm}\textcolor{white}{working}}
	 \colorbox[RGB]{100.09509348861266, 147.9228756833728, 250}{\rule[-.5mm]{0pt}{2.5mm}\textcolor{white}{for}}
	 \colorbox[RGB]{99.4371592911811, 147.46808245934176, 250}{\rule[-.5mm]{0pt}{2.5mm}\textcolor{white}{their}}
	 \colorbox[RGB]{94.53826111084521, 144.08174731164416, 250}{\rule[-.5mm]{0pt}{2.5mm}\textcolor{white}{own}}
	 \colorbox[RGB]{82.98113806875676, 136.09295258208994, 250}{\rule[-.5mm]{0pt}{2.5mm}\textcolor{white}{things}}
	 \colorbox[RGB]{99.96921332889355, 147.8358617480831, 250}{\rule[-.5mm]{0pt}{2.5mm}\textcolor{white}{,}}
	 \colorbox[RGB]{99.16181796263231, 147.27775435204998, 250}{\rule[-.5mm]{0pt}{2.5mm}\textcolor{white}{instead}}
	 \colorbox[RGB]{98.87911976450212, 147.08234085103834, 250}{\rule[-.5mm]{0pt}{2.5mm}\textcolor{white}{of}}
	 \colorbox[RGB]{92.85828596729772, 142.92047417094312, 250}{\rule[-.5mm]{0pt}{2.5mm}\textcolor{white}{trying}}
	 \colorbox[RGB]{94.48659618864858, 144.04603423178474, 250}{\rule[-.5mm]{0pt}{2.5mm}\textcolor{white}{to}}
        \caption{\textbf{FN, Negation-FN Latinos:} Perspective API on ToxiGen}
        \label{fig:FN-NE}
    \end{subfigure}
    \hfill
    \begin{subfigure}[b]{0.45\textwidth}
        \Description{This figure (Fig. b) is a SHAP value visualization analyzing antisemitic bias in a sentence that generalizes negative traits about Jewish people.

Key Points:

Red-colored words: Words like "Jews," "control," "greed," and "jealousy" are highlighted in red, indicating their contribution to hate speech classification.
Blue-colored words: Words like "in most countries" and "everything" are highlighted in blue, mitigating the classification.

Interpretation:

Despite the explicit antisemitic stereotypes, the model fails to fully classify the sentence as hate speech due to the mitigating weight placed on neutral structural words. The SHAP analysis exposes the model's inability to consistently handle hate speech that is expressed through sweeping generalizations and coded language, failing to adequately prioritize the harmful context of the statement.}
	 \colorbox[RGB]{88.1011400516222, 139.6321244596467, 250}{\rule[-.5mm]{0pt}{2.5mm}\textcolor{white}{All}}
	 \colorbox[RGB]{97.87945833315055, 146.39133064503494, 250}{\rule[-.5mm]{0pt}{2.5mm}\textcolor{white}{the}}
	 \colorbox[RGB]{255, 13.0, 87.0}{\rule[-.5mm]{0pt}{2.5mm}\textcolor{white}{jews}}
	 \colorbox[RGB]{52.61264707923297, 115.10090811928546, 250}{\rule[-.5mm]{0pt}{2.5mm}\textcolor{white}{in}}
	 \colorbox[RGB]{63.69374179681793, 122.7606510116253, 250}{\rule[-.5mm]{0pt}{2.5mm}\textcolor{white}{most}}
	 \colorbox[RGB]{66.68713246796014, 124.82981507001853, 250}{\rule[-.5mm]{0pt}{2.5mm}\textcolor{white}{countries}}
	 \colorbox[RGB]{73.5173056958426, 129.55113296947644, 250}{\rule[-.5mm]{0pt}{2.5mm}\textcolor{white}{are}}
	 \colorbox[RGB]{38.0, 105.0, 250}{\rule[-.5mm]{0pt}{2.5mm}\textcolor{white}{rich}}
	 \colorbox[RGB]{48.305458560837764, 112.12358886693855, 250}{\rule[-.5mm]{0pt}{2.5mm}\textcolor{white}{and}}
	 \colorbox[RGB]{99.1868824867916, 147.2950800599942, 250}{\rule[-.5mm]{0pt}{2.5mm}\textcolor{white}{control}}
	 \colorbox[RGB]{97.86147357390988, 146.37889878380867, 250}{\rule[-.5mm]{0pt}{2.5mm}\textcolor{white}{everything}}
	 \colorbox[RGB]{88.14490172397706, 139.6623744635786, 250}{\rule[-.5mm]{0pt}{2.5mm}\textcolor{white}{.}}
	 \colorbox[RGB]{68.83473309193076, 126.31433163036687, 250}{\rule[-.5mm]{0pt}{2.5mm}\textcolor{white}{It}}
	 \colorbox[RGB]{67.70347273277312, 125.5323544235759, 250}{\rule[-.5mm]{0pt}{2.5mm}\textcolor{white}{'}}
	 \colorbox[RGB]{73.96299126909796, 129.859210554676, 250}{\rule[-.5mm]{0pt}{2.5mm}\textcolor{white}{s}}
	 \colorbox[RGB]{88.86953084982613, 140.1632701726909, 250}{\rule[-.5mm]{0pt}{2.5mm}\textcolor{white}{all}}
	 \colorbox[RGB]{81.88156476293224, 135.33287886838633, 250}{\rule[-.5mm]{0pt}{2.5mm}\textcolor{white}{down}}
	 \colorbox[RGB]{88.54284700828262, 139.93745184904327, 250}{\rule[-.5mm]{0pt}{2.5mm}\textcolor{white}{to}}
	 \colorbox[RGB]{150.91985170420566, 183.0551970305569, 250}{\rule[-.5mm]{0pt}{2.5mm}\textcolor{black}{greed}}
	 \colorbox[RGB]{86.40812503180825, 138.46183757959096, 250}{\rule[-.5mm]{0pt}{2.5mm}\textcolor{white}{and}}
	 \colorbox[RGB]{121.73512608682483, 162.88142356232132, 250}{\rule[-.5mm]{0pt}{2.5mm}\textcolor{white}{jealousy}}
	 \colorbox[RGB]{89.72594455419508, 140.75526121257724, 250}{\rule[-.5mm]{0pt}{2.5mm}\textcolor{white}{.}}
	 \colorbox[RGB]{101.04675033205896, 148.5807029945108, 250}{\rule[-.5mm]{0pt}{2.5mm}\textcolor{white}{They}}
	 \colorbox[RGB]{96.62308138654353, 145.52286731788723, 250}{\rule[-.5mm]{0pt}{2.5mm}\textcolor{white}{like}}
	 \colorbox[RGB]{77.90799049516178, 132.5861685450427, 250}{\rule[-.5mm]{0pt}{2.5mm}\textcolor{white}{to}}
	 \colorbox[RGB]{67.01584382016264, 125.05703489872994, 250}{\rule[-.5mm]{0pt}{2.5mm}\textcolor{white}{control}}
	 \colorbox[RGB]{73.52140382405247, 129.5539657769948, 250}{\rule[-.5mm]{0pt}{2.5mm}\textcolor{white}{and}}
	 \colorbox[RGB]{88.90331566488527, 140.18662373148751, 250}{\rule[-.5mm]{0pt}{2.5mm}\textcolor{white}{run}}
	 \colorbox[RGB]{91.7160369271746, 142.13090110173357, 250}{\rule[-.5mm]{0pt}{2.5mm}\textcolor{white}{countries}}
	 \colorbox[RGB]{65.68180548854494, 124.1348885865518, 250}{\rule[-.5mm]{0pt}{2.5mm}\textcolor{white}{because}}
        \caption{\textbf{FN, SOS Bias} Perspective on ToxiGen}
        \label{fig:TG7622}
    \end{subfigure}
    
    \begin{subfigure}[b]{0.45\textwidth}
        \Description{This figure (Fig. c) shows a SHAP value visualization, which demonstrates how individual words in the sentence contribute to the model's decision-making process for evaluating content for bias or toxicity.

Key Points:

Words highlighted in red significantly increase the likelihood of the content being flagged as toxic. In this sentence, the word "competent" is shaded bright red, indicating that it is the most influential in driving the model’s prediction of toxicity.
Words shaded in blue reduce the likelihood of the content being flagged as toxic. For example, words like "would," "probably," and "have" are highlighted in blue, suggesting they decrease the overall toxicity assessment.

Interpretation:

The word "competent" is identified as a key driver of the model’s prediction due to its association with biased or harmful context, particularly in relation to disability. The blue-colored words act as moderating influences, lowering the perceived toxicity of the sentence. This visualization highlights how the model evaluates specific terms in context, particularly those tied to sensitive topics, to assess their contribution to potential harmfulness.}
	 \colorbox[RGB]{255, 228.94466418617552, 236.91199827800614}{\rule[-.5mm]{0pt}{2.5mm}\textcolor{black}{you}}
	 \colorbox[RGB]{38.0, 105.0, 250}{\rule[-.5mm]{0pt}{2.5mm}\textcolor{white}{would}}
	 \colorbox[RGB]{73.32107788551957, 129.4154916259352, 250}{\rule[-.5mm]{0pt}{2.5mm}\textcolor{white}{probably}}
	 \colorbox[RGB]{139.29699357352735, 175.02096329967327, 250}{\rule[-.5mm]{0pt}{2.5mm}\textcolor{black}{have}}
	 \colorbox[RGB]{208.86909178843644, 223.11227542979478, 250}{\rule[-.5mm]{0pt}{2.5mm}\textcolor{black}{to}}
	 \colorbox[RGB]{187.3567930499745, 208.24202284560448, 250}{\rule[-.5mm]{0pt}{2.5mm}\textcolor{black}{pay}}
	 \colorbox[RGB]{253.18074443126713, 253.74245006769618, 250}{\rule[-.5mm]{0pt}{2.5mm}\textcolor{black}{disabled}}
	 \colorbox[RGB]{255, 217.00387468978874, 228.62252457803518}{\rule[-.5mm]{0pt}{2.5mm}\textcolor{black}{people}}
	 \colorbox[RGB]{184.39605257269108, 206.19542804563898, 250}{\rule[-.5mm]{0pt}{2.5mm}\textcolor{black}{less}}
	 \colorbox[RGB]{110.47525321225189, 155.09810129879162, 250}{\rule[-.5mm]{0pt}{2.5mm}\textcolor{white}{than}}
	 \colorbox[RGB]{137.55428227238139, 173.8163241514157, 250}{\rule[-.5mm]{0pt}{2.5mm}\textcolor{black}{everyone}}
	 \colorbox[RGB]{86.3680755255044, 138.43415358905833, 250}{\rule[-.5mm]{0pt}{2.5mm}\textcolor{white}{else}}
	 \colorbox[RGB]{105.49772495145822, 151.6574135609158, 250}{\rule[-.5mm]{0pt}{2.5mm}\textcolor{white}{because}}
	 \colorbox[RGB]{152.23971740541273, 183.9675465936033, 250}{\rule[-.5mm]{0pt}{2.5mm}\textcolor{black}{they}}
	 \colorbox[RGB]{98.8110864721254, 147.03531322957977, 250}{\rule[-.5mm]{0pt}{2.5mm}\textcolor{white}{'}}
	 \colorbox[RGB]{186.71460953317552, 207.798117188831, 250}{\rule[-.5mm]{0pt}{2.5mm}\textcolor{black}{re}}
	 \colorbox[RGB]{255, 248.26884292777558, 250.3271306275467}{\rule[-.5mm]{0pt}{2.5mm}\textcolor{black}{not}}
	 \colorbox[RGB]{152.90925210196076, 184.4303585958254, 250}{\rule[-.5mm]{0pt}{2.5mm}\textcolor{black}{as}}
	 \colorbox[RGB]{255, 13.0, 87.0}{\rule[-.5mm]{0pt}{2.5mm}\textcolor{white}{competent}}
        \caption{\textbf{FN, Positive Terms:} Perspective on ToxiGen}
        \label{fig:FN-PT}
    \end{subfigure}
    \hfill
    \vspace{1em}
    \begin{subfigure}[b]{0.45\textwidth}
        \Description{This figure (Fig. d) is a SHAP value visualization analyzing a sentence that expresses bias against people with disabilities, suggesting eugenic-like ideas.

Key Points:

Red-colored words: The phrase "not be allowed" is highlighted in red, signaling a push toward hate speech classification.
Blue-colored words: Words like "disability" and "someone has" are highlighted in blue, mitigating the classification.

Interpretation:

The model fails to fully classify this sentence as hate speech, despite the explicit discriminatory sentiment expressed in the phrase "should not be allowed to have children." By giving undue weight to neutral or descriptive terms like "disability," the model overlooks the severity of the underlying bias. This highlights its difficulty in detecting hate speech that is framed as a statement of policy or opinion.}
	 \colorbox[RGB]{196.38472663088248, 214.48252992918142, 250}{\rule[-.5mm]{0pt}{2.5mm}\textcolor{black}{if}}
	 \colorbox[RGB]{202.28456914418805, 218.560762081236, 250}{\rule[-.5mm]{0pt}{2.5mm}\textcolor{black}{someone}}
	 \colorbox[RGB]{207.3654446939179, 222.07288803727045, 250}{\rule[-.5mm]{0pt}{2.5mm}\textcolor{black}{has}}
	 \colorbox[RGB]{155.96067221194556, 186.5396351695476, 250}{\rule[-.5mm]{0pt}{2.5mm}\textcolor{black}{a}}
	 \colorbox[RGB]{255, 179.28472307960462, 202.43732841889909}{\rule[-.5mm]{0pt}{2.5mm}\textcolor{black}{disability}}
	 \colorbox[RGB]{106.54628595950462, 152.38222531762992, 250}{\rule[-.5mm]{0pt}{2.5mm}\textcolor{white}{,}}
	 \colorbox[RGB]{38.0, 105.0, 250}{\rule[-.5mm]{0pt}{2.5mm}\textcolor{white}{then}}
	 \colorbox[RGB]{116.57234798137856, 159.3126829364368, 250}{\rule[-.5mm]{0pt}{2.5mm}\textcolor{white}{they}}
	 \colorbox[RGB]{168.70194706442567, 195.3469680168841, 250}{\rule[-.5mm]{0pt}{2.5mm}\textcolor{black}{should}}
	 \colorbox[RGB]{255, 13.0, 87.0}{\rule[-.5mm]{0pt}{2.5mm}\textcolor{white}{not}}
	 \colorbox[RGB]{187.09748157488636, 208.0627752821795, 250}{\rule[-.5mm]{0pt}{2.5mm}\textcolor{black}{be}}
	 \colorbox[RGB]{152.76728622821213, 184.33222550337246, 250}{\rule[-.5mm]{0pt}{2.5mm}\textcolor{black}{allowed}}
	 \colorbox[RGB]{84.97593345484395, 137.47184340196586, 250}{\rule[-.5mm]{0pt}{2.5mm}\textcolor{white}{to}}
	 \colorbox[RGB]{64.83619106395454, 123.55036248660451, 250}{\rule[-.5mm]{0pt}{2.5mm}\textcolor{white}{have}}
	 \colorbox[RGB]{255, 137.5435374696917, 173.45997642524054}{\rule[-.5mm]{0pt}{2.5mm}\textcolor{black}{children}}\\
        \caption{\textbf{FN, SOS Bias} Perspective on ToxiGen}
        \label{fig:TG316}
    \end{subfigure}
   
    \begin{subfigure}[b]{0.45\textwidth}
        \Description{This figure (Fig. e) is a SHAP value visualization showing how Perspectives content moderation model interprets a sentence containing implicit hate speech expressed through negation.

Key Points:

Red-colored words: The word "racist" and the phrase "get married" are highlighted in red, pushing the model toward classifying the sentence as hate speech. The word "black" also contributes to this classification.
Blue-colored words: Words like "not," "I," and "think" are highlighted in blue, indicating they push the model away from classifying the sentence as hate speech, as they provide mitigating context.

Interpretation:

Although the speaker denies being racist ("I am not a racist"), the model is influenced by implicit bias present in the latter part of the sentence, particularly targeting interracial marriage. The SHAP analysis highlights the challenge of detecting nuanced forms of implicit hate speech when denial or negation is used.}
	 \colorbox[RGB]{178.4191079629622, 202.06389951356834, 250}{\rule[-.5mm]{0pt}{2.5mm}\textcolor{black}{I}}
	 \colorbox[RGB]{187.37556623048476, 208.25499969849176, 250}{\rule[-.5mm]{0pt}{2.5mm}\textcolor{black}{am}}
	 \colorbox[RGB]{38.0, 105.0, 250}{\rule[-.5mm]{0pt}{2.5mm}\textcolor{white}{not}}
	 \colorbox[RGB]{179.880794512703, 203.07428192122327, 250}{\rule[-.5mm]{0pt}{2.5mm}\textcolor{black}{a}}
	 \colorbox[RGB]{255, 31.34875320938579, 99.73797743461492}{\rule[-.5mm]{0pt}{2.5mm}\textcolor{white}{racist}}
	 \colorbox[RGB]{168.9351537387996, 195.50817078718867, 250}{\rule[-.5mm]{0pt}{2.5mm}\textcolor{black}{,}}
	 \colorbox[RGB]{151.04336138180156, 183.1405723837338, 250}{\rule[-.5mm]{0pt}{2.5mm}\textcolor{black}{I}}
	 \colorbox[RGB]{165.5902915481551, 193.19605406554498, 250}{\rule[-.5mm]{0pt}{2.5mm}\textcolor{black}{have}}
	 \colorbox[RGB]{170.56269577817602, 196.63319984666546, 250}{\rule[-.5mm]{0pt}{2.5mm}\textcolor{black}{a}}
	 \colorbox[RGB]{174.17303309332164, 199.12882471888594, 250}{\rule[-.5mm]{0pt}{2.5mm}\textcolor{black}{lot}}
	 \colorbox[RGB]{162.5773884637229, 191.11340216386378, 250}{\rule[-.5mm]{0pt}{2.5mm}\textcolor{black}{of}}
	 \colorbox[RGB]{119.72228470183381, 161.49005854965472, 250}{\rule[-.5mm]{0pt}{2.5mm}\textcolor{white}{friends}}
	 \colorbox[RGB]{240.81534609422417, 245.1949396964683, 250}{\rule[-.5mm]{0pt}{2.5mm}\textcolor{black}{who}}
	 \colorbox[RGB]{255, 201.8046692527942, 218.07101005979104}{\rule[-.5mm]{0pt}{2.5mm}\textcolor{black}{are}}
	 \colorbox[RGB]{255, 13.0, 87.0}{\rule[-.5mm]{0pt}{2.5mm}\textcolor{white}{black}}
	 \colorbox[RGB]{191.8590487545281, 211.35418116672446, 250}{\rule[-.5mm]{0pt}{2.5mm}\textcolor{black}{,}}
	 \colorbox[RGB]{188.23238803560207, 208.84727283566963, 250}{\rule[-.5mm]{0pt}{2.5mm}\textcolor{black}{I}}
	 \colorbox[RGB]{191.06519733861967, 210.8054359483546, 250}{\rule[-.5mm]{0pt}{2.5mm}\textcolor{black}{just}}
	 \colorbox[RGB]{212.48487194184878, 225.61166263261438, 250}{\rule[-.5mm]{0pt}{2.5mm}\textcolor{black}{don}}
	 \colorbox[RGB]{225.45295810165743, 234.57577748962495, 250}{\rule[-.5mm]{0pt}{2.5mm}\textcolor{black}{'}}
	 \colorbox[RGB]{239.38028576408465, 244.20296250973593, 250}{\rule[-.5mm]{0pt}{2.5mm}\textcolor{black}{t}}
	 \colorbox[RGB]{219.66182021164028, 230.57268678223983, 250}{\rule[-.5mm]{0pt}{2.5mm}\textcolor{black}{think}}
	 \colorbox[RGB]{168.54362234775837, 195.23752696849658, 250}{\rule[-.5mm]{0pt}{2.5mm}\textcolor{black}{they}}
	 \colorbox[RGB]{232.10354729964064, 239.17295896288525, 250}{\rule[-.5mm]{0pt}{2.5mm}\textcolor{black}{should}}
	 \colorbox[RGB]{217.5447664417316, 229.10928555880065, 250}{\rule[-.5mm]{0pt}{2.5mm}\textcolor{black}{be}}
	 \colorbox[RGB]{217.0907021226408, 228.79541621380702, 250}{\rule[-.5mm]{0pt}{2.5mm}\textcolor{black}{able}}
	 \colorbox[RGB]{152.32902488655958, 184.02927987550203, 250}{\rule[-.5mm]{0pt}{2.5mm}\textcolor{black}{to}}
	 \colorbox[RGB]{255, 231.84938824743256, 238.92850093210194}{\rule[-.5mm]{0pt}{2.5mm}\textcolor{black}{get}}
	 \colorbox[RGB]{255, 213.85491191079927, 226.43646777278627}{\rule[-.5mm]{0pt}{2.5mm}\textcolor{black}{married}}
        \caption{\textbf{FN, Implicit hate speech and Negation} Perspective on ToxiGen}
        \label{fig:IMP-NE}
    \end{subfigure}
    \hfill
    \vspace{1em}
    \begin{subfigure}[b]{0.45\textwidth}
        \Description{This figure (Fig. f) is a SHAP value visualization showing how Perspective interprets a sentence with a claim connecting skin tone and intelligence.

Key Points:

Red-colored words: The word "intelligence" is highlighted in red, pushing the model toward classifying the sentence as hateful.
Blue-colored words: Words like "melanin" and "correlation" are highlighted in blue, indicating a weaker contribution toward hate classification.
Interpretation:
The sentence implies a racial bias through a pseudoscientific connection between melanin and intelligence. The SHAP analysis reveals the difficulty in detecting bias embedded in complex or factual-sounding phrasing.}
	 \colorbox[RGB]{196.11412010513084, 214.29547472704894, 250}{\rule[-.5mm]{0pt}{2.5mm}\textcolor{black}{there}}
	 \colorbox[RGB]{151.77682615297772, 183.64757568178186, 250}{\rule[-.5mm]{0pt}{2.5mm}\textcolor{black}{is}}
	 \colorbox[RGB]{179.38711232068033, 202.73302694977903, 250}{\rule[-.5mm]{0pt}{2.5mm}\textcolor{black}{a}}
	 \colorbox[RGB]{124.5070818311645, 164.7975220031091, 250}{\rule[-.5mm]{0pt}{2.5mm}\textcolor{white}{direct}}
	 \colorbox[RGB]{38.0, 105.0, 250}{\rule[-.5mm]{0pt}{2.5mm}\textcolor{white}{correlation}}
	 \colorbox[RGB]{173.74600507064577, 198.83364405805008, 250}{\rule[-.5mm]{0pt}{2.5mm}\textcolor{black}{between}}
	 \colorbox[RGB]{166.92567270329062, 194.11912859674467, 250}{\rule[-.5mm]{0pt}{2.5mm}\textcolor{black}{the}}
	 \colorbox[RGB]{204.55883627111825, 220.1328361321094, 250}{\rule[-.5mm]{0pt}{2.5mm}\textcolor{black}{amount}}
	 \colorbox[RGB]{202.679590501487, 218.83381831900024, 250}{\rule[-.5mm]{0pt}{2.5mm}\textcolor{black}{of}}
	 \colorbox[RGB]{196.08200148322572, 214.27327291467216, 250}{\rule[-.5mm]{0pt}{2.5mm}\textcolor{black}{mel}}
	 \colorbox[RGB]{199.76748608837465, 216.82084291823134, 250}{\rule[-.5mm]{0pt}{2.5mm}\textcolor{black}{ani}}
	 \colorbox[RGB]{141.11953542248045, 176.28078485424916, 250}{\rule[-.5mm]{0pt}{2.5mm}\textcolor{black}{n}}
	 \colorbox[RGB]{136.39549557442677, 173.01531952149315, 250}{\rule[-.5mm]{0pt}{2.5mm}\textcolor{black}{in}}
	 \colorbox[RGB]{117.42023036441796, 159.8987767495977, 250}{\rule[-.5mm]{0pt}{2.5mm}\textcolor{white}{a}}
	 \colorbox[RGB]{155.1685776988328, 185.99210440011484, 250}{\rule[-.5mm]{0pt}{2.5mm}\textcolor{black}{person}}
	 \colorbox[RGB]{156.28979165966535, 186.7671370919346, 250}{\rule[-.5mm]{0pt}{2.5mm}\textcolor{black}{'}}
	 \colorbox[RGB]{158.2191065257241, 188.1007648795328, 250}{\rule[-.5mm]{0pt}{2.5mm}\textcolor{black}{s}}
	 \colorbox[RGB]{170.85539628752076, 196.83552738768714, 250}{\rule[-.5mm]{0pt}{2.5mm}\textcolor{black}{skin}}
	 \colorbox[RGB]{112.25668856000306, 156.32950822120029, 250}{\rule[-.5mm]{0pt}{2.5mm}\textcolor{white}{and}}
	 \colorbox[RGB]{255, 213.8743066495194, 226.44993188892258}{\rule[-.5mm]{0pt}{2.5mm}\textcolor{black}{their}}
	 \colorbox[RGB]{255, 13.0, 87.0}{\rule[-.5mm]{0pt}{2.5mm}\textcolor{white}{intelligence}}
	 \colorbox[RGB]{117.78969482270915, 160.15416692813997, 250}{\rule[-.5mm]{0pt}{2.5mm}\textcolor{white}{.}}
        \caption{\textbf{FN: SOS Bias against black people.} Perspective for sentence from ToxiGen}
        \label{fig:TG6084}
    \end{subfigure}
    \vspace{1em}
    \begin{subfigure}[b]{0.45\textwidth}
        \Description{This figure (Fig. g) is a SHAP value visualization of Microsoft’s content moderation model analyzing a sentence implicitly targeting a religious identity.

Key Points:

Red-colored words: The word "bacon" and the phrase "toss bacon at it" are highlighted in red, indicating a push toward classifying the sentence as hateful.
Blue-colored words: Words like "next time" and "mos" (shortened from "moslim") are highlighted in blue, pushing the model away from classification.

Interpretation:

The sentence uses paraphrasing and coded language to target a religious group subtly. The SHAP analysis highlights the model’s difficulty in recognizing implicit bias when it is paraphrased or disguised.}
	 \colorbox[RGB]{38.0, 105.0, 250}{\rule[-.5mm]{0pt}{2.5mm}\textcolor{white}{next}}
	 \colorbox[RGB]{63.31954435433869, 122.50198918502676, 250}{\rule[-.5mm]{0pt}{2.5mm}\textcolor{white}{time}}
	 \colorbox[RGB]{255, 13.0, 87.0}{\rule[-.5mm]{0pt}{2.5mm}\textcolor{white}{you}}
	 \colorbox[RGB]{255, 36.06194048818787, 103.00994215708909}{\rule[-.5mm]{0pt}{2.5mm}\textcolor{white}{see}}
	 \colorbox[RGB]{167.8205733867802, 194.73772353924898, 250}{\rule[-.5mm]{0pt}{2.5mm}\textcolor{black}{a}}
	 \colorbox[RGB]{71.34892330521916, 128.05225113263998, 250}{\rule[-.5mm]{0pt}{2.5mm}\textcolor{white}{mo}}
	 \colorbox[RGB]{255, 248.15735480525035, 250.24973391438866}{\rule[-.5mm]{0pt}{2.5mm}\textcolor{black}{os}}
	 \colorbox[RGB]{164.82143510433716, 192.6645864776524, 250}{\rule[-.5mm]{0pt}{2.5mm}\textcolor{black}{lim}}
	 \colorbox[RGB]{238.2331971310793, 243.4100440998244, 250}{\rule[-.5mm]{0pt}{2.5mm}\textcolor{black}{toss}}
	 \colorbox[RGB]{247.9776789256003, 250.14586100847947, 250}{\rule[-.5mm]{0pt}{2.5mm}\textcolor{black}{bacon}}
	 \colorbox[RGB]{93.25746670932551, 143.1964055594416, 250}{\rule[-.5mm]{0pt}{2.5mm}\textcolor{white}{at}}
	 \colorbox[RGB]{84.04680465017816, 136.8295884678651, 250}{\rule[-.5mm]{0pt}{2.5mm}\textcolor{white}{it}}
        \caption{\textbf{FN, Implicit Bias and Paraphrasing Target Identity} Microsoft on  HateXplain.}
        \label{fig:FN-IMPPAR2}
    \end{subfigure}
    \hfill
        \vspace{1em}
    \begin{subfigure}[b]{0.45\textwidth}
        \Description{This figure (Fig. h) is a SHAP value visualization showing how Amazon Comprehend interprets a sentence with antisemitic undertones.

Key Points:

Red-colored words: The word "Jews" and the phrase "fighting over a penny" are highlighted in red, pushing the model toward classifying the sentence as hateful.
Blue-colored words: Words like "history" and "how" are highlighted in blue, pushing the model away from classification.

Interpretation:

The sentence conveys a stereotype through a historical reference. The SHAP analysis demonstrates the model's sensitivity to explicit bias terms but struggles to contextualize historical or stereotypical implications fully.}
	 \colorbox[RGB]{183.47404744155705, 205.55809730983208, 250}{\rule[-.5mm]{0pt}{2.5mm}\textcolor{black}{the}}
	 \colorbox[RGB]{38.0, 105.0, 250}{\rule[-.5mm]{0pt}{2.5mm}\textcolor{white}{history}}
	 \colorbox[RGB]{170.75360883104588, 196.76516739473215, 250}{\rule[-.5mm]{0pt}{2.5mm}\textcolor{black}{of}}
	 \colorbox[RGB]{183.2167419803019, 205.38023639191377, 250}{\rule[-.5mm]{0pt}{2.5mm}\textcolor{black}{how}}
	 \colorbox[RGB]{135.84789401360135, 172.63679309695945, 250}{\rule[-.5mm]{0pt}{2.5mm}\textcolor{black}{copper}}
	 \colorbox[RGB]{227.34114017547276, 235.88097247152496, 250}{\rule[-.5mm]{0pt}{2.5mm}\textcolor{black}{wire}}
	 \colorbox[RGB]{209.50790436390145, 223.553850942789, 250}{\rule[-.5mm]{0pt}{2.5mm}\textcolor{black}{was}}
	 \colorbox[RGB]{120.62140160358408, 162.11156792874476, 250}{\rule[-.5mm]{0pt}{2.5mm}\textcolor{white}{invented}}
	 \colorbox[RGB]{184.41381761637183, 206.2077080297501, 250}{\rule[-.5mm]{0pt}{2.5mm}\textcolor{black}{in}}
	 \colorbox[RGB]{215.1241379054902, 227.43604002683654, 250}{\rule[-.5mm]{0pt}{2.5mm}\textcolor{black}{six}}
	 \colorbox[RGB]{228.59416874232738, 236.74712125045673, 250}{\rule[-.5mm]{0pt}{2.5mm}\textcolor{black}{words}}
	 \colorbox[RGB]{255, 235.93257424794422, 241.76310939526707}{\rule[-.5mm]{0pt}{2.5mm}\textcolor{black}{two}}
	 \colorbox[RGB]{255, 13.0, 87.0}{\rule[-.5mm]{0pt}{2.5mm}\textcolor{white}{jews}}
	 \colorbox[RGB]{207.33188249911885, 222.0496883634462, 250}{\rule[-.5mm]{0pt}{2.5mm}\textcolor{black}{fighting}}
	 \colorbox[RGB]{224.1358547881298, 233.6653374111496, 250}{\rule[-.5mm]{0pt}{2.5mm}\textcolor{black}{over}}
	 \colorbox[RGB]{178.92064284601344, 202.4105826124517, 250}{\rule[-.5mm]{0pt}{2.5mm}\textcolor{black}{a}}
	 \colorbox[RGB]{255, 185.93244892531052, 207.05227859277755}{\rule[-.5mm]{0pt}{2.5mm}\textcolor{black}{penny}}
        \caption{\textbf{FN, Implicit} Amazon Comprehend on HateXplain}
        \label{fig:FN-IMP3}
    \end{subfigure}
    \vspace{1em}
    \begin{subfigure}[b]{0.45\textwidth}
        \Description{This figure (Fig. i) is a SHAP value visualization of Microsoft Moderators analyzing implicit bias in a sentence comparing historical figures.

Key Points:

Red-colored words: The phrase "Obama has watermelon" is highlighted in red, which pushes the model somewhat toward a hate speech classification.
Blue-colored words: Words like "George Washington" and "God on his side" are highlighted in blue, which significantly mitigate the classification.

Interpretation:

The model struggles to detect the hate speech in this sentence, as it overweights the neutral context provided by references to George Washington and "God on his side" while underweighting the racialized stereotype embedded in "Obama has watermelon." This highlights a significant limitation in the model's ability to capture implicit bias conveyed through stereotypes and paraphrasing, especially when surrounded by seemingly neutral content.}
	 \colorbox[RGB]{255, 13.0, 87.0}{\rule[-.5mm]{0pt}{2.5mm}\textcolor{white}{A}}
	 \colorbox[RGB]{154.8571565304693, 185.7768363113843, 250}{\rule[-.5mm]{0pt}{2.5mm}\textcolor{black}{picture}}
	 \colorbox[RGB]{251.56732250690985, 252.62718145638928, 250}{\rule[-.5mm]{0pt}{2.5mm}\textcolor{black}{of}}
	 \colorbox[RGB]{43.94729119545778, 109.11103078027034, 250}{\rule[-.5mm]{0pt}{2.5mm}\textcolor{white}{George}}
	 \colorbox[RGB]{188.61809446427912, 209.11389018268142, 250}{\rule[-.5mm]{0pt}{2.5mm}\textcolor{black}{Washington}}
	 \colorbox[RGB]{101.70063738066628, 149.03269865023015, 250}{\rule[-.5mm]{0pt}{2.5mm}\textcolor{white}{crossing}}
	 \colorbox[RGB]{101.70063738066628, 149.03269865023015, 250}{\rule[-.5mm]{0pt}{2.5mm}\textcolor{white}{the}}
	 \colorbox[RGB]{121.94869036483178, 163.0290486392846, 250}{\rule[-.5mm]{0pt}{2.5mm}\textcolor{white}{Delaware}}
	 \colorbox[RGB]{214.61797829962302, 227.08616011494678, 250}{\rule[-.5mm]{0pt}{2.5mm}\textcolor{black}{,}}
	 \colorbox[RGB]{97.69251309909043, 146.26210582886435, 250}{\rule[-.5mm]{0pt}{2.5mm}\textcolor{white}{text}}
	 \colorbox[RGB]{38.0, 105.0, 250}{\rule[-.5mm]{0pt}{2.5mm}\textcolor{white}{states}}
	 \colorbox[RGB]{175.49834681995608, 200.0449401981263, 250}{\rule[-.5mm]{0pt}{2.5mm}\textcolor{black}{that}}
	 \colorbox[RGB]{175.49834681995608, 200.0449401981263, 250}{\rule[-.5mm]{0pt}{2.5mm}\textcolor{black}{while}}
	 \colorbox[RGB]{255, 244.89061961616167, 247.98191775006265}{\rule[-.5mm]{0pt}{2.5mm}\textcolor{black}{Washington}}
	 \colorbox[RGB]{255, 221.24769490535195, 231.56864770288897}{\rule[-.5mm]{0pt}{2.5mm}\textcolor{black}{had}}
	 \colorbox[RGB]{121.47443066498693, 162.7012193536776, 250}{\rule[-.5mm]{0pt}{2.5mm}\textcolor{white}{God}}
	 \colorbox[RGB]{116.8039981952347, 159.47280981237424, 250}{\rule[-.5mm]{0pt}{2.5mm}\textcolor{white}{on}}
	 \colorbox[RGB]{61.77333000915482, 121.43317742568306, 250}{\rule[-.5mm]{0pt}{2.5mm}\textcolor{white}{his}}
	 \colorbox[RGB]{127.58306427207076, 166.92377714659267, 250}{\rule[-.5mm]{0pt}{2.5mm}\textcolor{black}{side}}
	 \colorbox[RGB]{155.63056617397524, 186.3114512723331, 250}{\rule[-.5mm]{0pt}{2.5mm}\textcolor{black}{,}}
	 \colorbox[RGB]{255, 56.39777441026786, 117.1273805823347}{\rule[-.5mm]{0pt}{2.5mm}\textcolor{white}{Obama}}
	 \colorbox[RGB]{220.3301514046999, 231.03466686960823, 250}{\rule[-.5mm]{0pt}{2.5mm}\textcolor{black}{has}}
	 \colorbox[RGB]{170.05604668350833, 196.28298157846197, 250}{\rule[-.5mm]{0pt}{2.5mm}\textcolor{black}{water}}
	 \colorbox[RGB]{240.94714860372656, 245.28604742193082, 250}{\rule[-.5mm]{0pt}{2.5mm}\textcolor{black}{mel}}
	 \colorbox[RGB]{255, 232.15220476192297, 239.13872066116966}{\rule[-.5mm]{0pt}{2.5mm}\textcolor{black}{on}}
	 \colorbox[RGB]{255, 22.43213178326647, 93.54792619664781}{\rule[-.5mm]{0pt}{2.5mm}\textcolor{white}{.}}
        \caption{\textbf{FN, Implicit and Paraphrasing Identity Token} Microsoft Moderators on ToxiGen.}
        \label{fig:FN-IMP2}
    \end{subfigure}
        \hfill
        \vspace{1em}
    \begin{subfigure}[b]{0.45\textwidth}
            \Description{This figure (Fig. j) is a SHAP value visualization of GPT's content moderation model analyzing a sentence containing implicit hate against a religious identity.
    
    Key Points:
    
    Red-colored words: Terms like "muzrat," "soldiers," and "enemy lines" are highlighted in red, strongly pushing the model toward classification as hate speech.
    Blue-colored words: Words like "houses of" and "peace" are highlighted in blue, pushing the model away from classification.
    
    Interpretation:
    
    The sentence uses derogatory language and generalizations to promote hostility against a religious group. The SHAP analysis illustrates the challenges in detecting hate speech when coded language and indirect phrasing are used.}
	 \colorbox[RGB]{38.0, 105.0, 250}{\rule[-.5mm]{0pt}{2.5mm}\textcolor{white}{if}}
	 \colorbox[RGB]{255, 132.9771163401821, 170.2898989469033}{\rule[-.5mm]{0pt}{2.5mm}\textcolor{black}{mu}}
	 \colorbox[RGB]{255, 238.040981799773, 243.22679728248704}{\rule[-.5mm]{0pt}{2.5mm}\textcolor{black}{z}}
	 \colorbox[RGB]{255, 238.040981799773, 243.22679728248704}{\rule[-.5mm]{0pt}{2.5mm}\textcolor{black}{rat}}
	 \colorbox[RGB]{255, 132.9771163401821, 170.2898989469033}{\rule[-.5mm]{0pt}{2.5mm}\textcolor{white}{s}}
	 \colorbox[RGB]{255, 132.9771163401821, 170.2898989469033}{\rule[-.5mm]{0pt}{2.5mm}\textcolor{black}{reckon}}
	 \colorbox[RGB]{255, 132.9771163401821, 170.2898989469033}{\rule[-.5mm]{0pt}{2.5mm}\textcolor{black}{mu}}
	 \colorbox[RGB]{255, 139.191863500498, 174.60426887637877}{\rule[-.5mm]{0pt}{2.5mm}\textcolor{black}{z}}
	 \colorbox[RGB]{255, 136.75297852079288, 172.9111586425339}{\rule[-.5mm]{0pt}{2.5mm}\textcolor{black}{rat}}
	 \colorbox[RGB]{255, 146.68609057945167, 179.80687279895818}{\rule[-.5mm]{0pt}{2.5mm}\textcolor{black}{countries}}
	 \colorbox[RGB]{255, 122.04718958357293, 162.70218119851344}{\rule[-.5mm]{0pt}{2.5mm}\textcolor{black}{are}}
	 \colorbox[RGB]{255, 127.39657061230982, 166.41580108623162}{\rule[-.5mm]{0pt}{2.5mm}\textcolor{black}{houses}}
	 \colorbox[RGB]{255, 128.7717004674019, 167.37043668811373}{\rule[-.5mm]{0pt}{2.5mm}\textcolor{black}{of}}
	 \colorbox[RGB]{154.8571565304693, 185.7768363113843, 250}{\rule[-.5mm]{0pt}{2.5mm}\textcolor{black}{peace}}
	 \colorbox[RGB]{154.8571565304693, 185.7768363113843, 250}{\rule[-.5mm]{0pt}{2.5mm}\textcolor{black}{and}}
	 \colorbox[RGB]{255, 103.78275830772111, 150.02274130453367}{\rule[-.5mm]{0pt}{2.5mm}\textcolor{white}{non}}
	 \colorbox[RGB]{255, 132.9771163401821, 170.2898989469033}{\rule[-.5mm]{0pt}{2.5mm}\textcolor{black}{mu}}
	 \colorbox[RGB]{255, 132.9771163401821, 170.2898989469033}{\rule[-.5mm]{0pt}{2.5mm}\textcolor{black}{z}}
	 \colorbox[RGB]{255, 132.9771163401821, 170.2898989469033}{\rule[-.5mm]{0pt}{2.5mm}\textcolor{black}{rat}}
	 \colorbox[RGB]{255, 132.9771163401821, 170.2898989469033}{\rule[-.5mm]{0pt}{2.5mm}\textcolor{white}{countries}}
	 \colorbox[RGB]{255, 100.77948210537849, 147.93782228803133}{\rule[-.5mm]{0pt}{2.5mm}\textcolor{white}{are}}
	 \colorbox[RGB]{255, 238.040981799773, 243.22679728248704}{\rule[-.5mm]{0pt}{2.5mm}\textcolor{black}{houses}}
	 \colorbox[RGB]{255, 131.1927668092622, 169.05117695849606}{\rule[-.5mm]{0pt}{2.5mm}\textcolor{black}{of}}
	 \colorbox[RGB]{255, 132.9771163401821, 170.2898989469033}{\rule[-.5mm]{0pt}{2.5mm}\textcolor{black}{war}}
	 \colorbox[RGB]{255, 145.75858457273534, 179.1629843314857}{\rule[-.5mm]{0pt}{2.5mm}\textcolor{black}{then}}
	 \colorbox[RGB]{255, 97.47029001804981, 145.64053191335688}{\rule[-.5mm]{0pt}{2.5mm}\textcolor{white}{every}}
	 \colorbox[RGB]{255, 111.31255737882233, 155.25003983323202}{\rule[-.5mm]{0pt}{2.5mm}\textcolor{black}{mu}}
	 \colorbox[RGB]{255, 162.04251749285484, 190.4675328049571}{\rule[-.5mm]{0pt}{2.5mm}\textcolor{black}{z}}
	 \colorbox[RGB]{255, 68.81745776969419, 125.74930952606869}{\rule[-.5mm]{0pt}{2.5mm}\textcolor{black}{rat}}
	 \colorbox[RGB]{255, 137.24421711968978, 173.25218378556977}{\rule[-.5mm]{0pt}{2.5mm}\textcolor{black}{in}}
	 \colorbox[RGB]{154.8571565304693, 185.7768363113843, 250}{\rule[-.5mm]{0pt}{2.5mm}\textcolor{black}{non}}
	 \colorbox[RGB]{255, 87.42067419123106, 138.66393910796205}{\rule[-.5mm]{0pt}{2.5mm}\textcolor{black}{mu}}
	 \colorbox[RGB]{255, 113.78616798150108, 156.9672571111247}{\rule[-.5mm]{0pt}{2.5mm}\textcolor{black}{z}}
	 \colorbox[RGB]{255, 238.040981799773, 243.22679728248704}{\rule[-.5mm]{0pt}{2.5mm}\textcolor{black}{rat}}
	 \colorbox[RGB]{255, 238.040981799773, 243.22679728248704}{\rule[-.5mm]{0pt}{2.5mm}\textcolor{black}{countries}}
	 \colorbox[RGB]{255, 238.040981799773, 243.22679728248704}{\rule[-.5mm]{0pt}{2.5mm}\textcolor{black}{should}}
	 \colorbox[RGB]{255, 238.040981799773, 243.22679728248704}{\rule[-.5mm]{0pt}{2.5mm}\textcolor{black}{be}}
	 \colorbox[RGB]{255, 238.040981799773, 243.22679728248704}{\rule[-.5mm]{0pt}{2.5mm}\textcolor{black}{considered}}
	 \colorbox[RGB]{255, 238.040981799773, 243.22679728248704}{\rule[-.5mm]{0pt}{2.5mm}\textcolor{black}{as}}
	 \colorbox[RGB]{255, 238.040981799773, 243.22679728248704}{\rule[-.5mm]{0pt}{2.5mm}\textcolor{black}{soldiers}}
	 \colorbox[RGB]{255, 149.06419365925876, 181.4577873337003}{\rule[-.5mm]{0pt}{2.5mm}\textcolor{black}{or}}
	 \colorbox[RGB]{255, 238.040981799773, 243.22679728248704}{\rule[-.5mm]{0pt}{2.5mm}\textcolor{black}{spies}}
	 \colorbox[RGB]{255, 238.040981799773, 243.22679728248704}{\rule[-.5mm]{0pt}{2.5mm}\textcolor{black}{behind}}
	 \colorbox[RGB]{255, 238.040981799773, 243.22679728248704}{\rule[-.5mm]{0pt}{2.5mm}\textcolor{black}{enemy}}
	 \colorbox[RGB]{255, 238.040981799773, 243.22679728248704}{\rule[-.5mm]{0pt}{2.5mm}\textcolor{black}{lines}}
        \caption{\textbf{FN, Paraphrased target and Implicit} GPT on HateXplain}
        \label{fig:FN-PAR}
    \end{subfigure}
    \vspace{1em}
    \begin{subfigure}[b]{0.45\textwidth}
        \Description{This figure (Fig. k) is a SHAP value visualization showing how Perspective analyzes hate speech in a sentence with antisemitic content.

Key Points:

Red-colored words: The phrase "Hitler get the Jews" and "gas chambers" are highlighted in red, contributing toward classification as hate speech.
Blue-colored words: Words like "how" and "into" are highlighted in blue, significantly mitigating the classification.

Interpretation:

The model fails to fully classify this sentence as hate speech, despite explicit antisemitic references like "Hitler" and "gas chambers." By placing undue weight on neutral or structural words like "how" and "into," the model struggles to account for the harmful context of the sentence. This demonstrates a critical shortcoming in handling sentences that explicitly include hate speech but are structured in a conversational or indirect tone, making it difficult for the model to respond appropriately.}
	 \colorbox[RGB]{38.0, 105.0, 250}{\rule[-.5mm]{0pt}{2.5mm}\textcolor{white}{How}}
	 \colorbox[RGB]{51.188609022986185, 114.11655001588906, 250}{\rule[-.5mm]{0pt}{2.5mm}\textcolor{white}{did}}
	 \colorbox[RGB]{236.46644706666135, 242.1887882949272, 250}{\rule[-.5mm]{0pt}{2.5mm}\textcolor{black}{hitler}}
	 \colorbox[RGB]{172.47638766660972, 197.95602834097446, 250}{\rule[-.5mm]{0pt}{2.5mm}\textcolor{black}{get}}
	 \colorbox[RGB]{90.41034674868234, 141.22835028710762, 250}{\rule[-.5mm]{0pt}{2.5mm}\textcolor{white}{the}}
	 \colorbox[RGB]{255, 13.0, 87.0}{\rule[-.5mm]{0pt}{2.5mm}\textcolor{white}{jews}}
	 \colorbox[RGB]{132.8922051601405, 170.5936902028621, 250}{\rule[-.5mm]{0pt}{2.5mm}\textcolor{black}{into}}
	 \colorbox[RGB]{104.72735226093673, 151.12489787622354, 250}{\rule[-.5mm]{0pt}{2.5mm}\textcolor{white}{the}}
	 \colorbox[RGB]{217.4628870416874, 229.0526868951756, 250}{\rule[-.5mm]{0pt}{2.5mm}\textcolor{black}{gas}}
	 \colorbox[RGB]{118.06407756017913, 160.34383241487035, 250}{\rule[-.5mm]{0pt}{2.5mm}\textcolor{white}{chambers}}
	 \colorbox[RGB]{70.49562304972181, 127.46241224635149, 250}{\rule[-.5mm]{0pt}{2.5mm}\textcolor{white}{?}}
	 \colorbox[RGB]{117.16513357077565, 159.722442560444, 250}{\rule[-.5mm]{0pt}{2.5mm}\textcolor{white}{He}}
	 \colorbox[RGB]{172.52651888792906, 197.99068125893714, 250}{\rule[-.5mm]{0pt}{2.5mm}\textcolor{black}{threw}}
	 \colorbox[RGB]{92.49962268363629, 142.6725502421449, 250}{\rule[-.5mm]{0pt}{2.5mm}\textcolor{white}{a}}
	 \colorbox[RGB]{100.64394125623119, 148.30226354117363, 250}{\rule[-.5mm]{0pt}{2.5mm}\textcolor{white}{penny}}
	 \colorbox[RGB]{111.91470343217023, 156.09311297154625, 250}{\rule[-.5mm]{0pt}{2.5mm}\textcolor{white}{in}}
	 \colorbox[RGB]{50.496406129973764, 113.63806875343809, 250}{\rule[-.5mm]{0pt}{2.5mm}\textcolor{white}{.}}
        \caption{\textbf{FN, Implicit} Perspective on ToxiGen.}
        \label{fig:IMP5}
    \end{subfigure}
    \hfill
    \begin{subfigure}[b]{0.45\textwidth}
    \Description{This figure (Fig. l) is a SHAP value visualization of HateXplain’s analysis of implicit bias in a sentence referencing racial dynamics.

Key Points:

Red-colored words: The word "coon" and the phrase "Philando Castile" are highlighted in red, pushing the model toward hate speech classification.
Blue-colored words: Words like "Minnesota" and "fuss over" are highlighted in blue, pushing the model away from classification.
Interpretation:
The sentence uses coded language to reference racial injustice subtly. The SHAP analysis underscores the difficulty of identifying implicit bias when context is subtle and coded terms are used.}
	 \colorbox[RGB]{108.32602784014975, 153.6124616406565, 250}{\rule[-.5mm]{0pt}{2.5mm}\textcolor{white}{there}}
	 \colorbox[RGB]{135.68434184489752, 172.52373860246374, 250}{\rule[-.5mm]{0pt}{2.5mm}\textcolor{black}{has}}
	 \colorbox[RGB]{77.02445869919998, 131.97543228055298, 250}{\rule[-.5mm]{0pt}{2.5mm}\textcolor{white}{not}}
	 \colorbox[RGB]{49.528547098077894, 112.9690417728649, 250}{\rule[-.5mm]{0pt}{2.5mm}\textcolor{white}{been}}
	 \colorbox[RGB]{171.0127797755145, 196.94431781717591, 250}{\rule[-.5mm]{0pt}{2.5mm}\textcolor{black}{this}}
	 \colorbox[RGB]{118.72262937401834, 160.79905256268546, 250}{\rule[-.5mm]{0pt}{2.5mm}\textcolor{white}{much}}
	 \colorbox[RGB]{255, 153.36394742041847, 184.44274035797645}{\rule[-.5mm]{0pt}{2.5mm}\textcolor{black}{fuss}}
	 \colorbox[RGB]{65.09849867987256, 123.73168111511927, 250}{\rule[-.5mm]{0pt}{2.5mm}\textcolor{white}{over}}
	 \colorbox[RGB]{163.5177431019721, 191.76341689076412, 250}{\rule[-.5mm]{0pt}{2.5mm}\textcolor{black}{a}}
	 \colorbox[RGB]{255, 209.09652106747242, 223.13312206336929}{\rule[-.5mm]{0pt}{2.5mm}\textcolor{black}{minnesota}}
	 \colorbox[RGB]{255, 30.112042308661955, 98.8794343299802}{\rule[-.5mm]{0pt}{2.5mm}\textcolor{white}{co}}
	 \colorbox[RGB]{255, 13.0, 87.0}{\rule[-.5mm]{0pt}{2.5mm}\textcolor{white}{on}}
	 \colorbox[RGB]{38.0, 105.0, 250}{\rule[-.5mm]{0pt}{2.5mm}\textcolor{white}{since}}
	 \colorbox[RGB]{98.44052292429572, 146.77916331172514, 250}{\rule[-.5mm]{0pt}{2.5mm}\textcolor{white}{phil}}
	 \colorbox[RGB]{91.2696199577386, 141.82231794313728, 250}{\rule[-.5mm]{0pt}{2.5mm}\textcolor{white}{ando}}
	 \colorbox[RGB]{46.86739479291191, 111.12953557113727, 250}{\rule[-.5mm]{0pt}{2.5mm}\textcolor{white}{cast}}
	 \colorbox[RGB]{138.87969196792378, 174.7325059686109, 250}{\rule[-.5mm]{0pt}{2.5mm}\textcolor{black}{ille}}
	 \colorbox[RGB]{155.07855487289393, 185.92987664024926, 250}{\rule[-.5mm]{0pt}{2.5mm}\textcolor{black}{was}}
	 \colorbox[RGB]{255, 114.46810511092613, 157.44066801089087}{\rule[-.5mm]{0pt}{2.5mm}\textcolor{black}{shot}} \\
    \caption{\textbf{FN: Implicit} Perspective on HateXplain}
        \label{fig:FN-MP-3}
    \end{subfigure}
    \vspace{-0.4cm}
    \caption{FN examples: SHAP value visualizations for examples from the ToxiGen and HateXplain datasets. Red indicates a strong contribution to deciding hate speech; blue indicates a strong contribution to deciding non-hate speech. For visualization, we added some tokens together and averaged contribution of both.}
    \label{fig:}
\end{figure*}

\end{document}